\documentclass{aa}

\usepackage{epsfig,amsmath,amssymb}

\def\ga{\,\hbox{\hbox{$ > $}\kern -0.8em \lower 1.0ex\hbox{$\sim$}}\,}
\def\la{\,\hbox{\hbox{$ < $}\kern -0.8em \lower 1.0ex\hbox{$\sim$}}\,}
\def\beq{\begin{equation}}
\def\eeq{\end{equation}}

\titlerunning{On the origin of filaments in the ISM}
\authorrunning{Hennebelle}

\begin{document}

\title{On the origin of non self-gravitating filaments in the ISM}

\author{Patrick Hennebelle\inst{1,2}}

\institute{
Laboratoire AIM, 
Paris-Saclay, CEA/IRFU/SAp - CNRS - Universit\'e Paris Diderot, 91191, 
Gif-sur-Yvette Cedex, France \\
\and
LERMA (UMR CNRS 8112), Ecole Normale Sup\'erieure, 75231 Paris Cedex, France}

\abstract{Filaments are ubiquitous in the interstellar medium as recently emphasized 
by Herschel, yet their physical origin remains elusive}
{It is therefore important to understand the physics of molecular 
clouds to investigate how filaments form and what is the  
role played by various processes such as turbulence and magnetic field.}
{We use  ideal MHD 
simulations to study the formation of clumps in various conditions
including different magnetization and Mach numbers as well as two 
completely different setup. We then perform several analysis to 
compute the shape of the clumps and their link to velocities and forces
 using various  approaches.}
{We find that on average, clumps in MHD simulations are more filamentary 
that clumps in hydrodynamical simulations. Detailed analyses reveal that the 
filaments are in general preferentially aligned with the strain
which  means that these structures simply  result from the strech induced by  turbulence. 
Moreover filaments  tend to be confined by the Lorentz force which therefore 
lead them to survive longer in magnetized flows. We show that they
have in all simulations a typical thickness equal to a few grid cells
suggesting that they  are primarily associated to the  energy dissipation 
within the flow. We estimate the order of magnitude of 
the dissipation length associated to the ion-neutral friction and
 conclude that in well UV shielded regions it is of the order of 0.1 pc
and therefore could possibly set the typical size of non self-gravitating
filaments.} 
{Filaments are ubiquitous 
because they are the results of the very generic turbulent strain
and because magnetic field help to keep them coherent. 
We suggest that energy dissipation is playing a determinant role in their 
formation.}

\keywords{Turbulence - Stars: formation - Interstellar medium:structure -
Magnetic field}

\maketitle

\section{Introduction}

For more than three decades,
  the evidences for the filamentary structure of the 
molecular clouds seen through their CO
emission has become  clear (e.g. Ungerechts \&
Thaddeus 1987, Bally et al. 1987).  The far-IR all-sky
IRAS survey (Low et al. 1984) also revealed the ubiquitous filamentary
structure of the ISM, and discovered the cirrus clouds, i.e. the
filamentary structure of the diffuse interstellar medium (ISM).  

Thanks to its unprecedented sensitivity and large-scale mapping
capabilities, Herschel has now provided a unique view of these
filamentary structures of cold dust (e.g. Miville-Desch\^enes et al.
2010, Ward-Thompson et al. 2010, Andr\'e et al. 2010). One of the main
and intriguing findings  is the very large range of column
densities -- a factor of 100 between the most tenuous ($N_{H_2} =2
\times 10^{20}$ cm$^{-2}$) and most opaque ($N_{H_2} \sim 10^{23}$ 
cm$^{-2}$) of
the observed filaments in several fields 
contrasting with a narrow range of filament thickness (between 0.03 and
0.2 pc) barely increasing with the central column density (Arzoumanian
et al. 2011). The present study is largely motivated by this
result though as described below is too premature to 
directly address it. 

While the  filamentary nature of molecular cloud 
is probably linked to their turbulence, the exact 
mechanism by which this happens remains to be understood.
In many published numerical simulations, filaments 
are clearly present (e.g. de Avillez \& Breitschwerdt 2005, 
Heitsch et al. 2005,  Joung \& MacLow 2006, 
Padoan et al. 2007, Hennebelle et al. 2008,  Price \& Bate 2008, 
Banerjee et al. 2009, Inoue et al. 2009, 
Nakamura \& Li 2008 , Seifreid et al. 2011 , V\'azquez-Semadeni et al. 2011,  Federrath \& Klessen 2013) but 
again the exact reason of their formation mechanism is not very 
clearly analyzed.  Based on the evidence for 
higher velocities  in the outer part of the filaments, 
Padoan et al. (2001) proposed that they form through the collision 
of two shocked sheets. 
Another type of explanation invokes instabilities in self-gravitating sheets
(e.g. Nagai et al. 1998). 
Although it is clear that since filaments are denser
than their environment some compression must necessarily occurs, it is
important to understand in greater details the conditions in which 
the filament formation happens. 
In particular an important question is the origin of 
the elongation. Is the elongation produced by the contraction 
along two directions like what would happen in a shock  or in a converging flow ? Or is 
the elongation the result of the stretching of the fluid particles along 
one direction ? 
The purpose of this paper is to investigate these questions. 

The paper is organized as follows. In the second part we 
describe the analysis that will be performed to  
the clumps and the filaments  formed in the numerical simulations.
We also describe the various runs that we  carry out to 
understand the filament origin. In the third section, we 
present a simple but enlighting preliminary numerical 
experiment that clearly demonstrates that the mechanism 
we propose can actually work. In the fourth section we show
the numerical simulations and present the clump statistics 
such as their aspect ratio, length and thickness distributions.
In the fifth section we analyze the link between the filament
orientation, the velocity field and the forces. The sixth 
section provides further discussion regarding the filament thickness
as well as simple orders of magnitude which suggest that ion-neutral friction 
could possibly explain their thickness.  The seventh section concludes the paper.

\section{General analysis and Numerical simulations}

\subsection{Structure analysis}
 In the present paper we perform numerical simulations relevant for 
the diffuse  and moderately dense ISM. The mean density 
we consider goes from 5 cm$^{-3}$ to about 100 cm$^{-3}$ while the 
largest densities is larger than a few 10$^3$ cm$^{-3}$.

\subsubsection{Definitions}
The first difficulty when discussing clouds is actually 
to define them. 
To identify structures in this work, we follow
a simple approach. The first step is to choose
a density threshold (two of them will be used through the 
paper namely $n_{thres}=50$ and 200 cm$^{-3}$) and to clip 
the density field. Then using a friend of friend analysis, 
the connected cells are associated to form a clump. Once 
this is done, the structure properties can easily be computed.
The structures obtained through this process are called 
clumps.  In many occasions we will also refer to filaments. 
Filaments as clumps are not well identified objects
and although it could be easy to adopt an arbitary criterion like
an aspect ratio above some values, that would not bring 
much information. When we use the word {\it filament}
it will simply refer to a clump that is sufficiently elongated  say by 
a factor of about 5 or more.

\subsubsection{Inertia Matrix}
\label{inertia_mat}
To estimate the shape of a structure, it is convenient to 
compute the Inertia Matrix and its eigenvalues and eigenvectors.
The inertia matrix is a three by three matrix defined as:
$I_{ij}= \int x_i x_j dm$ where $x_i$ are the coordinates 
of the cells belonging to the structure associated to its
center of mass. The three eigenvectors give the three main
axis of the structure which  correspond to symmetry axis
if the structure admits some while the eigenvalues represent 
the moment of inertia of the structure with respect to 
the three eigenvectors. From the inertia momentum, $I_i$, 
we can get an estimate of the structure size $\mu _i = \sqrt{I_i/M}$
where $M$ is the structure mass. 
In the case of filaments 
for example, the eigenvector associated with the largest 
eigenvalue tends to be aligned with the main axis of the filament while 
the two other eigenvectors tend to be perpendicular to the filament 
axis. In the following, we will refer to the eigenvector associated
to the largest eigenvalue as the main axis. We will also quantify 
the aspect ratio of structures by computing the ratio of eigenvalues.
We will in particular consider $\mu _1/\mu _3$ and 
$\mu _2/\mu _3$, the ratios of the smallest over largest 
structure size and the intermediate over largest respectively.

Note that one of the difficulties with this approach is that 
a thin but curved filaments will have moments of inertia 
that reflect the curvature rather than the effective thickness.
For this reason, we will use in the paper a second method to 
characterize their shape.

\subsection{Strain tensor}
\label{strain}
The strain tensor is another useful quantity that we will use to 
perform our analysis. It is obtained by considering 
the velocity difference between two fluid particles
located in $\bf{r}$ and $\bf{r+dr}$.
One gets $v_i(\bf{r+dr})-v_i(\bf{r})= \partial_j v_i dr_j$
where summation over repeating indices is used.
The three by three matrix, $\partial_j v_i$, can be 
splitted in its anti-symmetric part 
$A_{i,j}=(\partial_j v_i - \partial_i v_j)/2$, which described
the rotation of the fluid element  
and its symmetric part $S_{i,j}=(\partial_j v_i + \partial_i v_j)/2$, 
which describes the shape modification of the fluid element and is 
called the infinitesimal strain tensor.
The trace $S_{i,i}$, which is equal to the divergence of  ${\bf v}$,
describes the change of  volume. The symmetric part can 
be diagolised leading to three eigenvalues, $s_i$, 
where we assume that $s_3 > s_2 > s_1$. 
The  eigenvector associated to the largest eigenvalue, $s_3$,  
describes the axis along which the fluid particle is mostly elongated, below
we call it the strain. Note 
that in principle since the divergence of the fluid is non zero, all
eigenvalues could be negative which would then correspond to a global 
contraction. In practice, in our simulations this is almost never the case
at the scale of the clumps. The two other eigenvectors  associated to 
the two other eigenvalues correspond to the directions along which 
the shape of the fluid particle is either stretched or compressed 
depending on their signs. 

Note that computing the strain tensor is not straightforward since it requires 
to take the velocity gradients between cells. Moreover, because of the 
numerical diffusion, the gradients at the scale of the mesh are 
artificially smoothed. To limit this problem, we smooth the simulation
by a factor three computing the mean density weighted velocity within
the smoothed cells. Then we compute all velocity gradients using 
simple finite differences and we compute the mean gradients by 
summing over all cells which belong to the structure. Finally, we
use these values to compute $S_{ij}$.

\subsubsection{A simple skeleton-like approach}
\label{skeleton}

\setlength{\unitlength}{1cm}
\begin{figure}[h!]
\includegraphics[width=7.5cm]{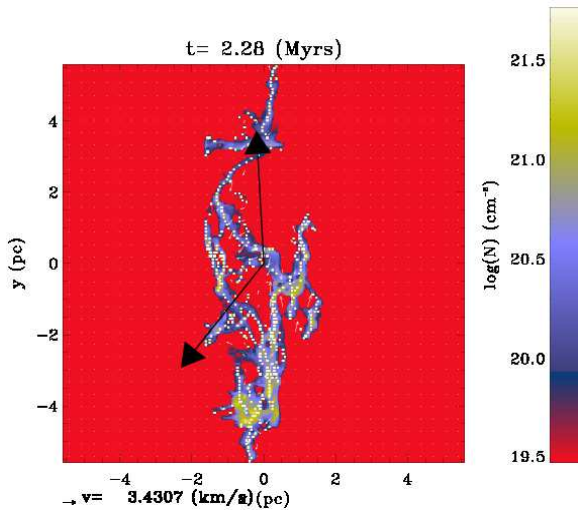}
\caption{Column density and velocity 
field for one of the clumps formed in the MHD simulations. The 
white points represent the local mass center $G_{i^j}$ and 
constitute the skeleton of the clump. The upwarding arrow represents the 
filament main axis as computed with the inertia matrix. The downwarding 
arrow represents the main direction of the strain as computed 
with the strain tensor.}
\label{fil_example}
\end{figure}

Another useful approach is to determine an average 
 line characteristic of the clump
shape. Such algorithm has been developed in the context 
of Cosmology to reconstruct the filaments leading 
to the so-called skeleton (Sousbie et al. 2009). Here 
we follow a simpler approach which is well suited for our analysis. 
The first step is to 
select the direction ($x$, $y$ or $z$) along which 
the structure is the longest. Then, we can subdivide it 
into a number of slice, $n_{sl}$, of a given length in 
the selected direction. Each slice, $i$, can itself be divided
into $n_{sr}$ connect regions, that is to say regions in which all 
cells are connected to each other through their neighbours. 
Each of these sub-regions, $j$, belongs to a different 
branch within the structure.
Then the center of mass, $G_i^j$ of each of these 
sub-regions within each slice, 
can be computed leading to an ensemble of points 
$G_i^j$. For each of them we look for the nearest neighbour, ${\cal G}_i^j$
of $G_i^j$
belonging to the slices $i-1$ or $i+1$. Finally, we calculate 
the vector 
${\bf u_i^j}= \pm {\bf G_i^j{\cal G}_i^j} / G_i^j{\cal G}_i^j$ 
which 
gives the local direction of the branch to which $G_i^j$ belongs.
The sign is then chosen in order to insure ${\bf u_i^j.X}>0$
where $X$ represents the selected axis, $x$, $y$ or $z$. 
Connecting all $G_i^j$ to their respective neighbours, we obtain
  a curve that represents the mean local direction. 
Figure~\ref{fil_example} shows 
an example of a clump extracted from the fiducial 
MHD simulation presented below. The white cells represent the 
position of the $G_i^j$, as can be seen they follow well 
each branch of the clump. The two arrows represent 
the clump main axis (computed with the inertia matrix as explained 
above) and the strain (computed with the strain tensor). Note that the clump is quite filamentary and that the 
main axis represented by the vertical arrow follows well the filament direction.

Once the ${\bf u_i^j}$ are known, it is an easy task to estimate the 
distance, $r_M$, from a given  filament cell to the skeleton.
Since  any cell belonging to the structure, is associated to a 
sub-region, one can calculate
$r_M= | {\bf G_i^j M} \times {\bf u_i^j} |$, where $M$ is the cell center. 
In the paper, we make use of the vectors ${\bf u_i^j}$ to 
compute various quantities as the mean component of various 
forces. 
 The mean  radius 
is then defined as $r_c=\Sigma r_M dm / \Sigma dm  $ where $dm$ is the mass
within the cell and where the sum is taken over all clump cells.
We stress that the thickness, $r_c$ calculated by this definition
is really measuring the thickness of the clump substructures and not the mean 
size of the clumps which would take into account the distance between the 
various branches. This is particularly clear in 
Fig.~\ref{fil_example} where the distance between the branches is of the order
of a few pc while in most regions of the clump, the size of the 
individual branches is typically  ten times smaller.

\subsection{Description of numerical simulations}

\subsubsection{Code and resolution}
We use the Ramses code (Teyssier 2002, Fromang et al. 2006) to 
perform the simulations. Ramses uses adaptive mesh refinement 
and solve the  ideal MHD equations using the Riemann HLLD solver (Miyoshi \& Kuzano 2005).
It uses constraint transport method to ensure that 
${\rm div} {\bf B}$ is maintained to zero. 
The cooling corresponds to the standard ISM cooling (e.g. Wolfire et al. 
1995) as described in Audit \& Hennebelle (2005) and includes
Lyman $\alpha$, $C^+$ and $O$ lines. The heating is due to the 
photo-electric effect on small dust grains.   
Here we carry simulations which have a based grid of 512$^3$ cells. 
Then depending on the simulations we either add another 
or two other AMR levels leading to an effective resolution in the 
most refined areas of $1024^3$ to $2048^3$ cells. 
The refinement criteria is based on density. Any cell which 
has a density larger than 50 cm$^{-3}$ is refined to resolution
1024$^3$ and when it is allowed, cells denser than 200$^{-3}$ are 
refined to resolution 2048$^3$.  The high resolution run allows
us to check for numerical convergence and to verify by comparison with the 
coarser runs that the AMR does not introduce any major bias.

\subsubsection{Initial conditions}
\label{initial}
Our initial conditions for the fiducial simulations 
consist in a uniform medium in density, 
temperature and magnetic field on which a turbulent velocity field 
has been super-imposed. The latter is generated using random phases and 
presents a powerspectrum of $k^{-5/3}$. No forcing 
is applied and the turbulence is therefore decaying.
The initial density is equal to $5$ cm$^{-3}$
and the initial temperature $T=1600$ K leading to a pressure of 
8000 K cm$^{-3}$ typical of the ISM. The magnetic field is 
initially aligned along the x-axis and has an intensity of about 5$\mu$G 
(or 0 in the hydrodynamical case). The initial rms velocity 
is equal to 10 km s$^{-1}$.
Since these simulations have no energy injection, the turbulence
is decaying in a few crossing times which is thus an important quantity 
to estimate.  It is however not straightforward since it is evolving 
with time. The total velocity to be considered is the sum of the rms 
velocity and the wave velocity (sound and Alfv\'en waves). Initially
both are of the order of 3-4 km s$^{-1}$ but they raise to about 
10 km s$^{-1}$ in the diffuse gas when the gas breaks up into warm and 
cold phase leading to a total velocity of the order of 20 km s$^{-1}$. 
Thus we estimate the crossing time to be of the order of 2-3 Myr.
It is worth stressing that the crossing time at the scale of the clumps 
is obviously much shorter thus it is probably the case that their properties 
are setup much quicker than a box crossing time.
 
We have performed several runs. The run that we consider as being 
fiducial has an effective resolution of $1024^3$ cells and is magnetized. 
The initial velocity dispersion is 10 km s$^{-1}$ which corresponds to 
a typical Mach number with respect to the cold gas of about ${\cal M}=10$
since its sound speed is about 1 km s$^{-1}$.
To investigate the effect that the magnetic field has on the 
medium structure, we have performed an hydrodynamical run at the same 
resolution.
Next we have explored the influence of the Mach number ${\cal M}$ by 
dividing the initial velocity amplitude by 3 and then by 10. We refer 
to these two runs as Mach ${\cal M=}$ 3 and 1 respectively keeping 
in mind that this corresponds to the initial rms velocity.
Then to investigate the influence of the resolution, we have repeated the 
fiducial run (magnetized and ${\cal M}=10$) with an effective resolution of 
$2048^3$ cells. Note that in order to compare well this simulation with the fiducial run, 
we have identified the clumps at the {\it same} resolution which means that cells having an effective resolution of $2048^3$
have been smoothed before performing the analysis.
Below the results are  given for these 5 simulations. To show that they 
do not strongly depend on time evolution, we also present all statistics at 
two different timesteps of the hydrodynamical run, one after about 1/2-1  crossing times and 
one at about 1.5-2.  In order to verify that no spurious effect has been introduced by 
the magnetic field being initially aligned with the mesh, we have repeated the simulation 
with 5 $\mu$G and ${\cal M}=10$ but tilting the  initial magnetic field with respect to the mesh
from $45^\circ$. The corresponding result is shown in appendix~\ref{tilt}, no significant difference with the aligned 
case is seen.

Finally, in order to verify the robustness of our results, we have also
used another very different type of setup, namely  converging flow
type simulations which include self-gravity. These simulations 
are very similar to the ones presented in 
Hennebelle et al. (2008) and in Klessen \& Hennebelle (2010). 
They consist in imposing from the x-boundaries two streams 
of warm neutral medium having velocities of about 
$\pm 20$ km/s, density of 1 cm$^{-3}$ and temperature of
8000 K. The magnetic field is initially uniform and oriented along the x-axis. 
Unlike the decaying simulations, no
velocity field is initially imposed in the computational box.
 Moreover the turbulence which develops
is sustained by the energy due to the incoming flow. Also the 
mean density is typically ten times higher in the colliding flow 
simulations. 
Four simulations of this type have been performed.
Three simulations have an effective resolution of 1024$^3$ cells
 amongst which one is hydrodynamical, 
one has an initial magnetic field of 2.5 $\mu$G and one has 
5 $\mu$G.  The fourth one is identical to the intermediate resolution
simulation  with 2.5 $\mu$G  but has an effective resolution of $4096^3$ cells.
In spite of these important differences between the decaying  
and colliding simulations, 
the conclusions we inferred remain unchanged. All the trends 
which are inferred in the decaying runs are recovered in the 
colliding flow runs. Therefore, for the sake 
of conciseness we   present 
the corresponding results in the appendix~\ref{conv}.

\section{A simple preliminary numerical experiment}
\label{prelimi}

\setlength{\unitlength}{1cm}
\begin{figure}[h!] 
\begin{picture} (0,21)
\put(0,14){\includegraphics[width=8cm]{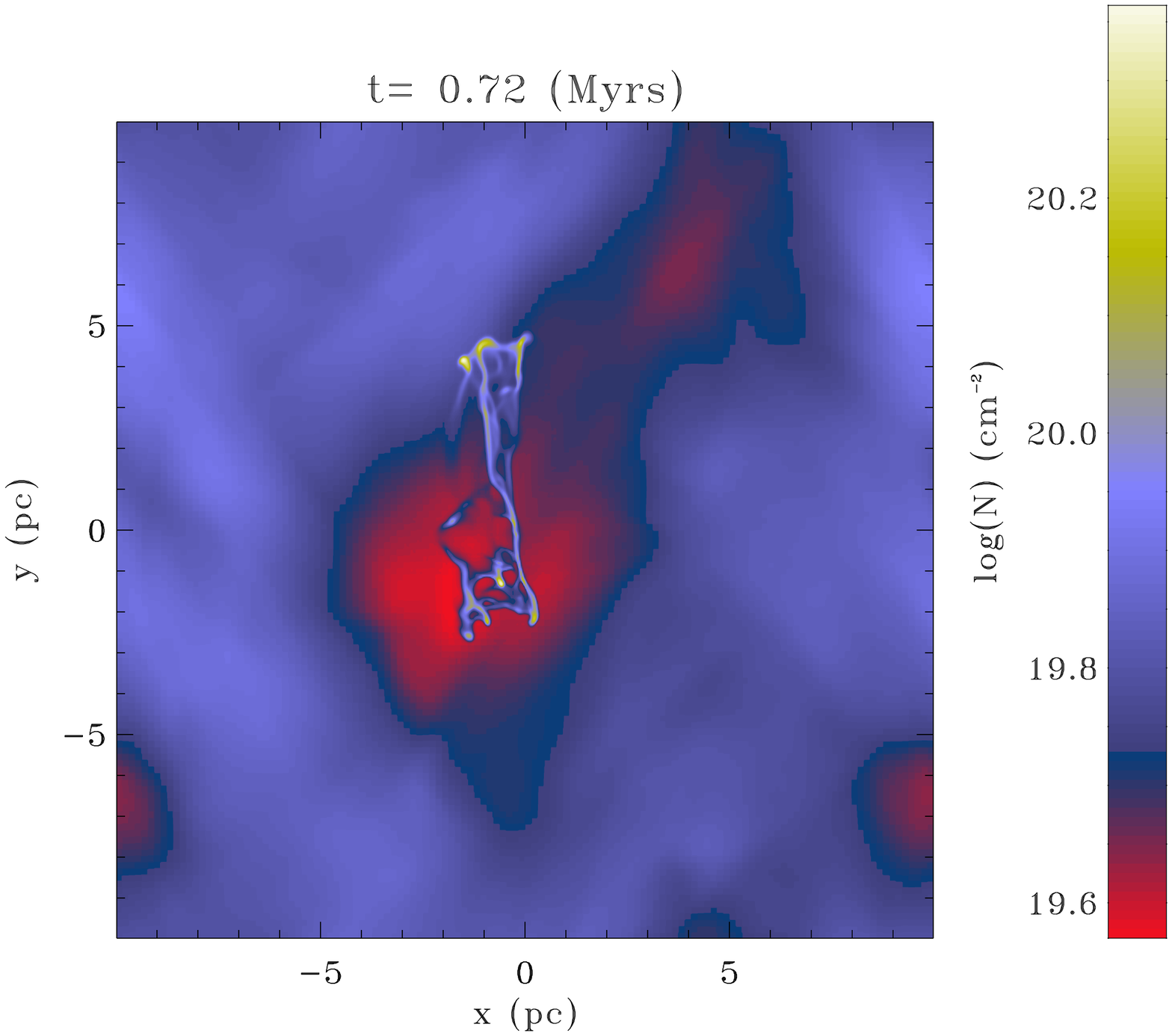}}
\put(0,7){\includegraphics[width=8cm]{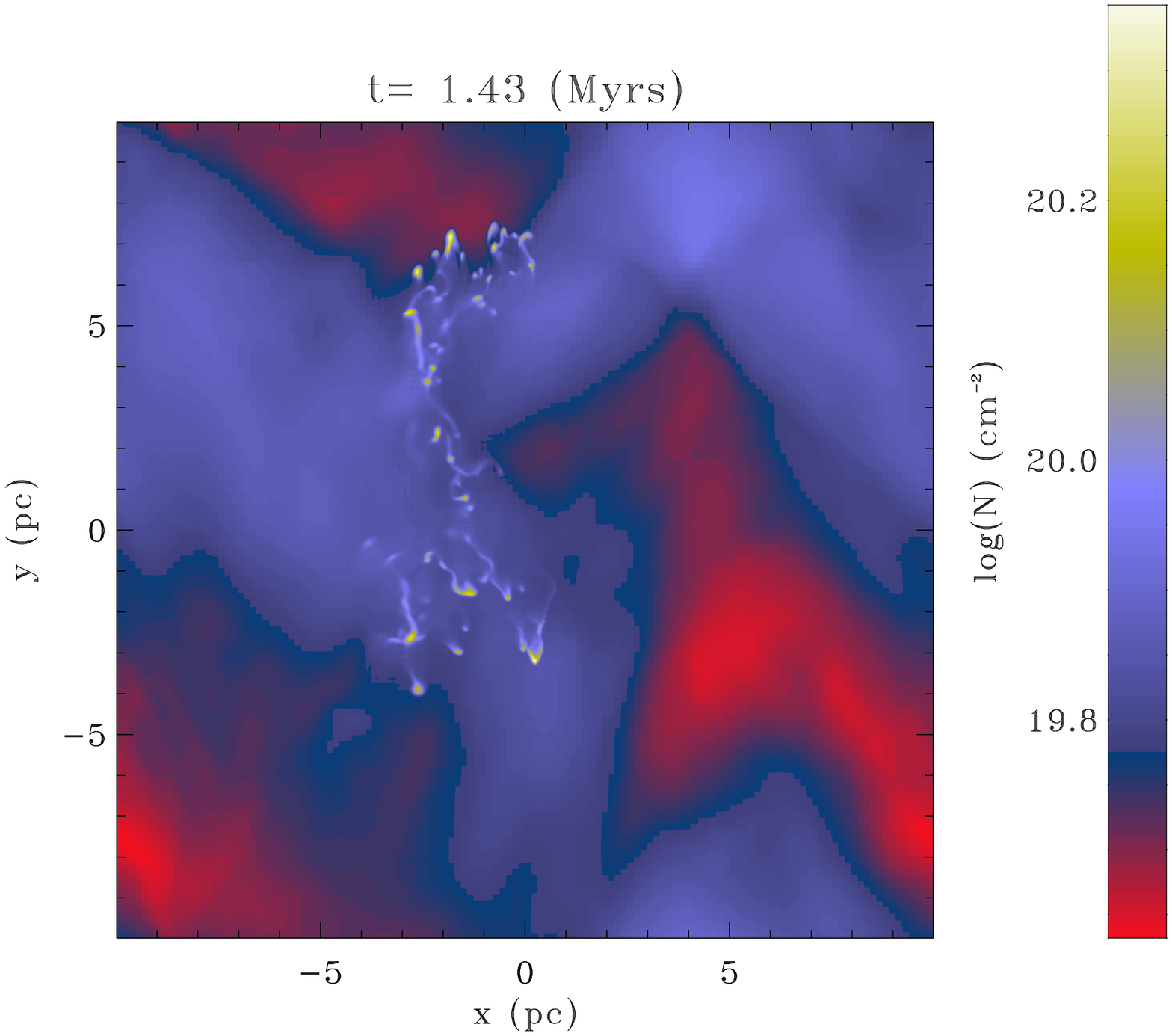}}
\put(0,0){\includegraphics[width=8cm]{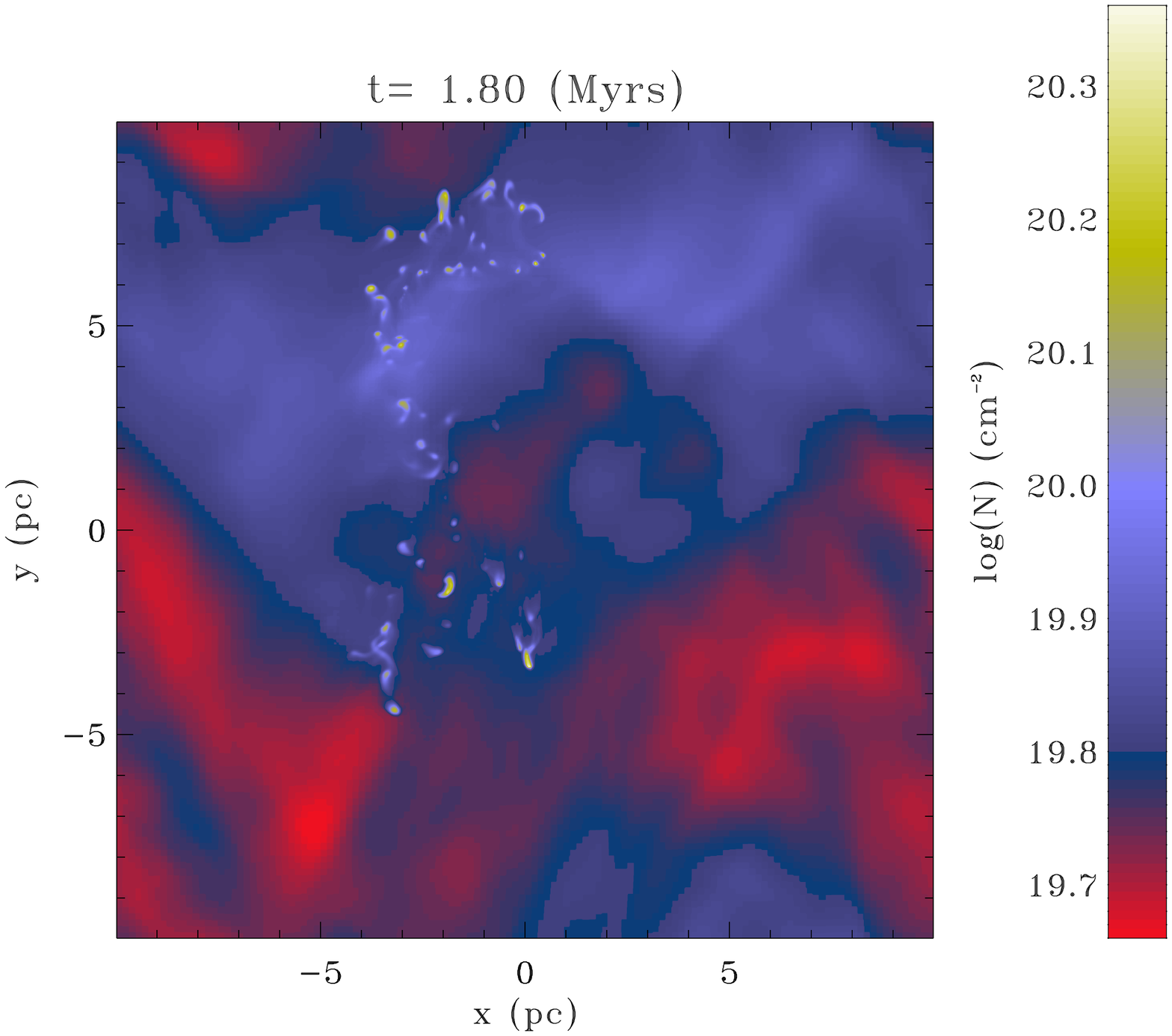}}
\end{picture}
\caption{Formation of a filament from a spherical 
cloud in the presence of shear, $v_y(x)$. 
The column density for three snapshots is displayed. Hydrodynamical case.
The filament quickly fragments in many cloudlets.}
\label{simpl_fil_hy}
\end{figure}

\setlength{\unitlength}{1cm}
\begin{figure}[h!] 
\begin{picture} (0,21)
\put(0,14){\includegraphics[width=8cm]{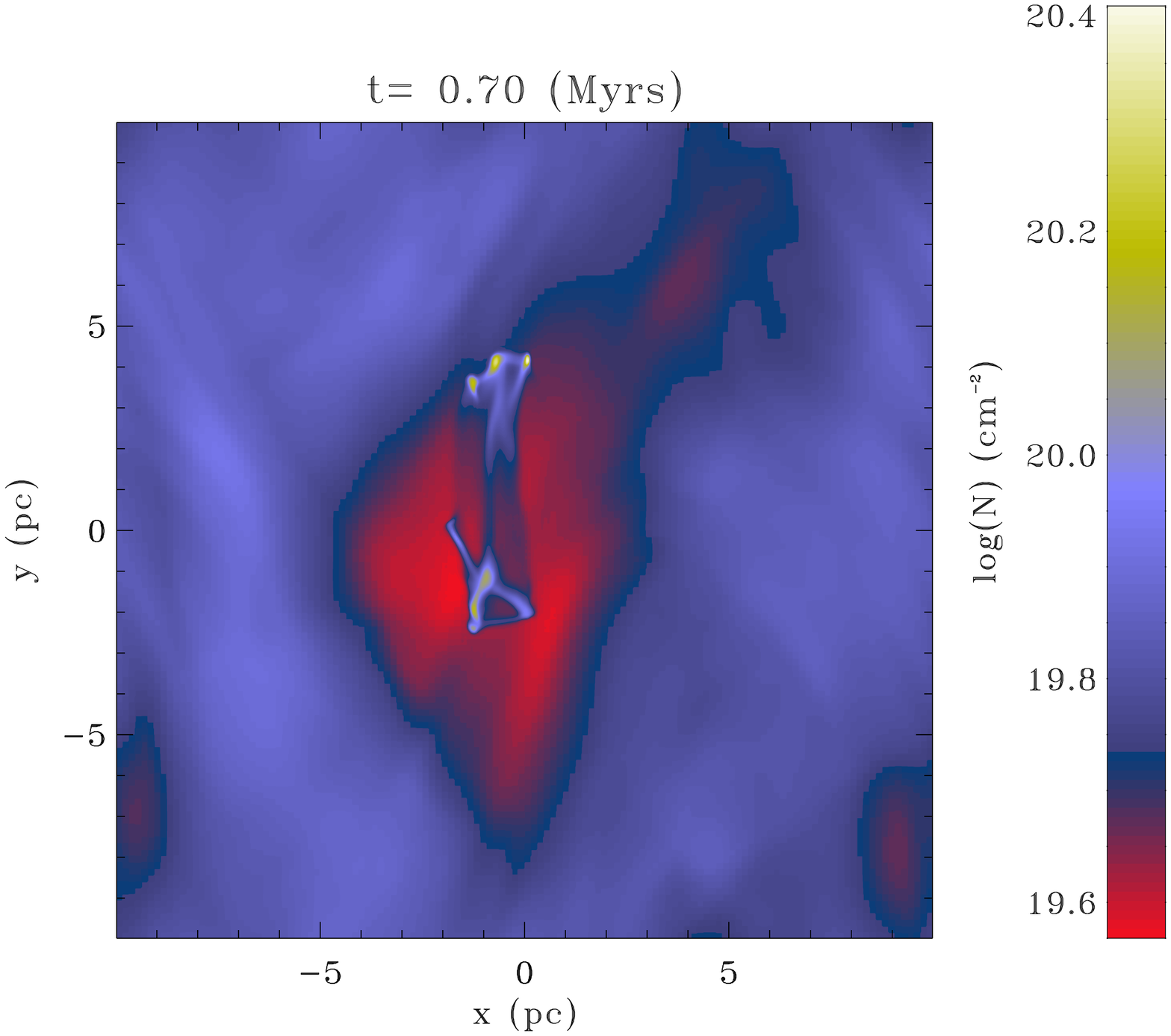}}
\put(0,7){\includegraphics[width=8cm]{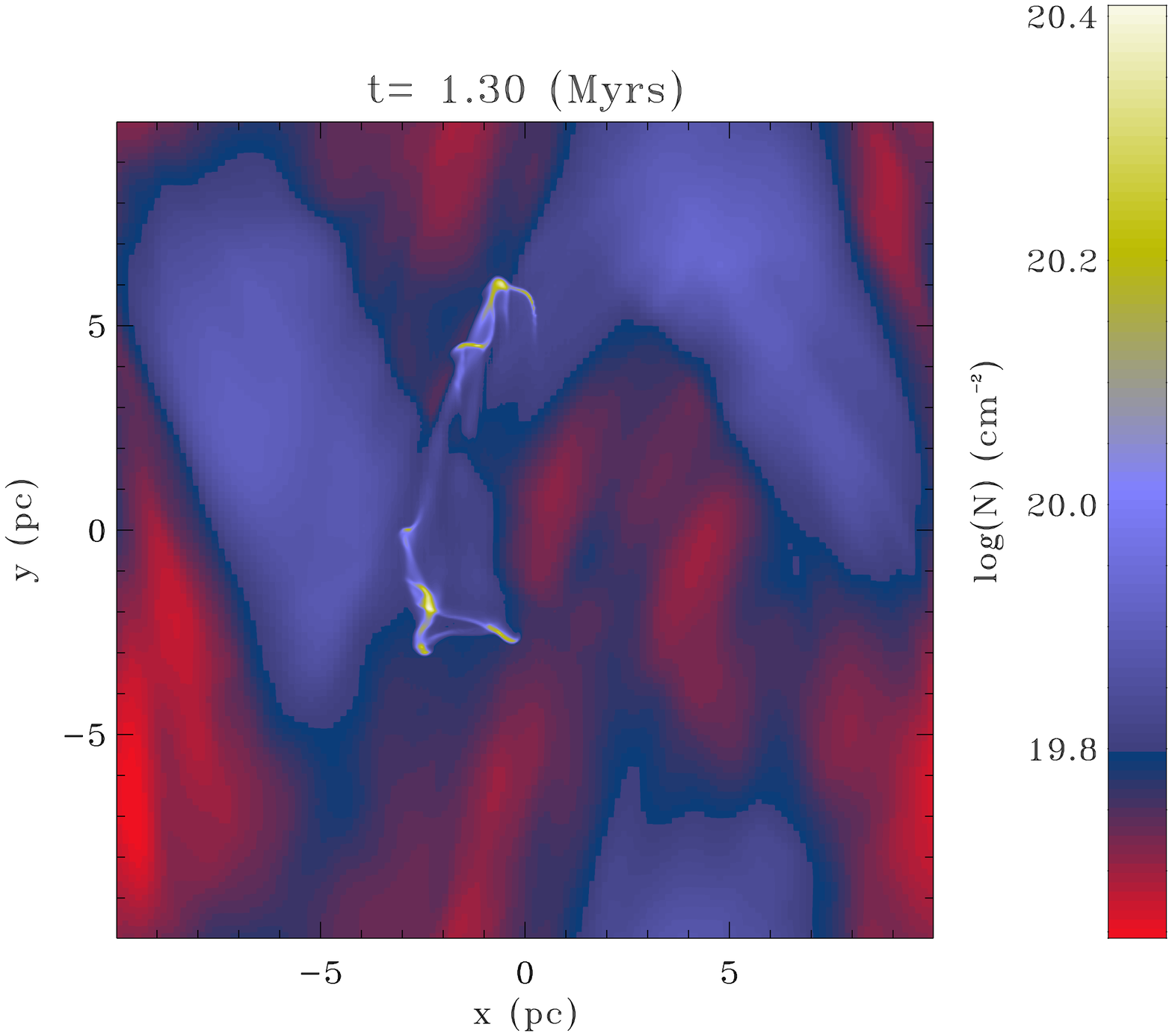}}
\put(0,0){\includegraphics[width=8cm]{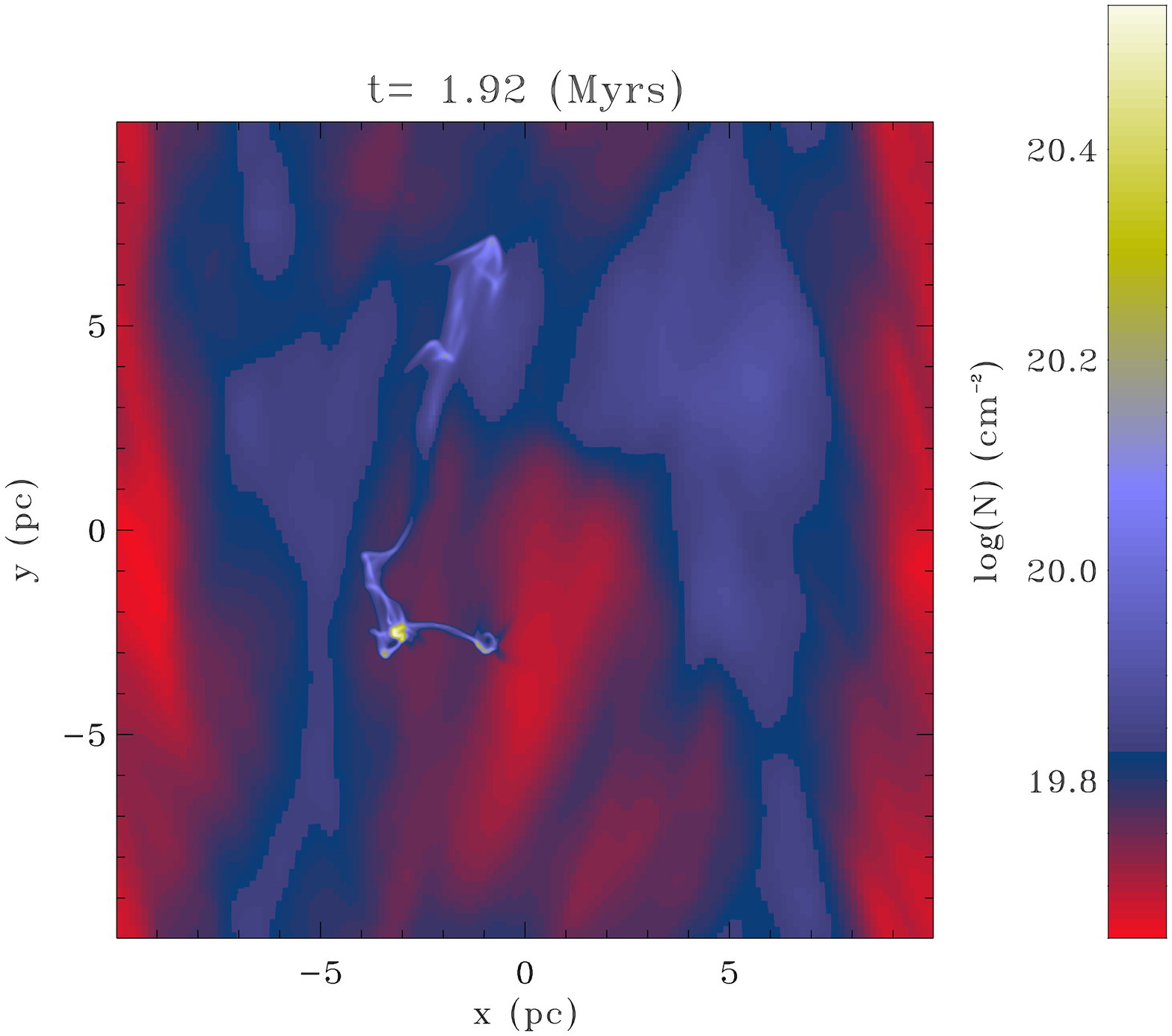}}
\end{picture}
\caption{Same as Fig.~\ref{simpl_fil_hy} in the MHD case.
A weak magnetic field (1 $\mu G$) along the x-axis is permeating
the cloud initially. The filament remains much more coherent
than in the hydrodynamical case.}
\label{simpl_fil_mhd}
\end{figure}

Before proceeding to the  complex turbulent simulations, we 
present two simple numerical simulations that illustrate 
some of the conclusions that will be drawn later. It consists 
in a spherical cloud which has a strong shear initially
 and therefore is prone to form a filament. More precisely,  
the spherical cloud of density 100 cm$^{-3}$, temperature 100 K 
and radius 
0.5 pc is {placed in the middle of the computational box}
and is  embedded into a diffuse  and warm medium 
of density 1 cm$^{-3}$ and temperature 8000 K. The total box size is 20 pc.
A transverse velocity gradient along the x-axis of 
1.5 km s$^{-1}$ pc$^{-1}$ is initially imprinted through the box.   
Finally, a turbulent velocity field having a total rms dispersion of 
5 km s$^{-1}$ is super-imposed in the box. The reason of 
superimposing such velocity field is to create self-consistently 
perturbations that disturb the forming filament. 
Two such simulations have been performed, the first one is 
purely hydrodynamical while the second one has a magnetic field
of 1 $\mu$G, uniform initially and oriented along the x-axis
 therefore perpendicular to the initial main component of 
the velocity field.

Figure~\ref{simpl_fil_hy} shows  the column density for three snapshots of the hydrodynamical 
simulation.  
As it is clear from the figure, the  initially spherical cloud
is stretched and evolves in a filament because of 
the shear. In the same time the non-linear fluctuations induced by the 
surrounding medium perturb the cloud and likely trigger the growth 
of various instabilities (like Kelvin-Helmholtz).
The complex pattern displayed in the three snapshots is the result 
of the uniform shear and the turbulent fluctuations present in the surrounding medium.
 After 1.6 Myr the
third panel shows that the filament is totally destroyed and 
broken in many cloudlets. 

Figure~\ref{simpl_fil_mhd} shows the magnetized run. The early evolution of 
the dense cloud is similar initially. Due to the initial shear, 
a filament forms. However, the late evolution is quite different. 
The filament although subject to large fluctuations induced 
by the surrounding turbulent medium, remains much more 
coherent. This is because since the cloud is threated 
by a magnetic field initially, the pieces of fluid are connected to each
others through the field lines. Moreover, the shear that tends to form 
the filament, amplifies the magnetic field making its 
influence stronger. Indeed, as the filament is getting 
stretched, the magnetic field is amplified along the y-axis and becomes 
largely parallel to the filament at the end of the simulation. 
This behaviour is qualitatively in good agreement with the studies
of the development of the Kelvin-Helmholz instability which have been 
performed by various tems (e.g. Frank et al. 1996, Ryu et al. 2000). 
In these studies it is found that even weak magentic field significantly 
modify the evolution of flow making it much less unstable. Stronger 
fields, on the other hand, can completely stabilize the flow against 
this instability.

This simple experiment suggests a scenario for the formation 
of filaments. The gas is compressed by converging 
motions but in the same time, the fluid particle possesses
solenoidal modes inherited from the turbulent environment
that tends to stretch it. In the absence of magnetic field, 
the pieces of fluid  can easily move away from each other.
When a magnetic field is present, there are more tight to each other
and the filament remains coherent for longer times.

\section{Clump geometry}

\setlength{\unitlength}{1cm}
\begin{figure*}
\begin{picture} (0,25)
\put(0,0){\includegraphics[width=14cm]{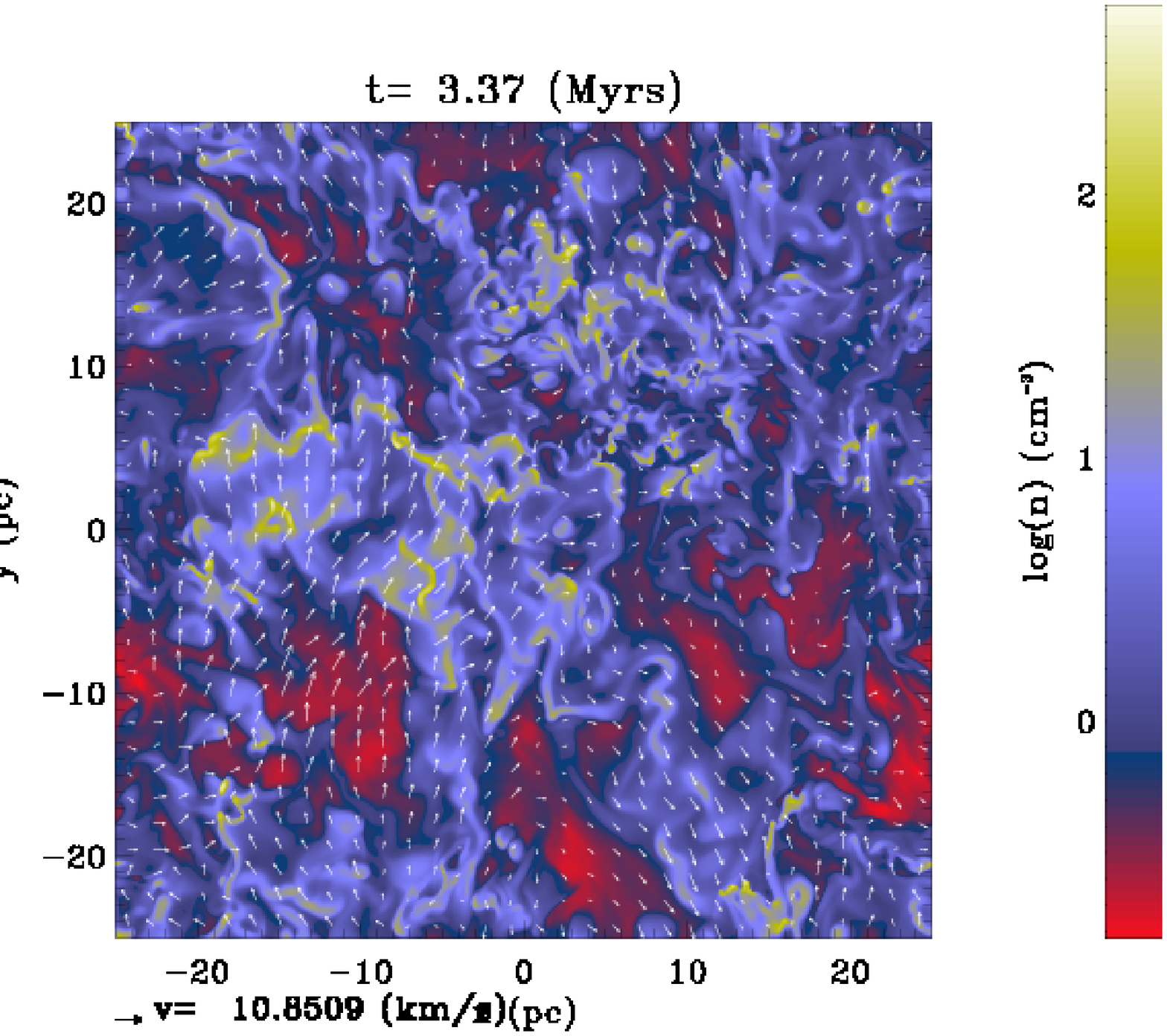}}
\put(0,12){\includegraphics[width=14cm]{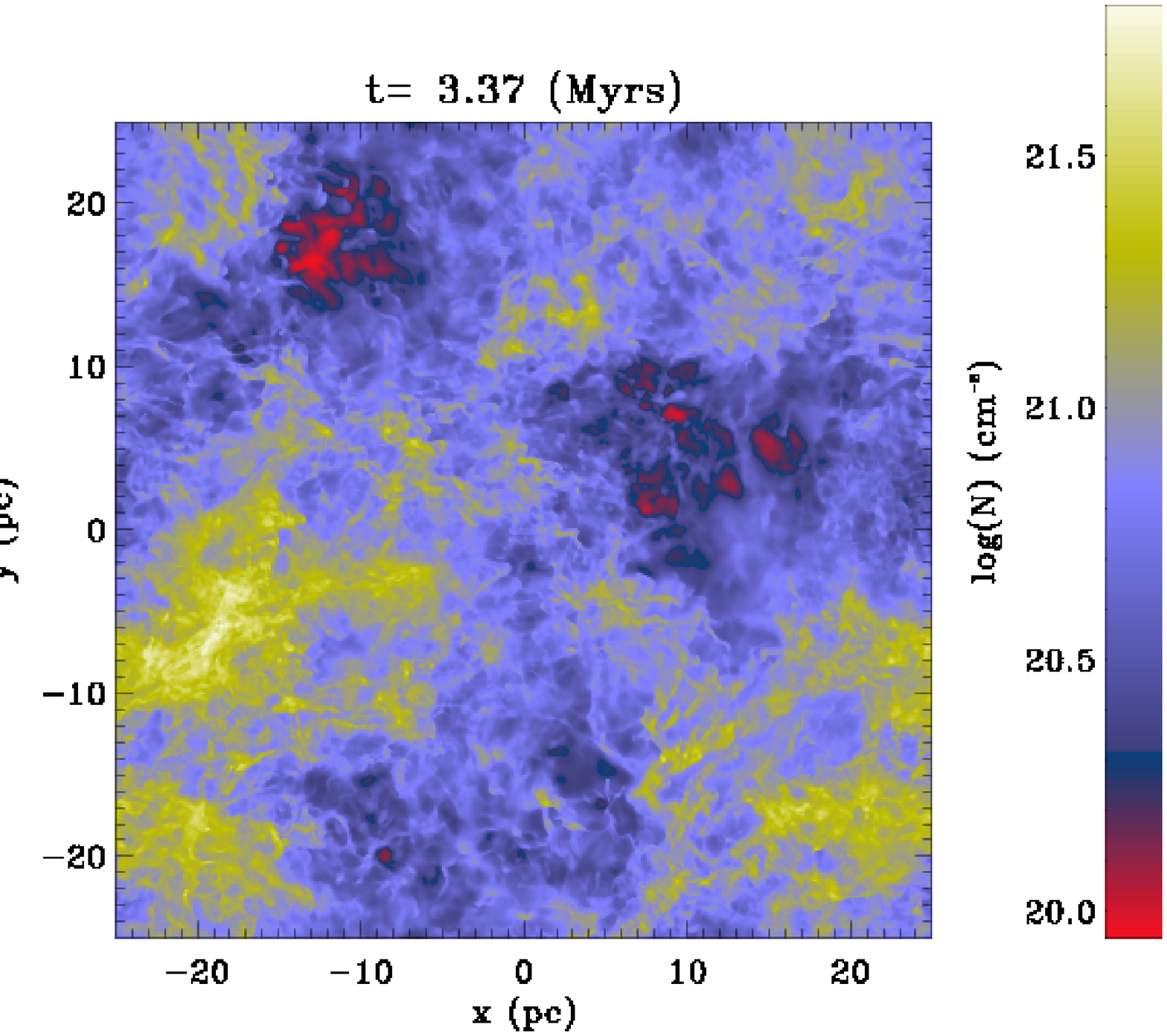}}
\end{picture}
\caption{Column density, density and velocity 
fields for one snapshot of the decaying
turbulence experiment in the hydrodynamical case.}
\label{decay_hydro}
\end{figure*}

\setlength{\unitlength}{1cm}
\begin{figure*} 
\begin{picture} (0,25)
\put(0,0){\includegraphics[width=14cm]{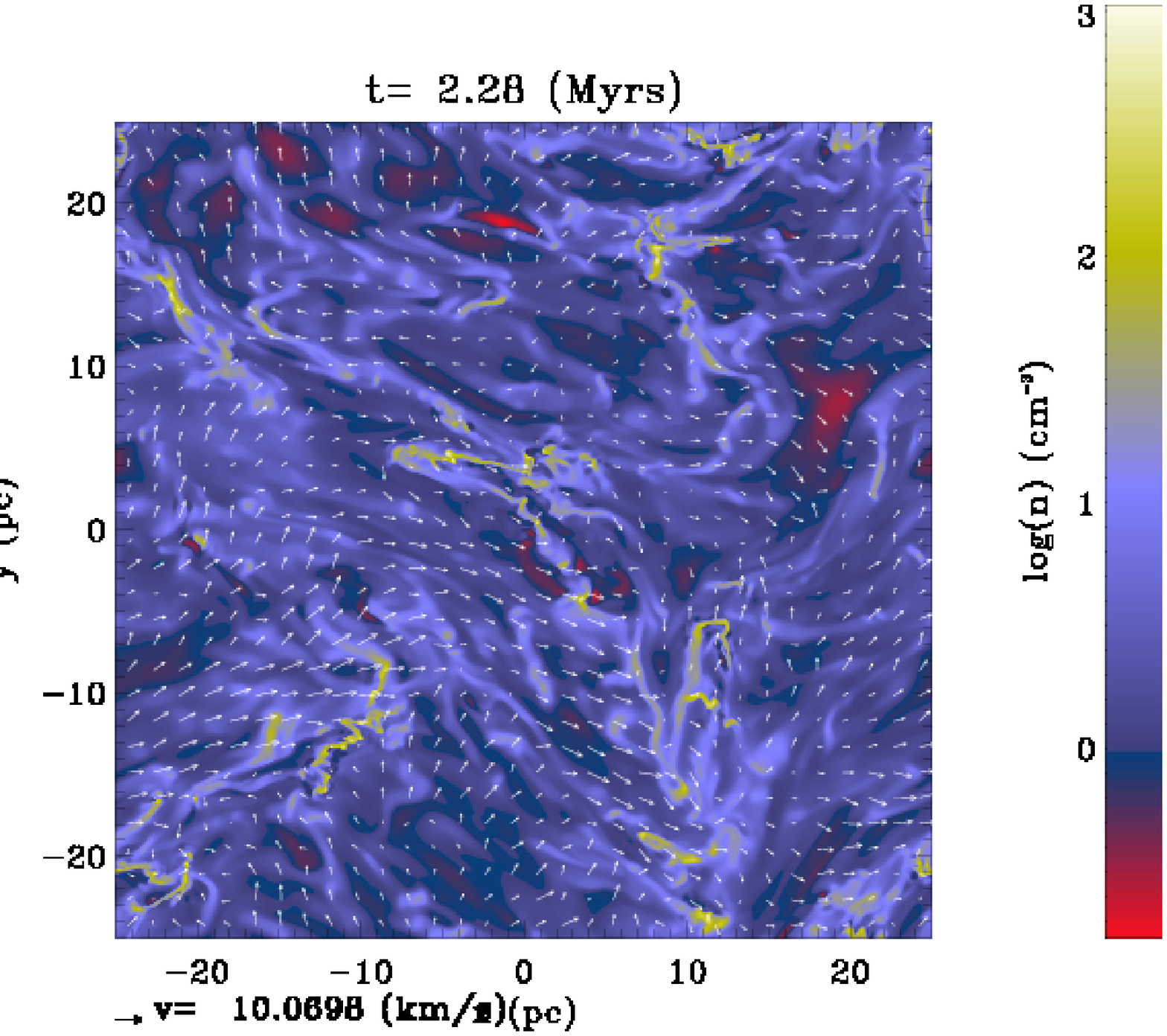}}
\put(0,12){\includegraphics[width=14cm]{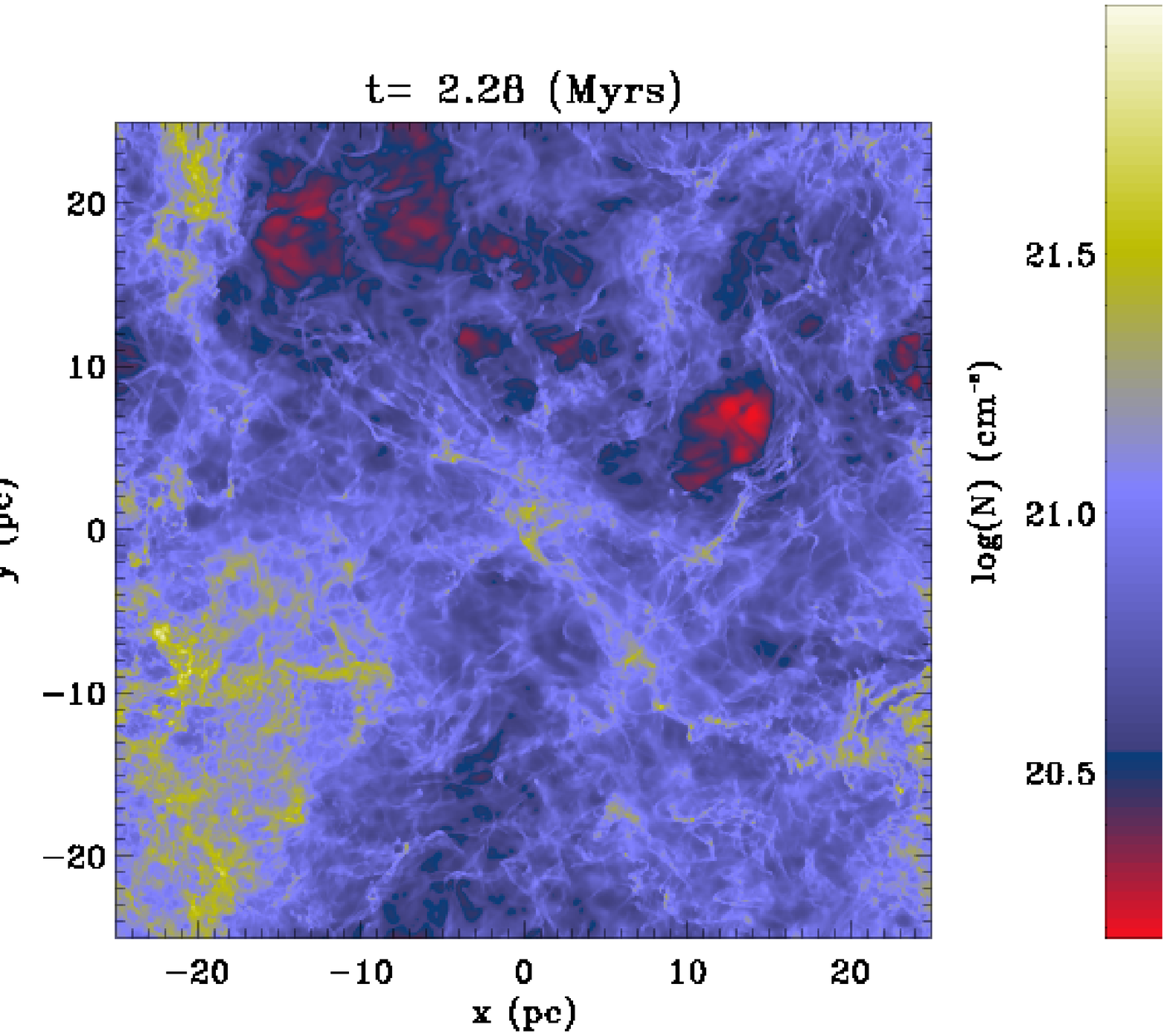}}
\end{picture}
\caption{Column density, density and velocity fields for one
 snapshot of the decaying
turbulence experiment in the MHD case.}
\label{decay_mhd}
\end{figure*}

\subsection{Qualitative description of decaying turbulence simulations}
Figures~\ref{decay_hydro} and~\ref{decay_mhd} show one snapshot
at roughly  one crossing time in the hydrodynamical and MHD cases.
Column density (top panels) and density together with velocity 
fields are shown (bottom panels). 
 The Column density is obtained by simple integration through the 
box and the density corresponds to the value in the $z=0$ plane.
As expected in both cases due to the large rms Mach numbers 
(about 3 initially to 1 in the WNM and 10
in the CNM), large density contrast develop partly because 
of the 2-phase structure and partly because of the 
supersonic motions.
The hydrodynamical and MHD cases present however obvious differences.
Overall the hydrodynamical case appears to be less filamentary 
than the MHD case in which very high aspect ratio structures
can be seen both in the column density and in the density. 
Some filamentary structures are also visible in the hydrodynamical
case but they have smaller aspect ratios. Moreover as seen from the 
density and velocity fields, it is often the case that the velocity
field is perpendicular to the elongated structure suggesting that 
shocks are triggering them. Indeed these structures are mainly sheets
as will be shown later.

To be complete, it should be said that at the beginning of our calculations
which we recall start with uniform density and a velocity field constructed
with ramdon phases, more high aspect ratio structures form in the hydrodynamical phase. 
However, this is a transient phenomenon due to our somehow arbitrary
initial conditions. These filaments quickly reexpand leading 
to the type of morphology seen in Fig.~\ref{decay_hydro}.
It is worth stressing that this visual impression of the MHD 
simulations being more filamentary is clearly visible in various 
other works (Padoan et al. 2007, Hennebelle et al. 2008, 
Federrath \& Klessen 2013). 

Beyond this visual impression, it is important to carefully quantify 
the aspect ratio which is the purpose of the following section.

\subsection{Axis ratio of clumps}
Here we attempt to quantify the clump aspect ratio using two 
different methods namely the inertia matrix and the skeleton
approach.

\subsubsection{Aspect ratio from inertia matrix}

\setlength{\unitlength}{1cm}
\begin{figure} 
\begin{picture} (0,8.5)
\put(0,0){\includegraphics[width=8cm]{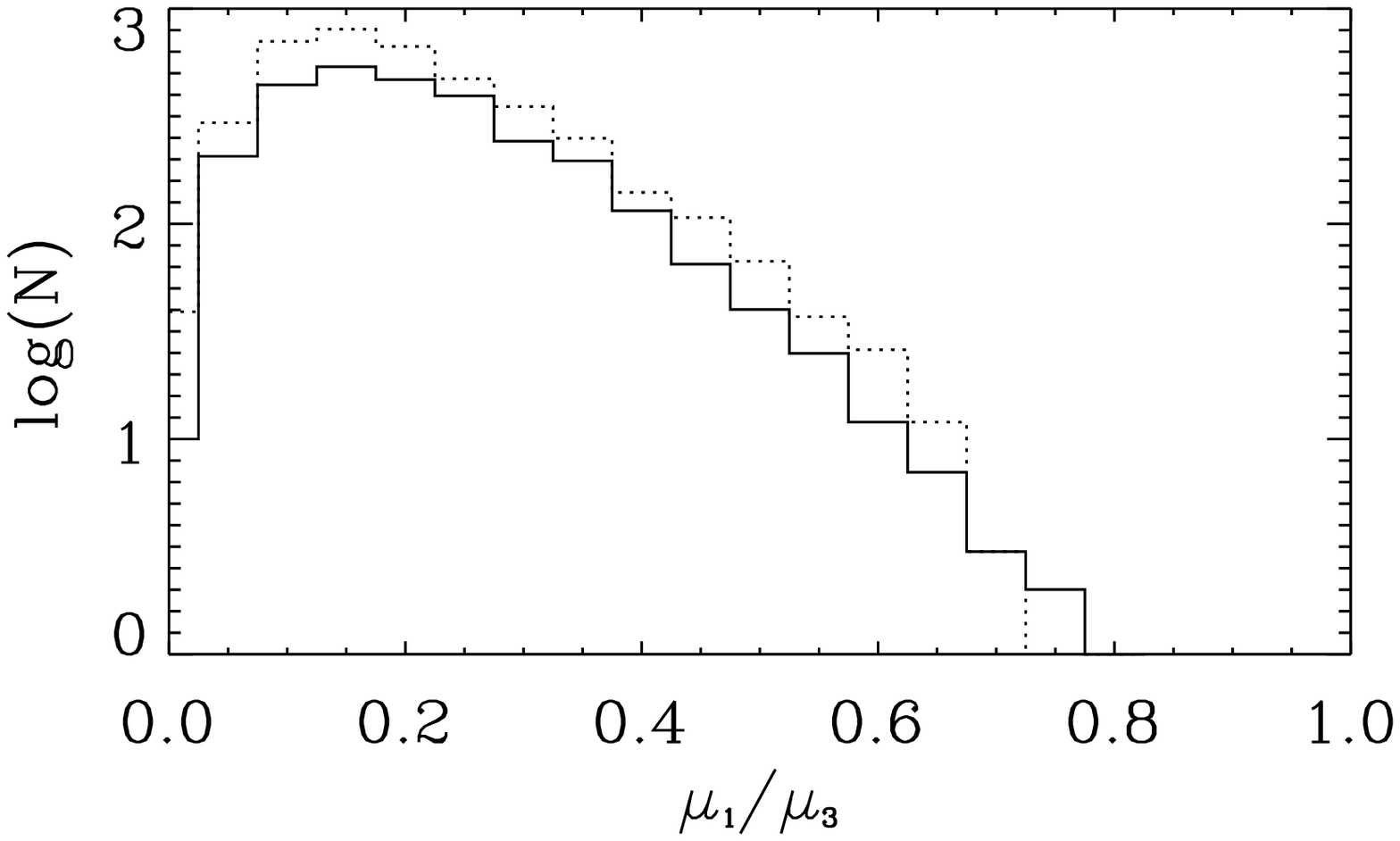}}
\put(0,4){\includegraphics[width=8cm]{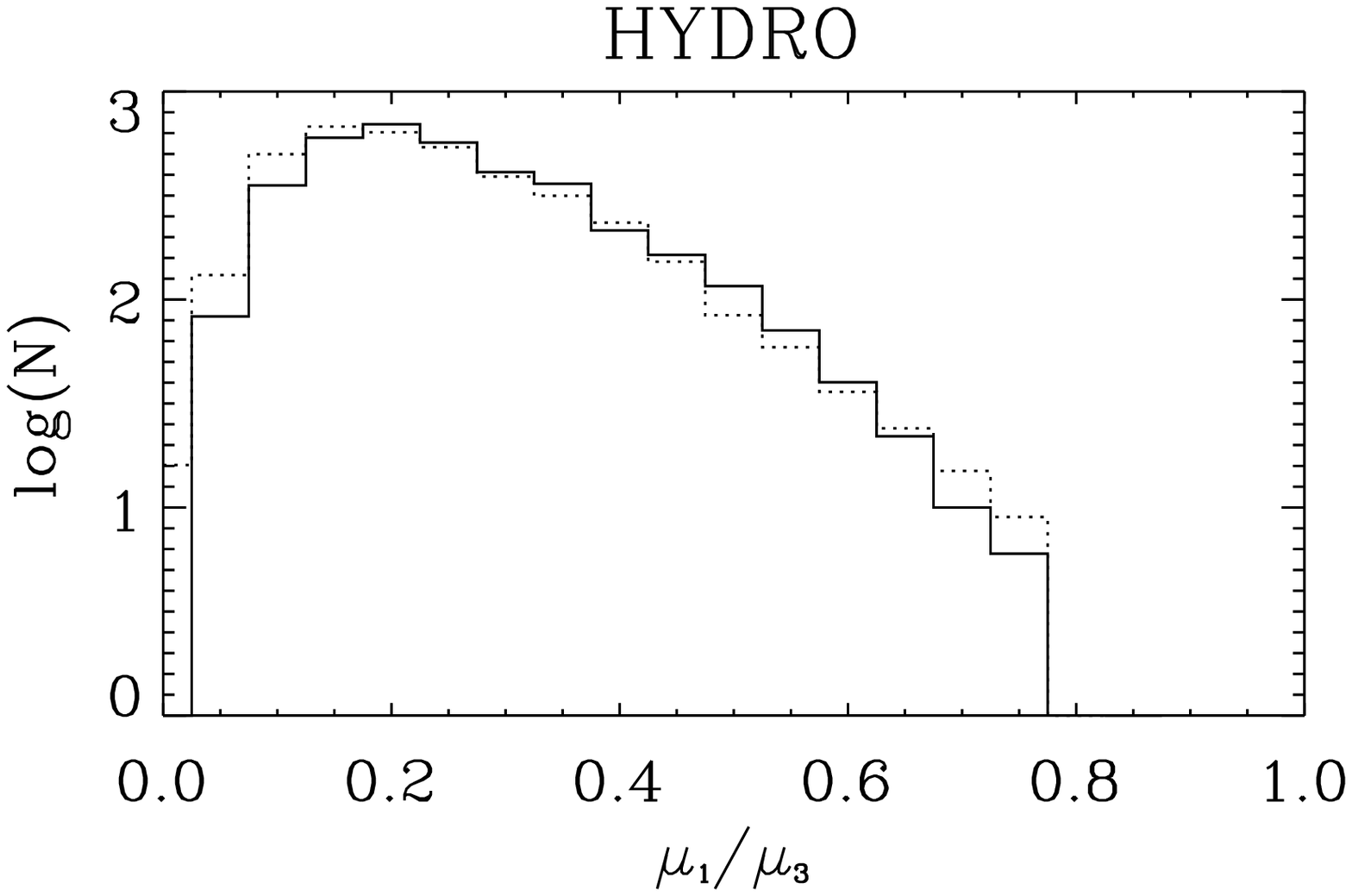}}
\end{picture}
\caption{Distribution of aspect ratio, $\mu_3/\mu_1$ of the clumps 
(threshold 50 cm$^{-3}$: upper panel and 200 cm$^{-3}$: lower panel) in the 
hydrodynamical simulation at time $t=1.52$ Myr (dotted lines) and $t=3.37$ Myr (solid lines).}
\label{aspect_ratio_hydro_decay1}
\end{figure}

\setlength{\unitlength}{1cm}
\begin{figure} 
\begin{picture} (0,13.5)
\put(0,5){\includegraphics[width=8cm]{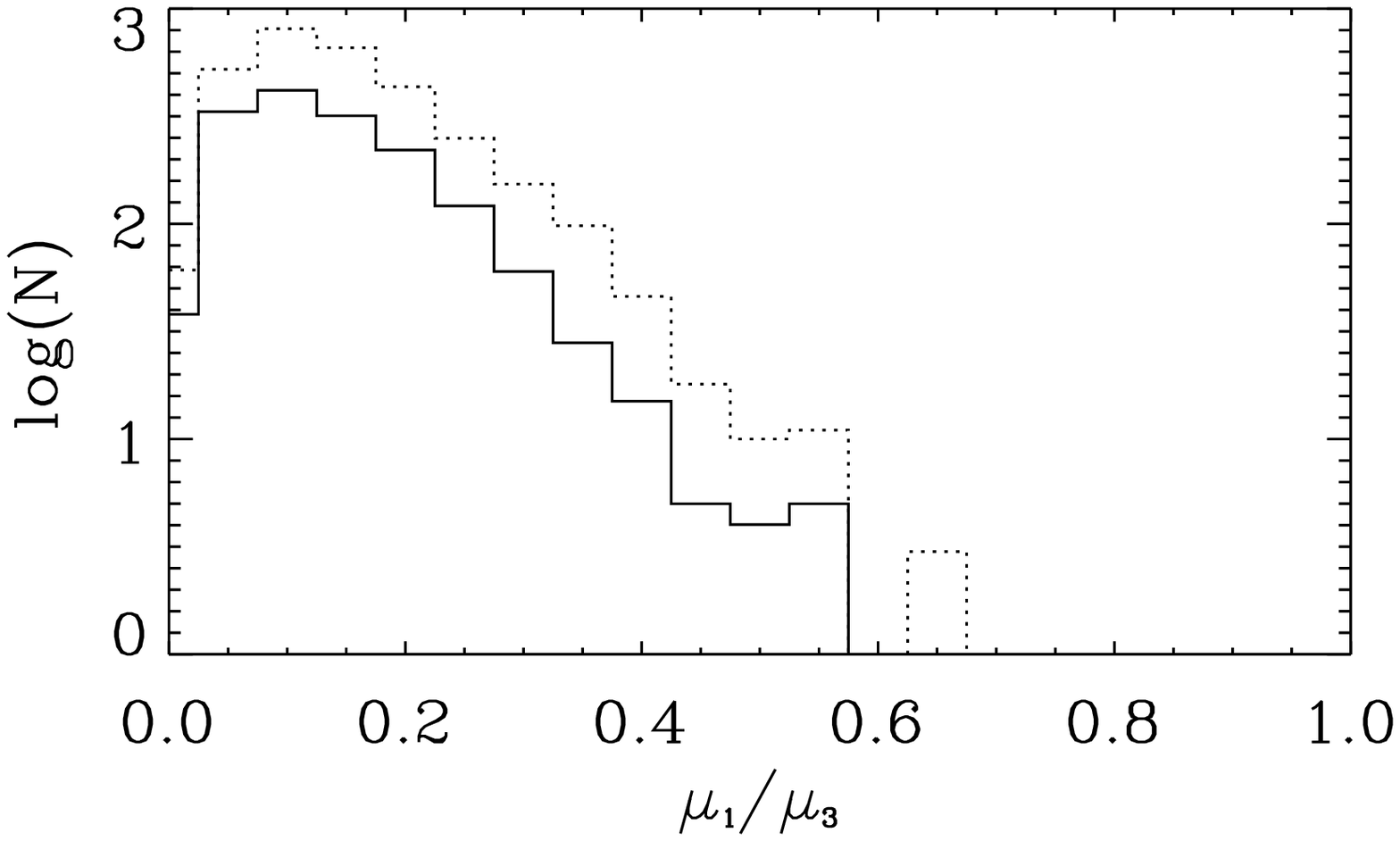}}
\put(0,9){\includegraphics[width=8cm]{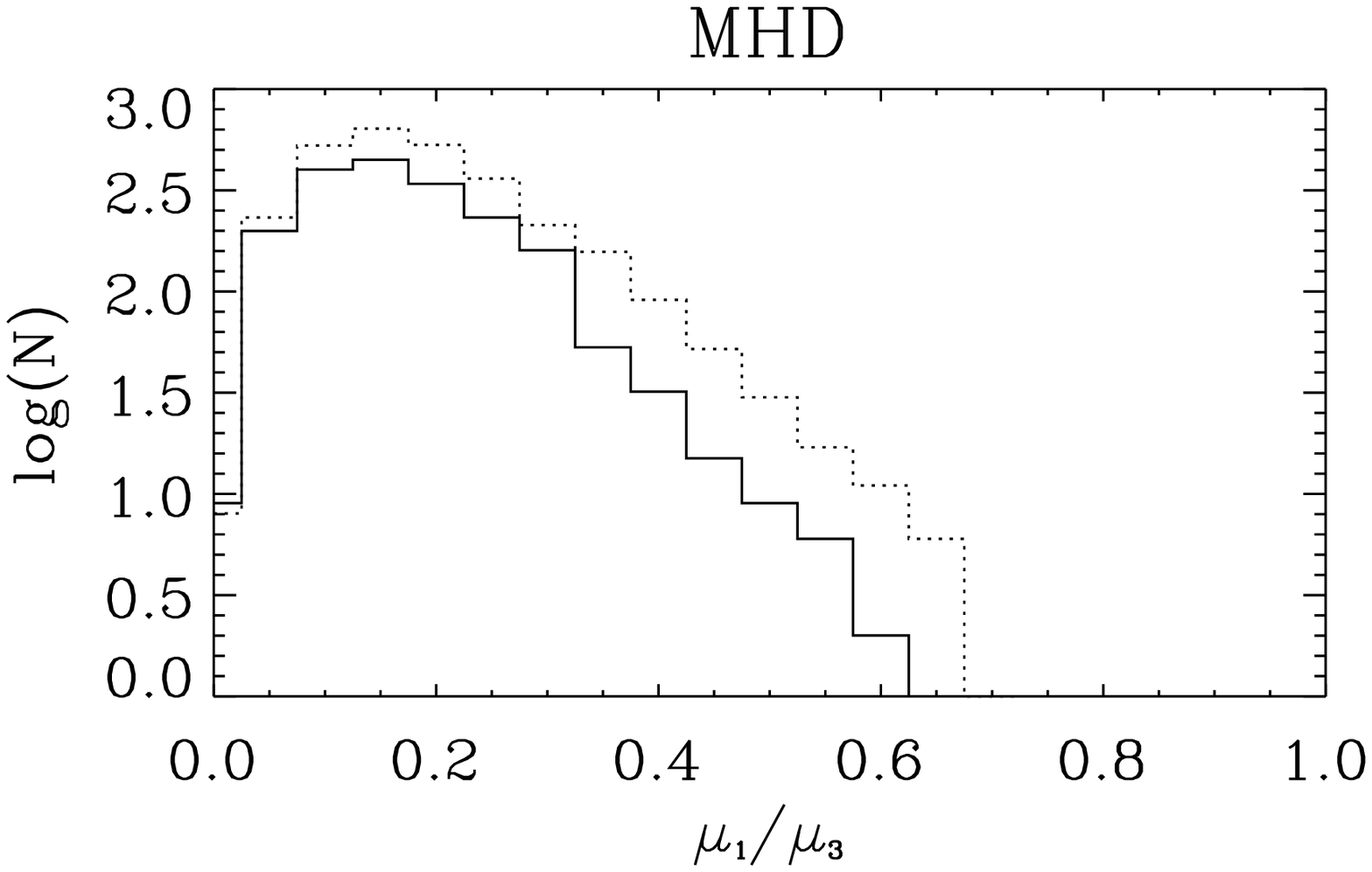}}
\put(0,0){\includegraphics[width=8cm]{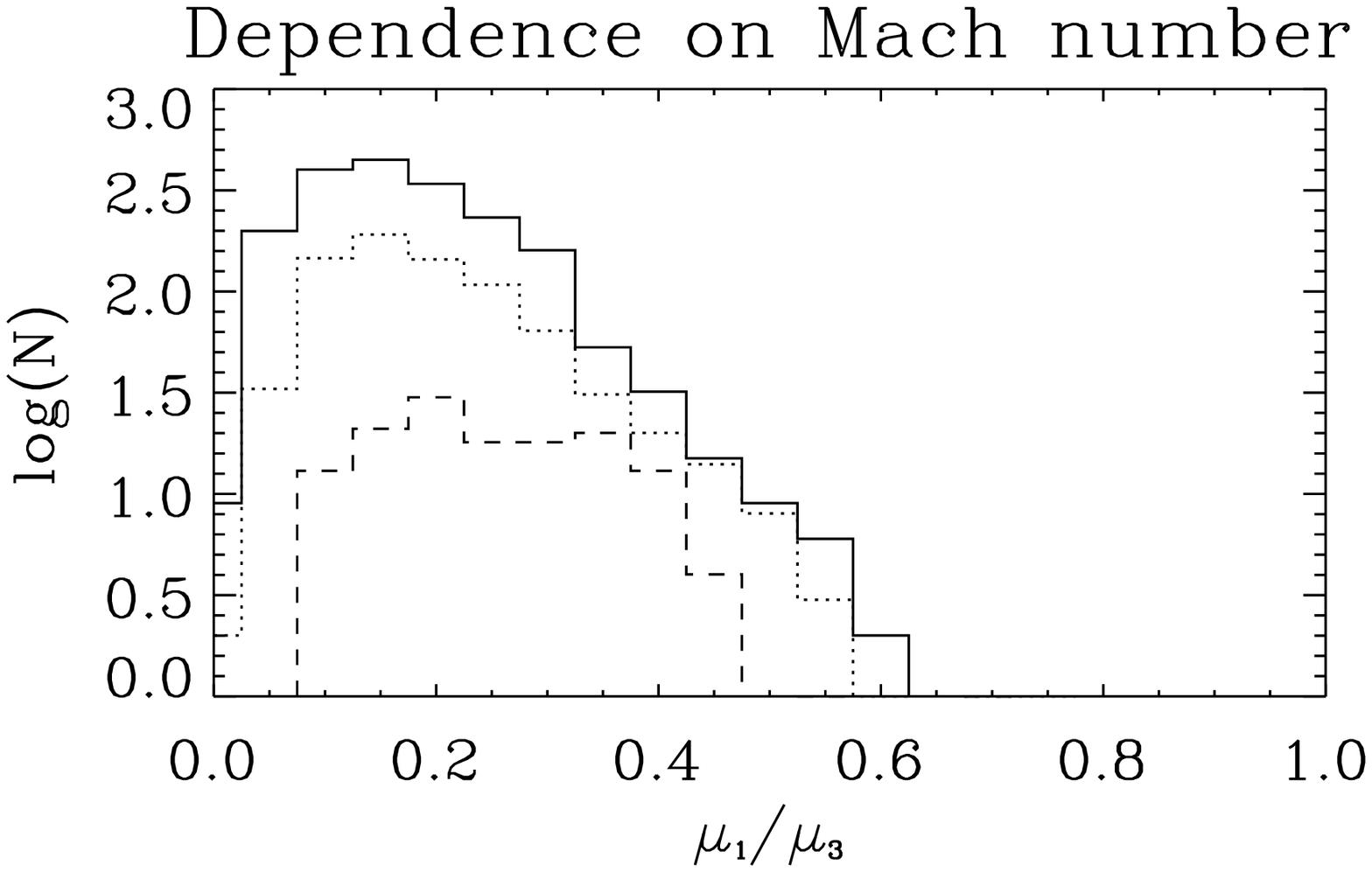}}
\end{picture}
\caption{Distribution of aspect ratio, $\mu_3/\mu_1$ of the clumps in 
the MHD simulations.
 Bottom panel shows the distribution for the threshold $n=50$ cm$^{-3}$
and for 3 Mach numbers (solid line: ${\cal M}=10$, dashed line: ${\cal M}=3$,
 dotted line: ${\cal M}=1$.
Middle and  top panels show the distribution at two different thresholds 
(middle: 50 cm$^{-3}$, bottom: 200 cm$^{-3}$) for the fiducial simulation
(magnetized, ${\cal M}=10$: solid line) at time 1.81 Myr and the high resolution simulation 
at time 2.26 Myr (dotted line). }
\label{aspect_ratio_mhd_decay1}
\end{figure}

As explained in Section~\ref{inertia_mat}, the inertia matrix
is computed for all clumps and the aspect ratio is estimated 
as the quantity $\sqrt{I_1/I_3}=\mu_1/\mu_3$ where $I_1$ and $I_3$ are the smallest 
and largest eigenvalues.

Figures~\ref{aspect_ratio_hydro_decay1} and~\ref{aspect_ratio_mhd_decay1} 
display the distribution of $\mu_1/\mu_3$ for the 
hydrodynamical and MHD simulations and for two thresholds 
50 (upper panels) and 200 cm$^{-3}$ (lower panels for Fig.~\ref{aspect_ratio_hydro_decay1}
and middle panel for Fig.~\ref{aspect_ratio_mhd_decay1}). Clearly, the 
aspect ratios in the MHD simulations are smaller by
a factor of $\simeq$1.5-2 than the aspect ratios in the 
hydrodynamical simulations. 
The threshold has only a modest influence on the resulting distribution. 

 Bottom panel of Fig.~\ref{aspect_ratio_mhd_decay1} shows that the Mach 
number, has only a modest influence on this result. There are only 
little differences between ${\cal M}=10$ and 3 runs (solid and dotted lines 
respectively). There are more differences in the  ${\cal M}=1$ run
but this could be due to the lack of statistics. The peak is may be shifted 
toward slightly higher values but stays below 0.2-0.3.

It is worth stressing that in all cases, most of the clumps
have an aspect ratio smaller than 0.3 and a good fraction 
of them have an aspect ratio smaller than 0.2 and even 0.1.
These latter can be called filaments.

\subsubsection{Triaxial clumps}

\setlength{\unitlength}{1cm}
\begin{figure} 
\begin{picture} (0,14)
\put(0,6.5){\includegraphics[width=8cm]{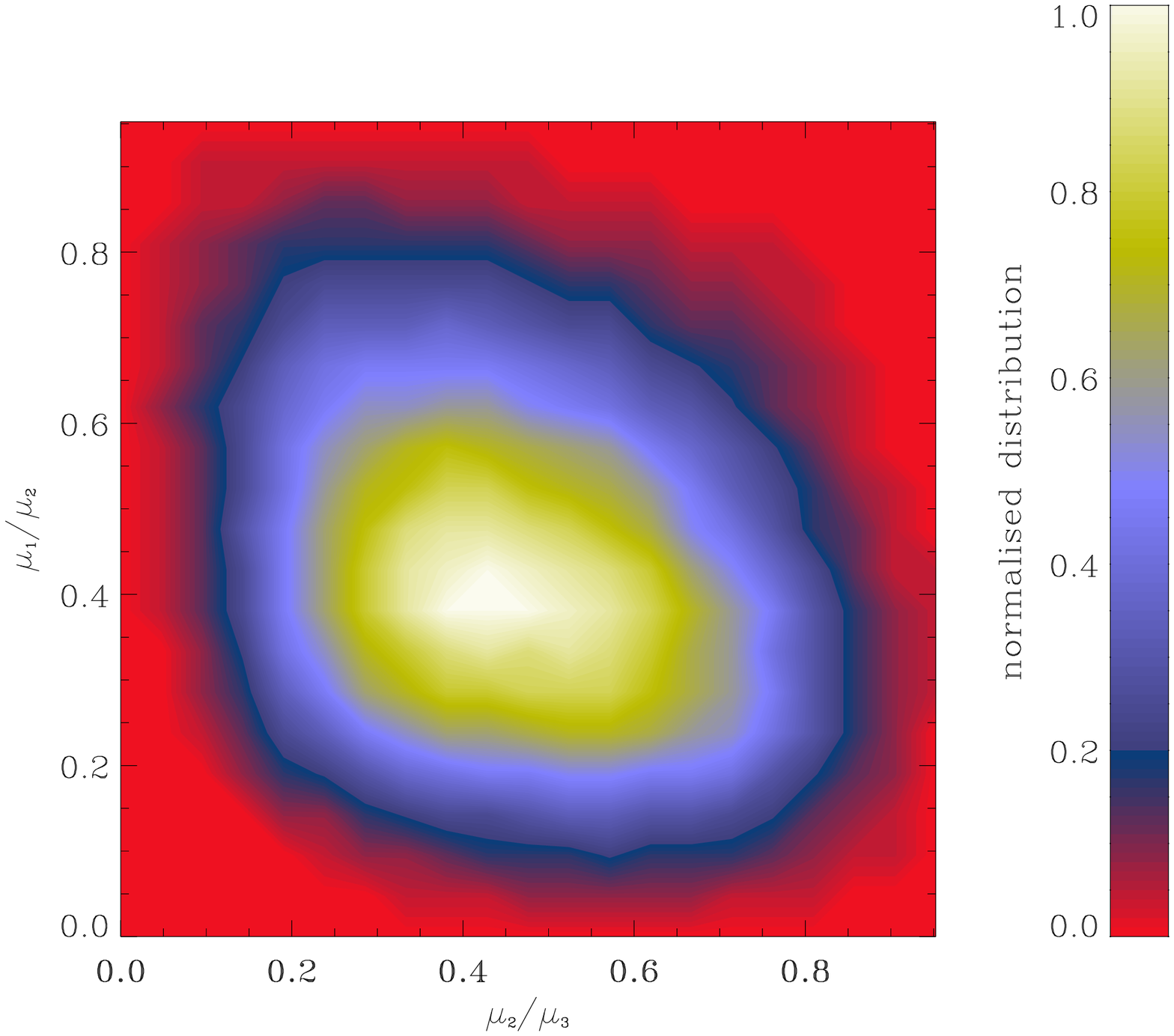}}
\put(0,0){\includegraphics[width=8cm]{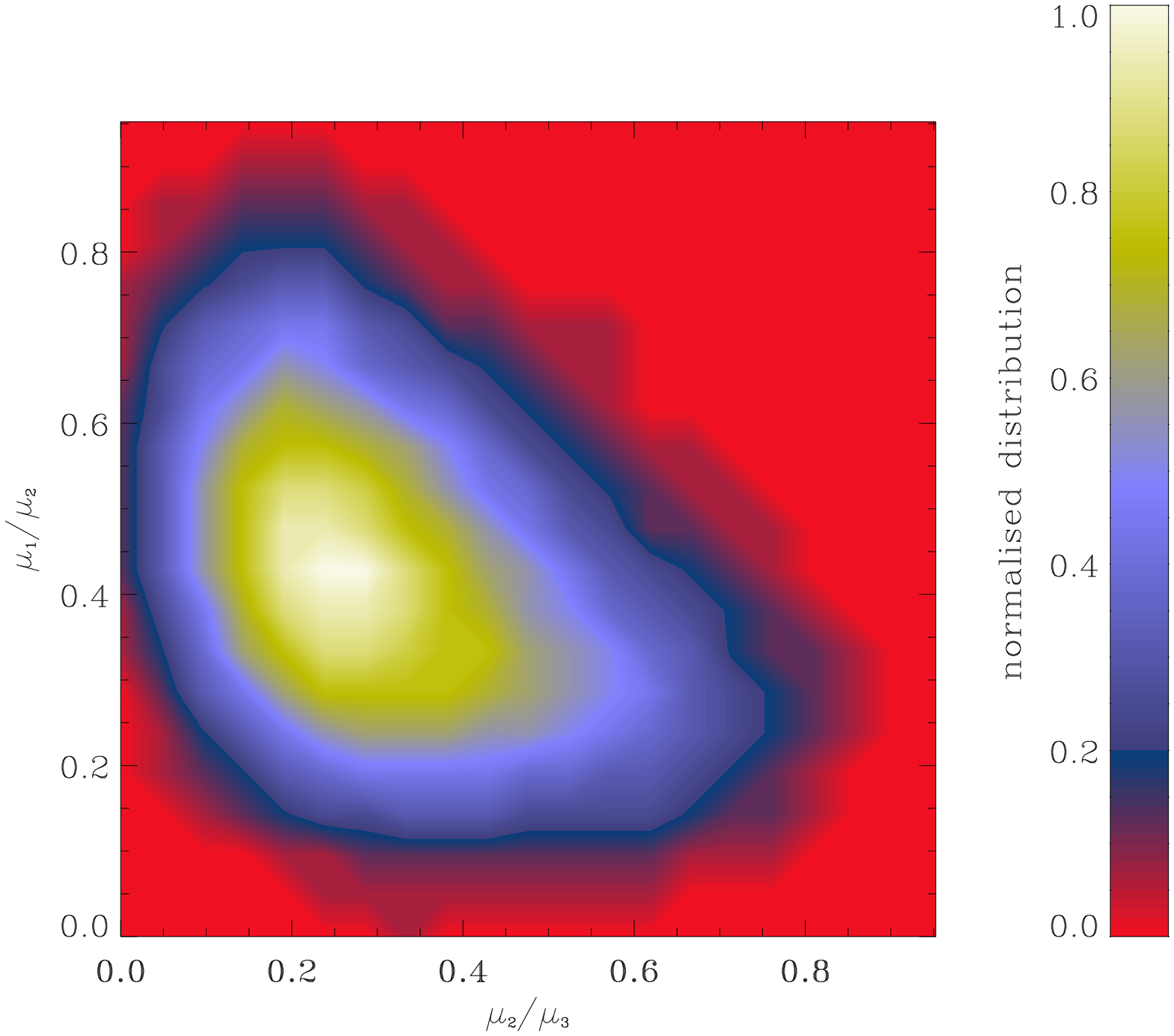}}
\end{picture}
\caption{Normalised bidimensional histogram displaying
$\mu_1/\mu_2$ as a function of $\mu_2/\mu_3$.
 Top panel: hydrodynamical simulation at time 3.37 Myrs.
 Bottom panel: MHD simulation at time 1.81 Myrs. }
\label{triaxis}
\end{figure}

The aspect ratio of the largest over smallest eigenvalues
gives only a partial description of the clump geometry. 
It is necessary, to get a more complete description, to investigate
the distribution of the two ratio $\mu_1/\mu_2$ and $\mu_2 / \mu_3$.
Figure~\ref{triaxis} displays the  normalised bidimensional 
histograms for the hydrodynamical 
and MHD simulations using the density threshold of 200 cm$^{-3}$. 
The two distributions present significant 
differences. The clumps from the hydrodynamical simulation
tend to cover more uniformly the  $\mu_1/\mu_2$-$\mu_2 / \mu_3$
plane. In particular, most of the points are located in a 
region with $\mu_1 / \mu_2 \simeq 0.2-0.6$ and $\mu_2 / \mu_3 \simeq 0.3-0.8$.
Such clumps can be described as ribbons or/and sheets.
More quantitatively, we find that the fraction of clumps 
having  $\mu_2 / \mu_3$ between 0.2 and 0.8 and 
$\mu_1 / \mu_2$ between 0.3 and 0.7 is 62$\%$ amongst which 
about half have their $\mu_2 / \mu_3$ lower than 0.5. 
The number of clumps with $\mu_2 / \mu_3$ smaller than 0.3 and
 $\mu_1 / \mu_2$ larger than 0.4 is only about 11$\%$.
In the MHD simulation, the values 
$\mu_1 / \mu_2 \simeq 0.5$ and $\mu_2 / \mu_3 \simeq 0.25$ are more 
typical. Such objects can be described as ribbons and/or filaments. 
The most important difference are the absence of spheroidal 
($\mu_1 \simeq \mu_2 \simeq \mu_3$)
clumps  and the scarcity of sheet like clumps 	
($\mu_1 \ll \mu_2 \simeq \mu_3$)
in the MHD simulations.
More quantitatively, the fraction of clumps 
having  $\mu_2 / \mu_3$ between 0.2 and 0.8 and 
$\mu_1 / \mu_2$ between 0.3 and 0.7 is 52$\%$ amongst which 
about 75$\%$ have their $\mu_2 / \mu_3$ lower than 0.5. 
The fraction of sheet like or spheroidal like objects 
($\mu_2 / \mu_3 >0.5$) is therefore two times smaller than 
in the hydrodynamical simulation.
The number of filamentary clumps with $\mu_2 / \mu_3$ smaller than 0.3 and
 $\mu_1 / \mu_2$ larger than 0.4 is  about 30$\%$ which 
is three times more than in the hydrodynamical simulation.

Since sheet-like objects are produced by shocks, 
this clearly suggests that while in the hydrodynamical simulations, 
shocks are important and numerous, they play a less 
important role in the MHD simulations. This is somehow 
expected because the magnetic field certainly reduces their 
ability to compress the gas.

\subsubsection{Aspect ratio from skeleton}

\setlength{\unitlength}{1cm}
\begin{figure} 
\begin{picture} (0,8.5)
\put(0,0){\includegraphics[width=8cm]{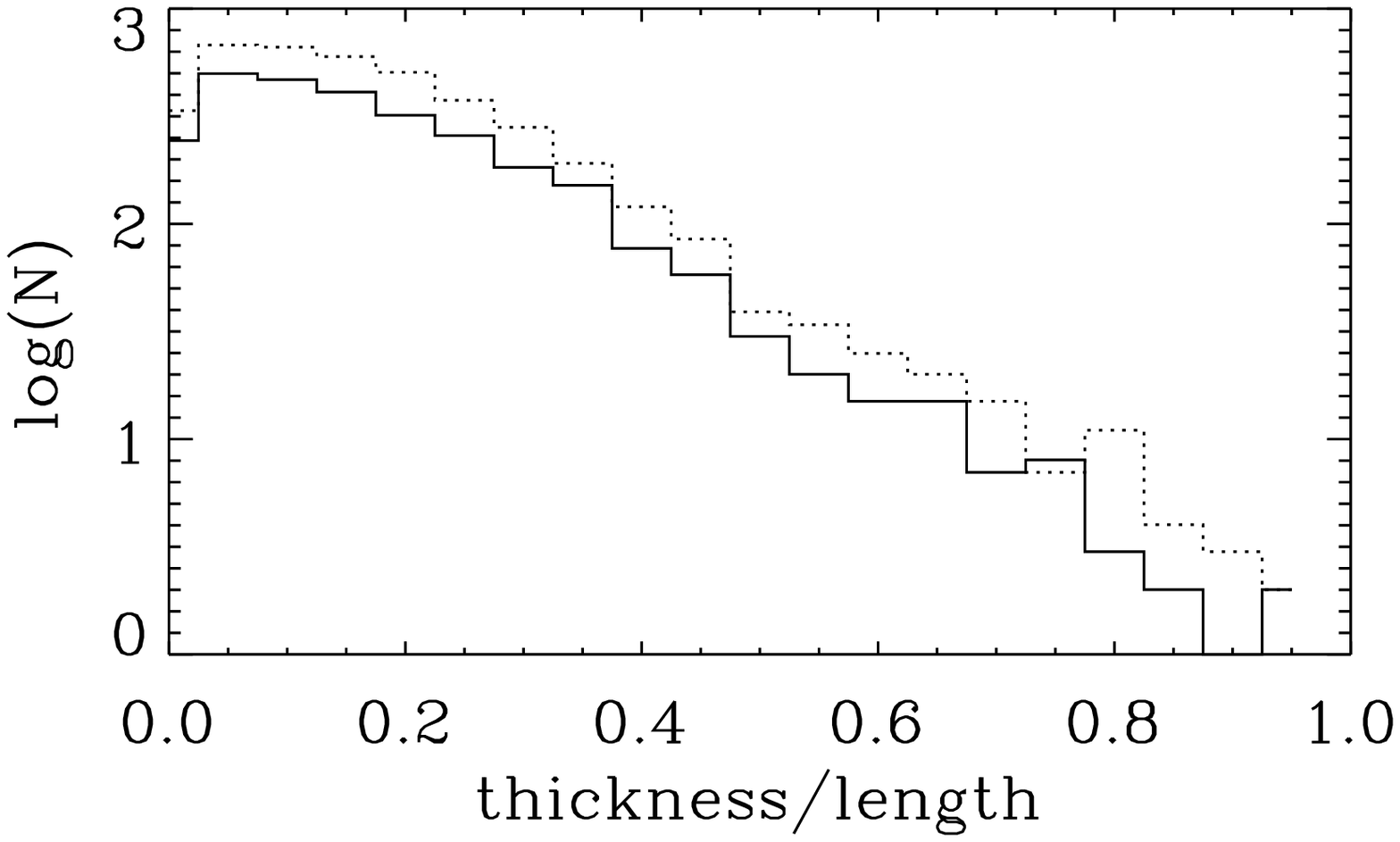}}
\put(0,4){\includegraphics[width=8cm]{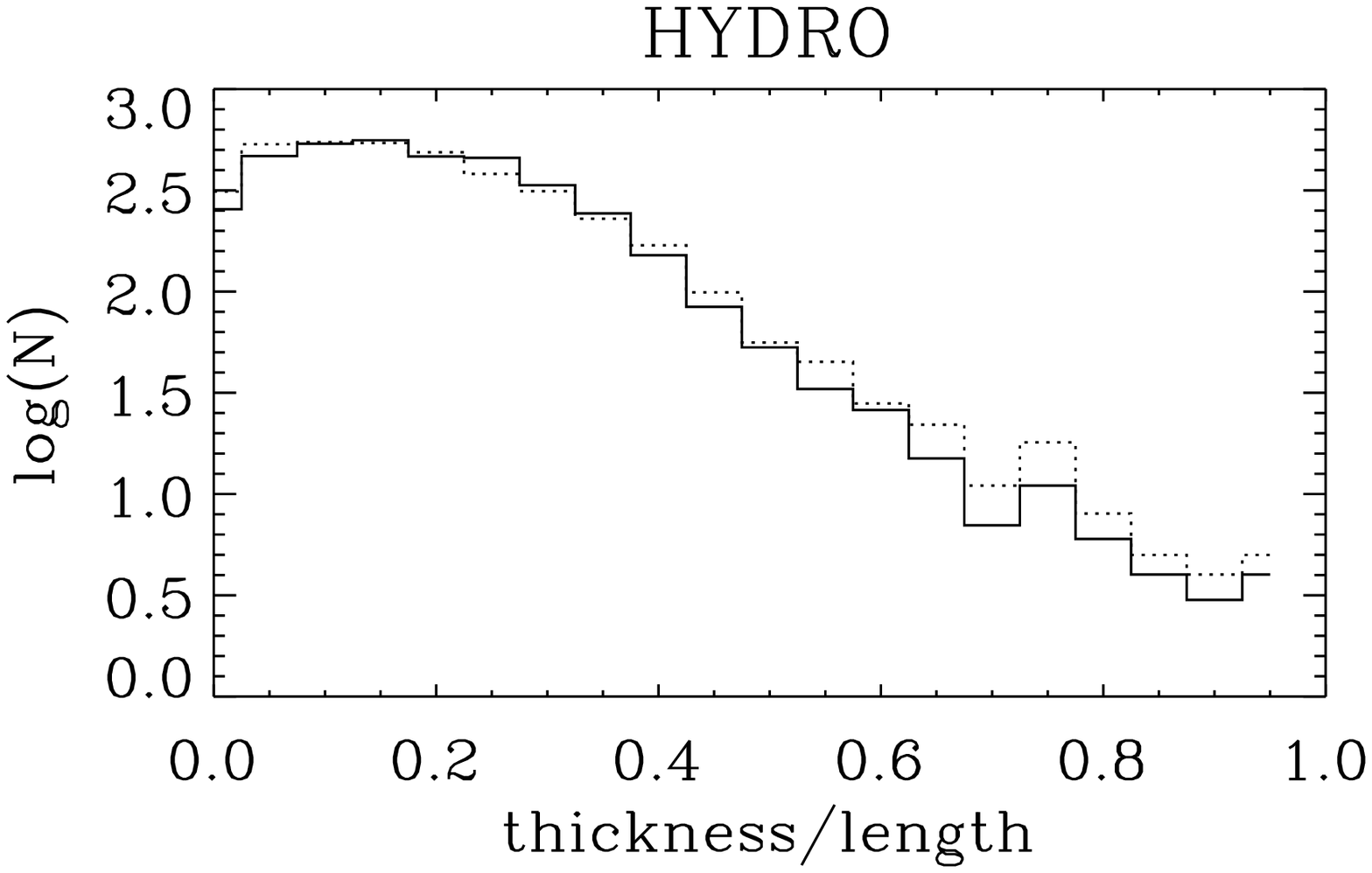}}
\end{picture}
\caption{Same as Fig.~\ref{aspect_ratio_hydro_decay1} for the distribution of the aspect ratio, $R/L$ of the clumps
 (threshold 50 cm$^{-3}$: upper panel and 200 cm$^{-3}$: lower panel) in the 
hydrodynamical simulation at time $t=1.52$ Myr (dotted lines) and $t=3.37$ Myr (solid lines).}
\label{aspect_ratio_hydro_decay2}
\end{figure}

\setlength{\unitlength}{1cm}
\begin{figure} 
\begin{picture} (0,13.5)
\put(0,5){\includegraphics[width=8cm]{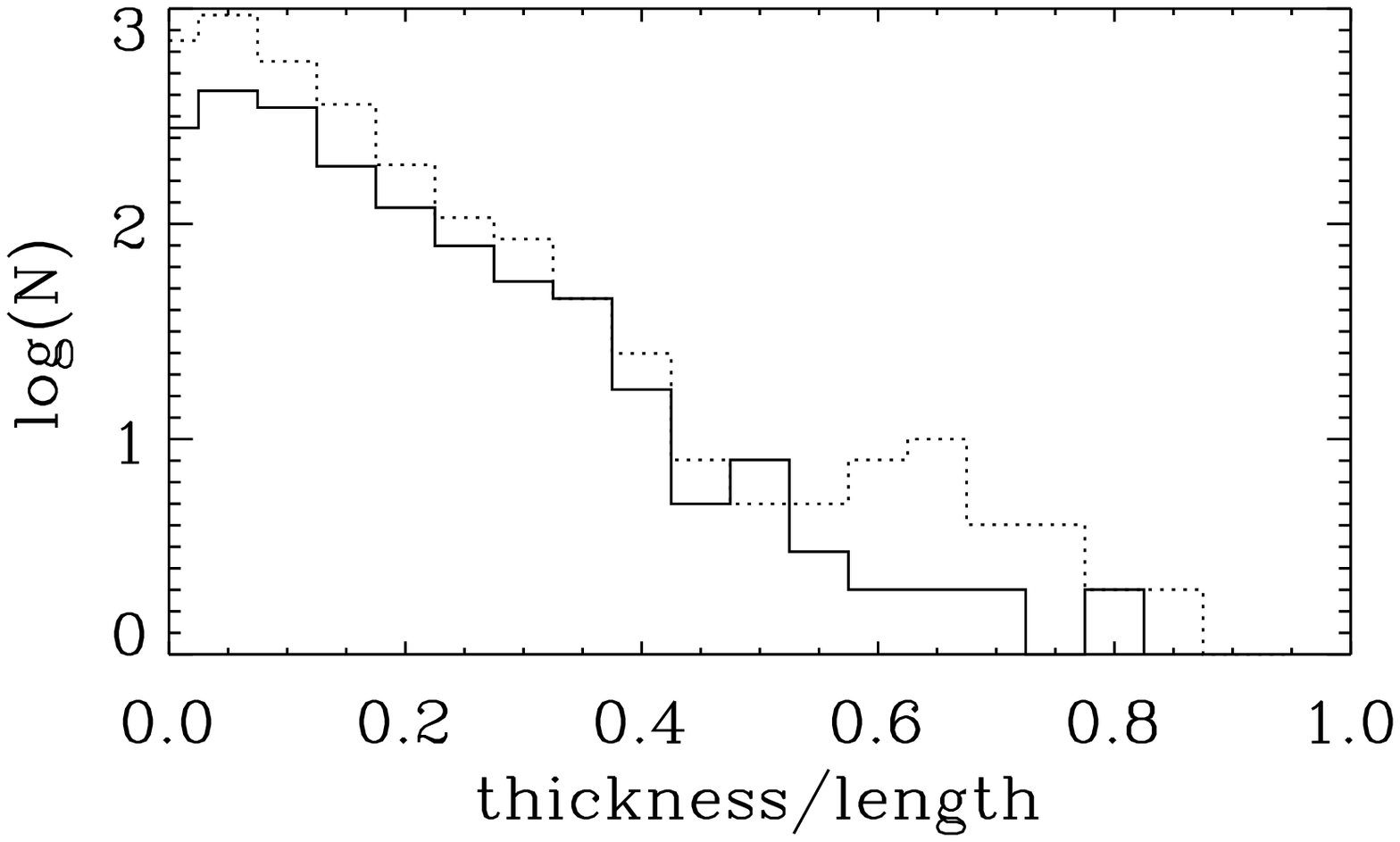}}
\put(0,9){\includegraphics[width=8cm]{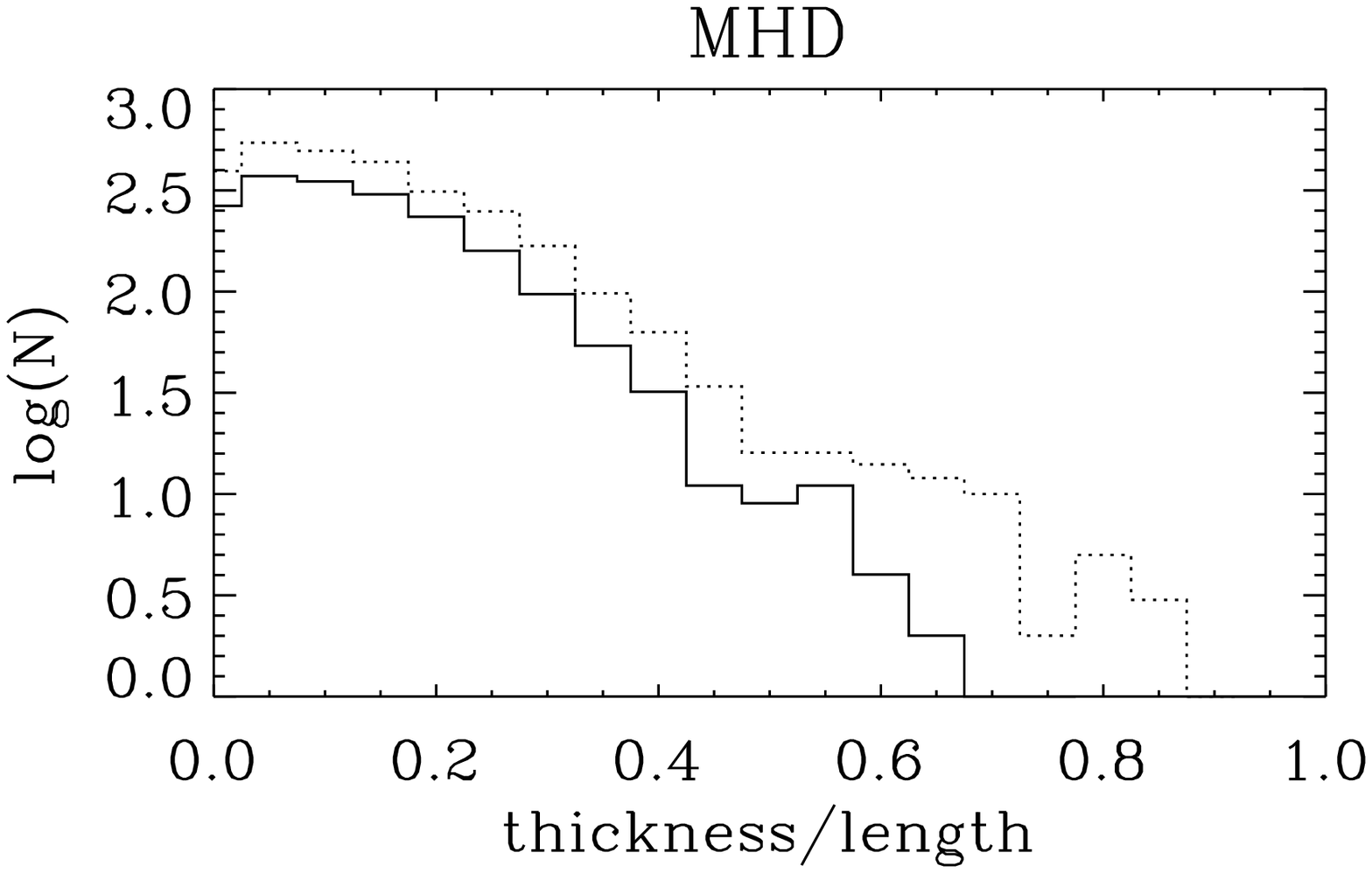}}
\put(0,0){\includegraphics[width=8cm]{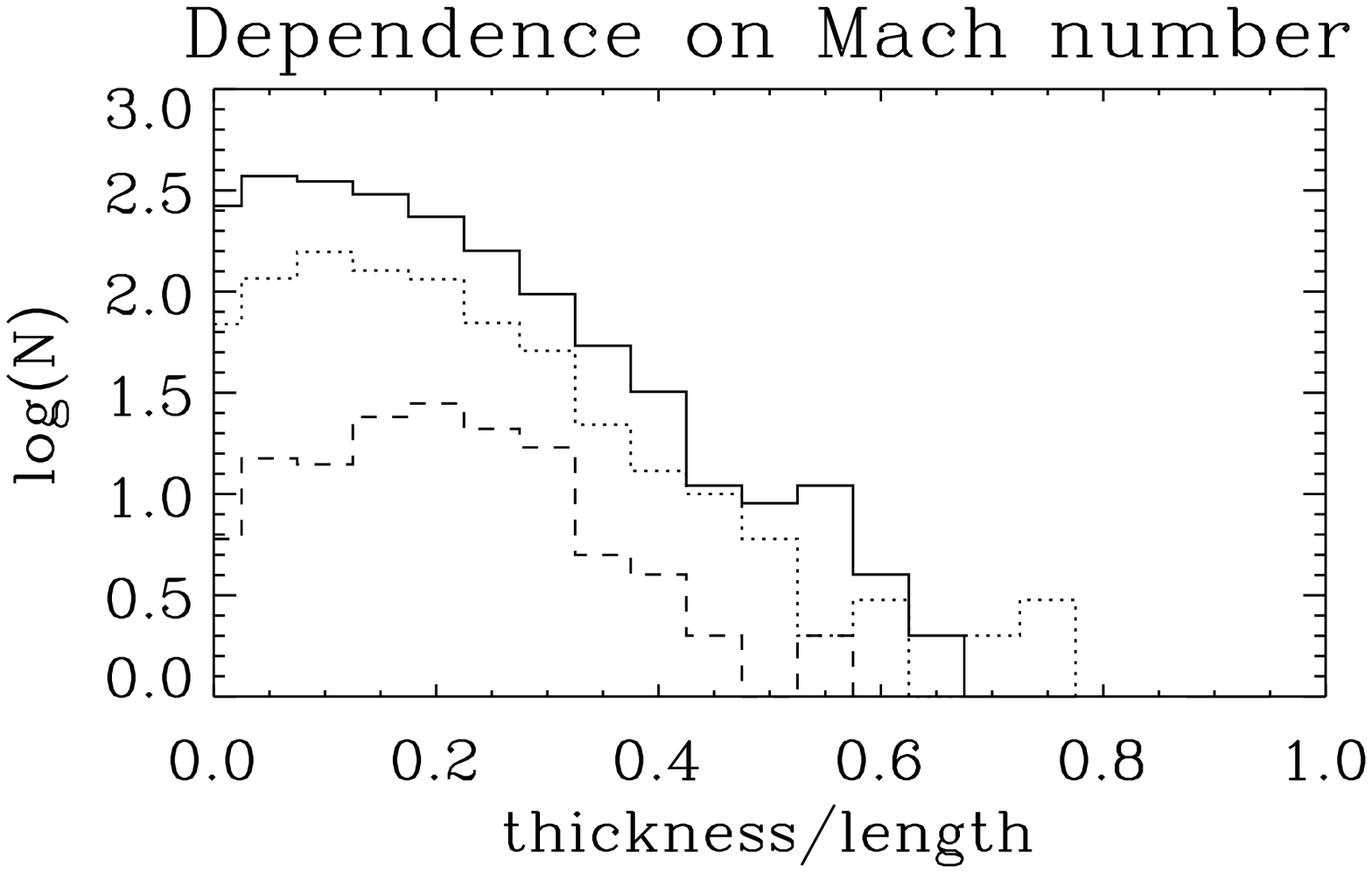}}
\end{picture}
\caption{Same as Fig.~\ref{aspect_ratio_mhd_decay1} for the distribution of the aspect ratio, $R/L$ of the clumps.
Bottom panel shows the distribution for the threshold $n=50$ cm$^{-3}$
and for 3 Mach numbers (solid line: ${\cal M}=10$, dashed line: ${\cal M}=3$,
 dotted line: ${\cal M}=1$.
Middle and  top panels show the distribution at two different thresholds 
(middle: 50 cm$^{-3}$, bottom: 200 cm$^{-3}$) for the fiducial simulation
(magnetized, ${\cal M}=10$: solid line) at time 1.81 Myr and the high resolution simulation 
at time 2.26 Myr (dotted line).}
\label{aspect_ratio_mhd_decay2}
\end{figure}

In order to acertain the trends inferred for the clump aspect ratio, 
we have also calculated  it  using the skeleton approach 
and the definitions given in Sect.~\ref{skeleton}. The results for the 
two snapshots of the hydrodynamical and MHD simulations are 
presented in Figs.~\ref{aspect_ratio_hydro_decay2} 
and~\ref{aspect_ratio_mhd_decay2}. As can be seen, the trends 
are very similar to what has been inferred from the inertia matrix
and the distribution are generally quite comparable. 
In particular, the clumps tend to be more elongated in the MHD case 
than in the hydrodynamical case. One difference however is the presence
of a tail of few weakly elongated clumps (aspect ratio 0.5-1) that 
is not apparent in the inertia matrix approach. This is likely 
due to the difference in the definition between the two methods. 

The nice similarity of the distributions obtained with two 
completely different methods, suggests however that the two methods are 
indeed reliable.

\subsection{Length and thickness}
We now turn to the study of the clump characteristic scales.
The length is defined as the sum over all $G_i^j{\cal G}_i^j$
within the clumps. This gives the sum of the length of all 
the branches which belong to the clumps and is therefore longer 
than the largest distance between two points in the clump. 
The thickness is defined as explained in Sect.~\ref{skeleton}, 
that is to say as the mean distance between the  clump cells
and the clump local direction (defined by the ${\bf u}_i^j$).

\subsubsection{Clump length}

\setlength{\unitlength}{1cm}
\begin{figure} 
\begin{picture} (0,8.5)
\put(0,0){\includegraphics[width=8cm]{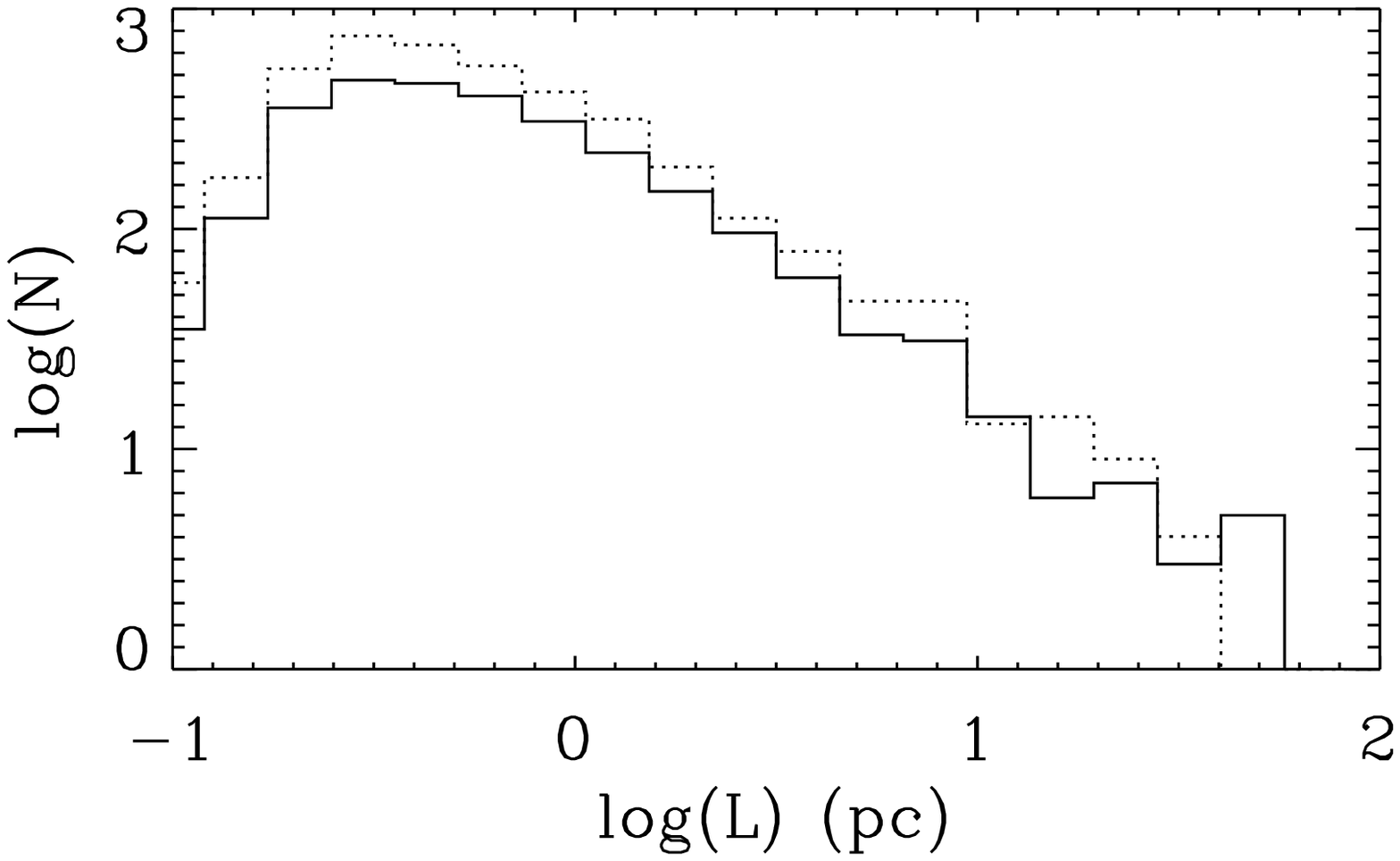}}
\put(0,4){\includegraphics[width=8cm]{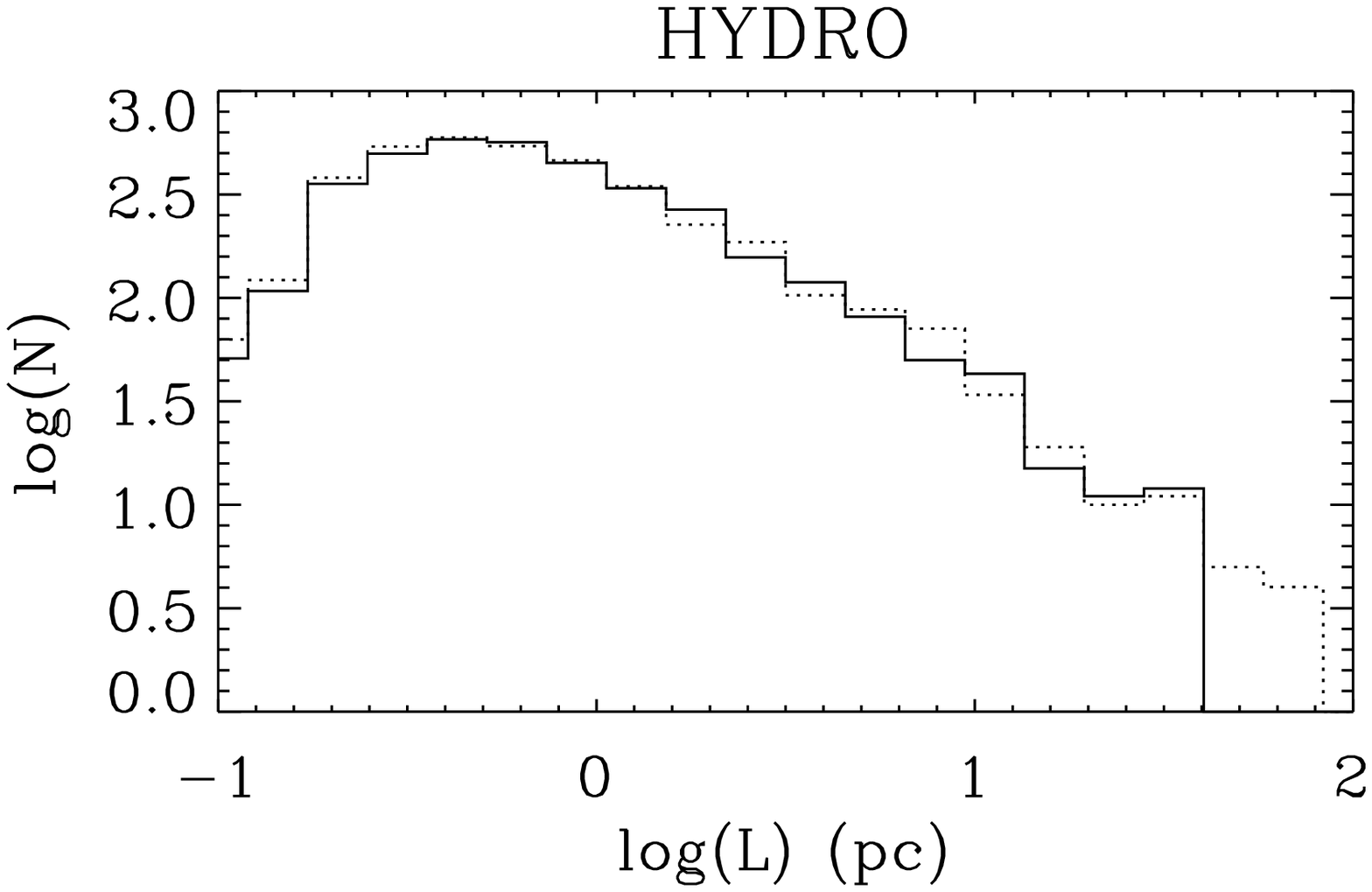}}
\end{picture}
\caption{Same as Fig.~\ref{aspect_ratio_hydro_decay1} for the distribution of the length of the clumps.}
\label{length_hydro}
\end{figure}

\setlength{\unitlength}{1cm}
\begin{figure} 
\begin{picture} (0,13.5)
\put(0,5){\includegraphics[width=8cm]{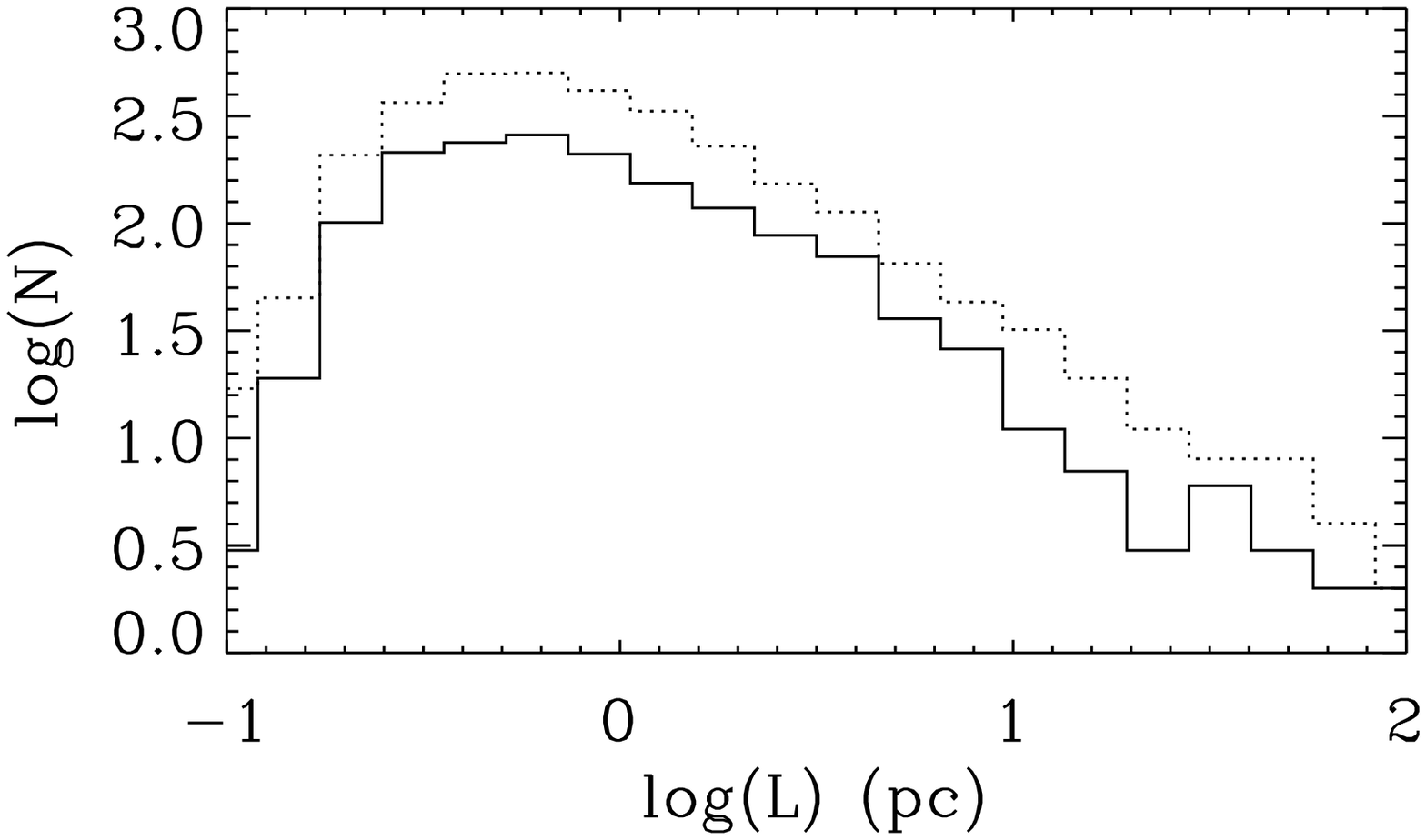}}
\put(0,9){\includegraphics[width=8cm]{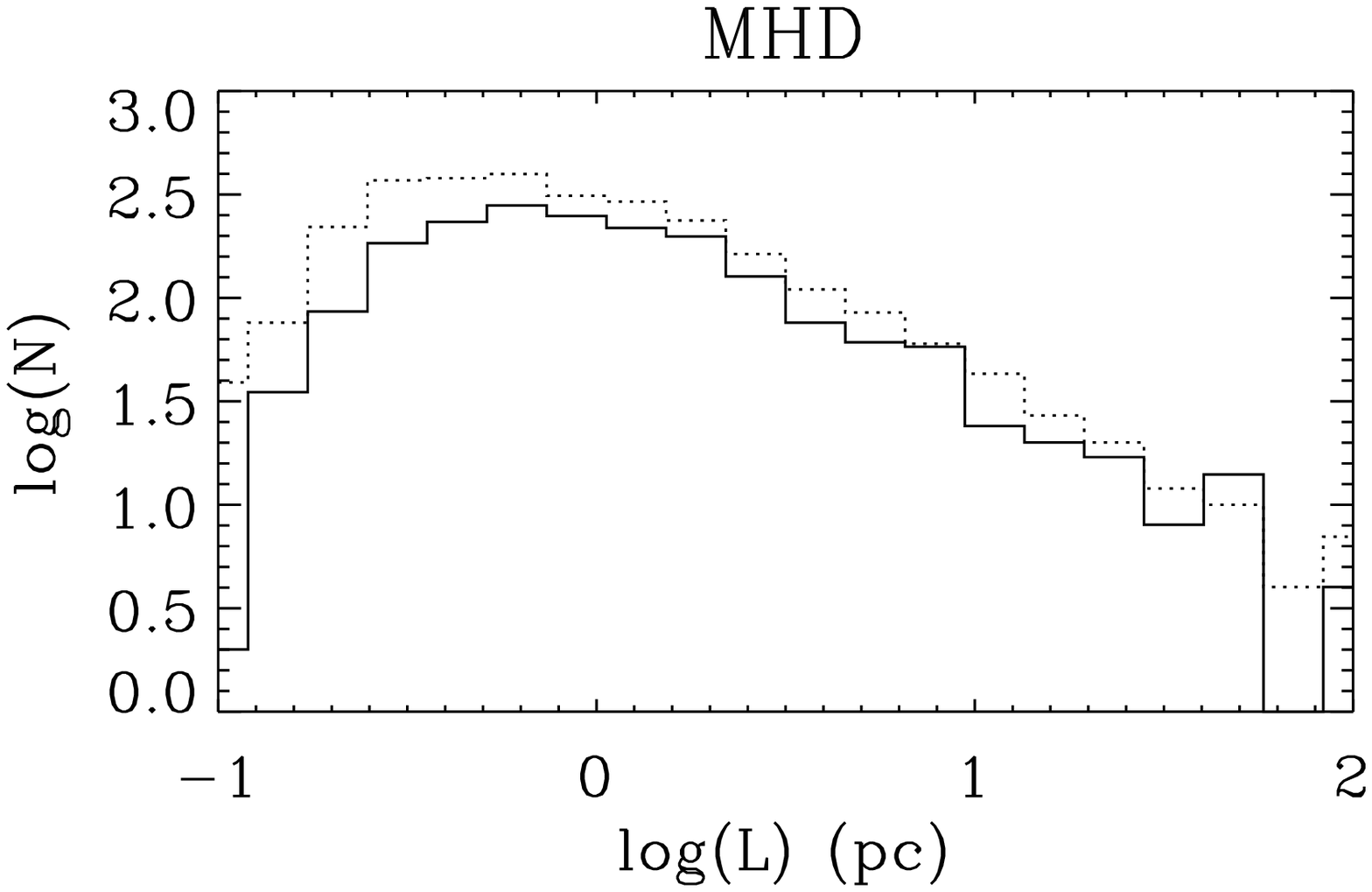}}
\put(0,0){\includegraphics[width=8cm]{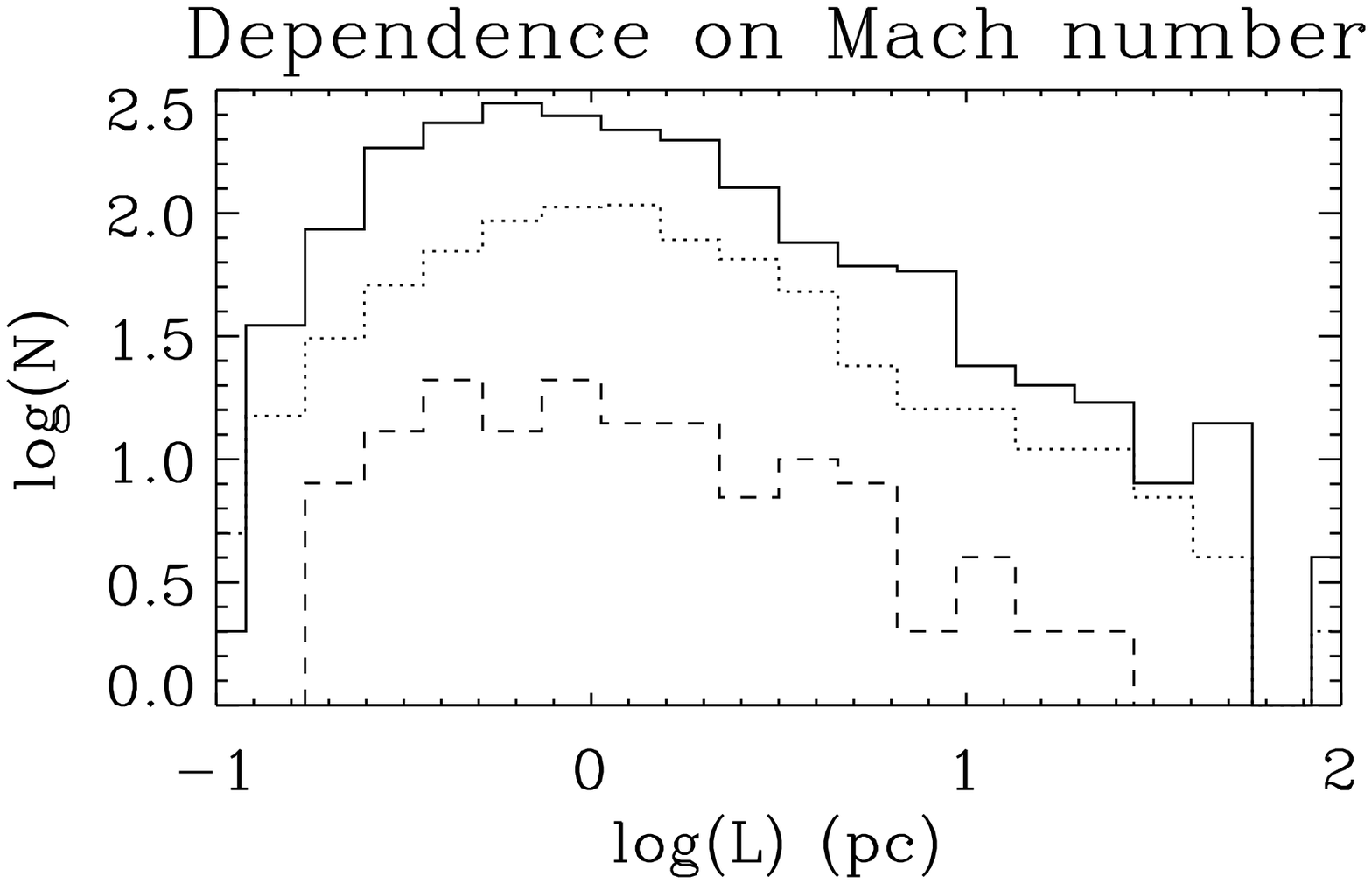}}
\end{picture}
\caption{Same as Fig.~\ref{aspect_ratio_mhd_decay1} for the distribution of the length of the clumps.}
\label{length_mhd}
\end{figure}

Figures~\ref{length_hydro} and~\ref{length_mhd} show the length distribution for the hydrodynamical and MHD 
simulations respectively. The distributions are similar. They 
peak at about $L \simeq 0.5$ pc and decrease with size for smaller value. This 
is very similar to the behaviour of the clump mass spectra 
(e.g. Hennebelle \& Audit 2007, Audit \& Hennebelle 2010) and is 
a clear consequence of the numerical diffusion induced by the finite size of the 
mesh. The comparison between the fiducial simulation (top panel of Fig.~\ref{length_mhd})
and the high resolution run shows that these peaks tend to shift toward the smaller
size in spite of the extraction being performed at the same physical resolution as explained
previously. 
At larger length, $L$, the distribution is close to a powerlaw. This is more 
obvious for the ${\cal M}=10$ simulations than for the 
${\cal M}=3$ and 1 cases (bottom panel of Fig.~\ref{length_mhd}) probably 
because there are less clumps in these simulations and the statistics are poorer.
Typically we get ${\cal N} = dN / d \log L \propto L^{-1}$ for the threshold $200$ cm$^{-3}$.
The exponent is slightly shallower for the threshold 50 cm$^{-3}$.
Apart from the fact that turbulence generally tends to generate powerlaws, it is 
worth to understand better the origin of this exponent. 

First of all, it is worth recalling 
that the mass spectra of clumps is found to be $dN/d \log M \propto M^{-\alpha _N +1}$ with 
$\alpha _N \simeq 1.8$ by various authors (Hennebelle \& Audit 2007, Heitsch et al. 2008, Dib et al. 2008, 
Audit \& Hennebelle 2010, Inoue \& Inutsuka 2012). This exponent is consistent with 
the value inferred by Hennebelle \& Chabrier (2008) for turbulent clumps. 
Anticipating on what will be shown in the next section, the thickness
 of the clumps, $r_c$,
stays roughly constant, i.e. is peaking toward a nearly constant value with a narrow 
distribution. But the mass of clumps is proportional to $L \times r_c^2 \rho$, thus 
since $r_c$ is found to be roughly constant and that the mean density in most 
of the clumps is of the order of the density threshold, one finds that the mass
of the clumps is proportional to its length $M \propto L$ (we recall that $L$ is the integrated length 
through all branches). Consequently, it is not surprising to find that 
$dN/d \log L \propto L^{-1}$ which is close enough to the 
clump mass spectrum.

\subsubsection{Clump thickness}

\setlength{\unitlength}{1cm}
\begin{figure} 
\begin{picture} (0,8.5)
\put(0,0){\includegraphics[width=8cm]{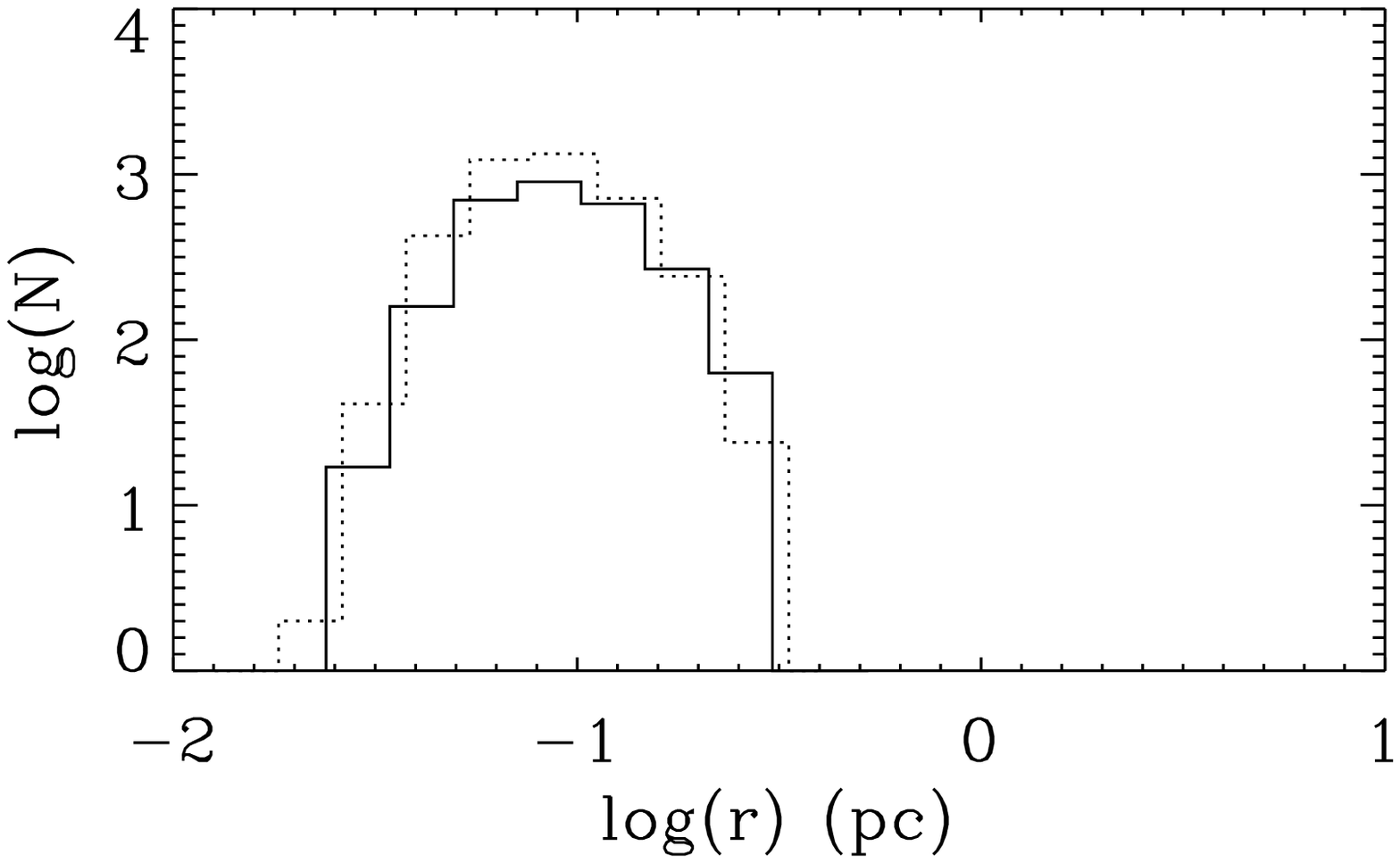}}
\put(0,4){\includegraphics[width=8cm]{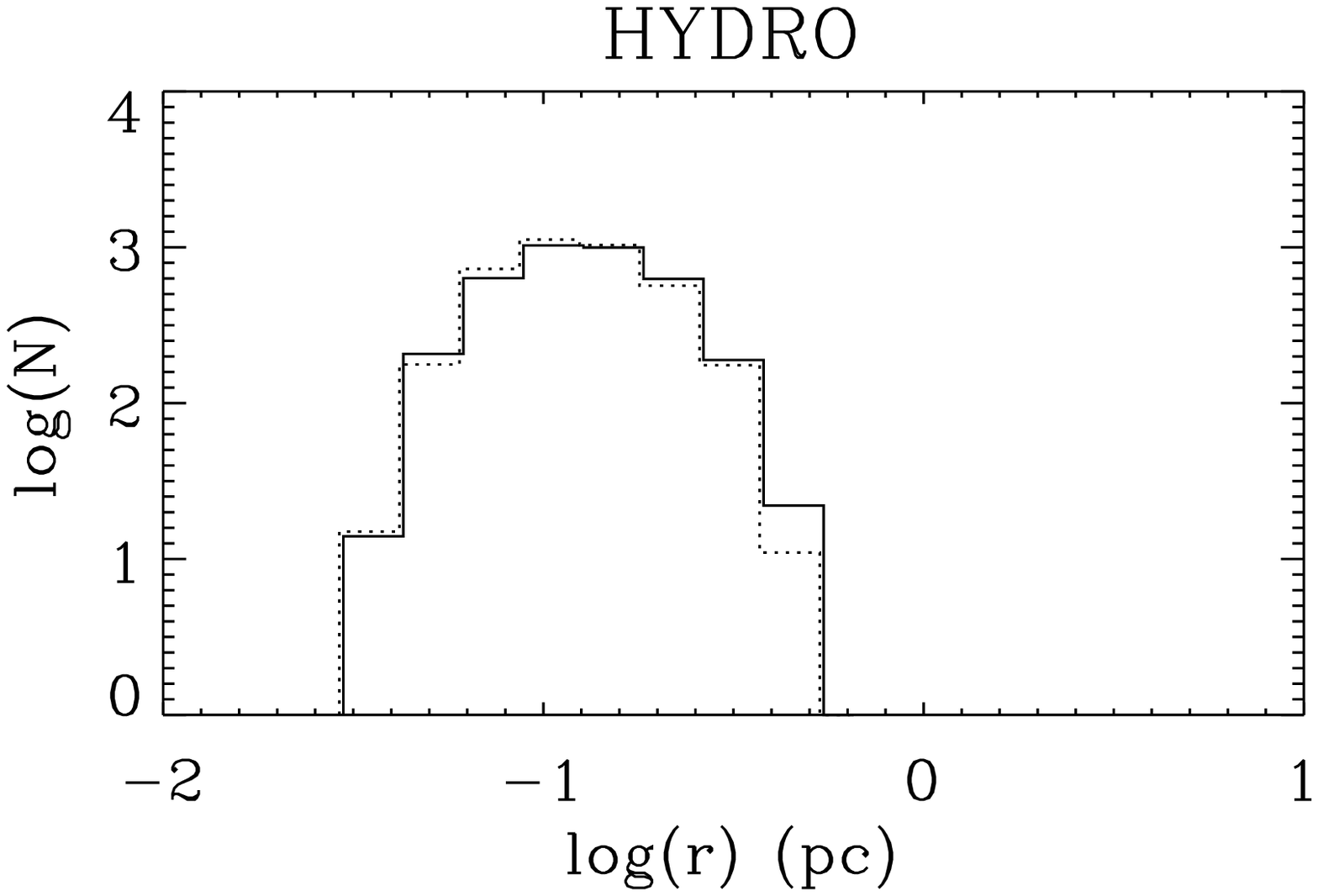}}
\end{picture}
\caption{Same as Fig.~\ref{aspect_ratio_hydro_decay1} for the distribution of the thickness of the clumps.}
\label{rad_hydro}
\end{figure}

\setlength{\unitlength}{1cm}
\begin{figure} 
\begin{picture} (0,13.5)
\put(0,5){\includegraphics[width=8cm]{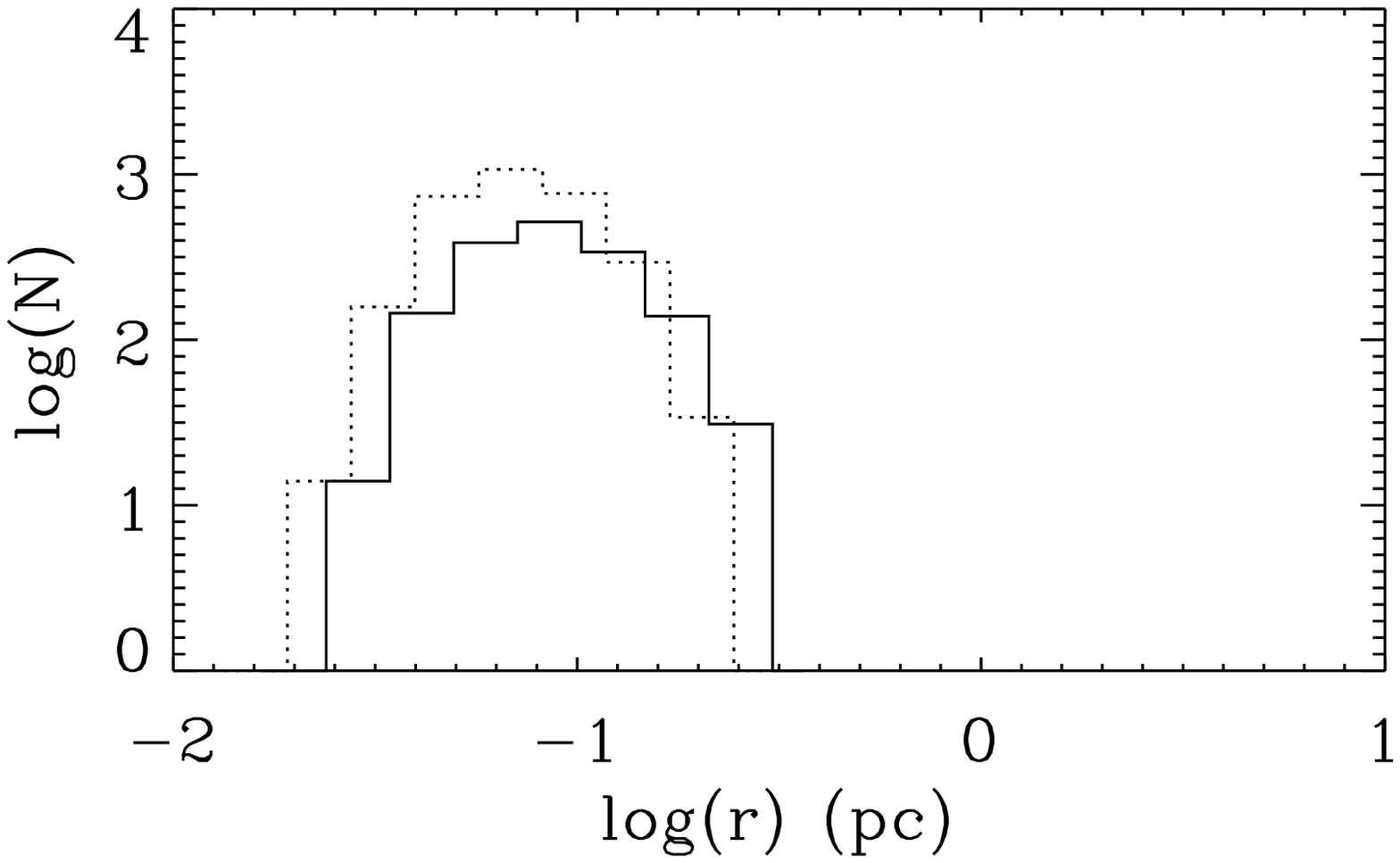}}
\put(0,9){\includegraphics[width=8cm]{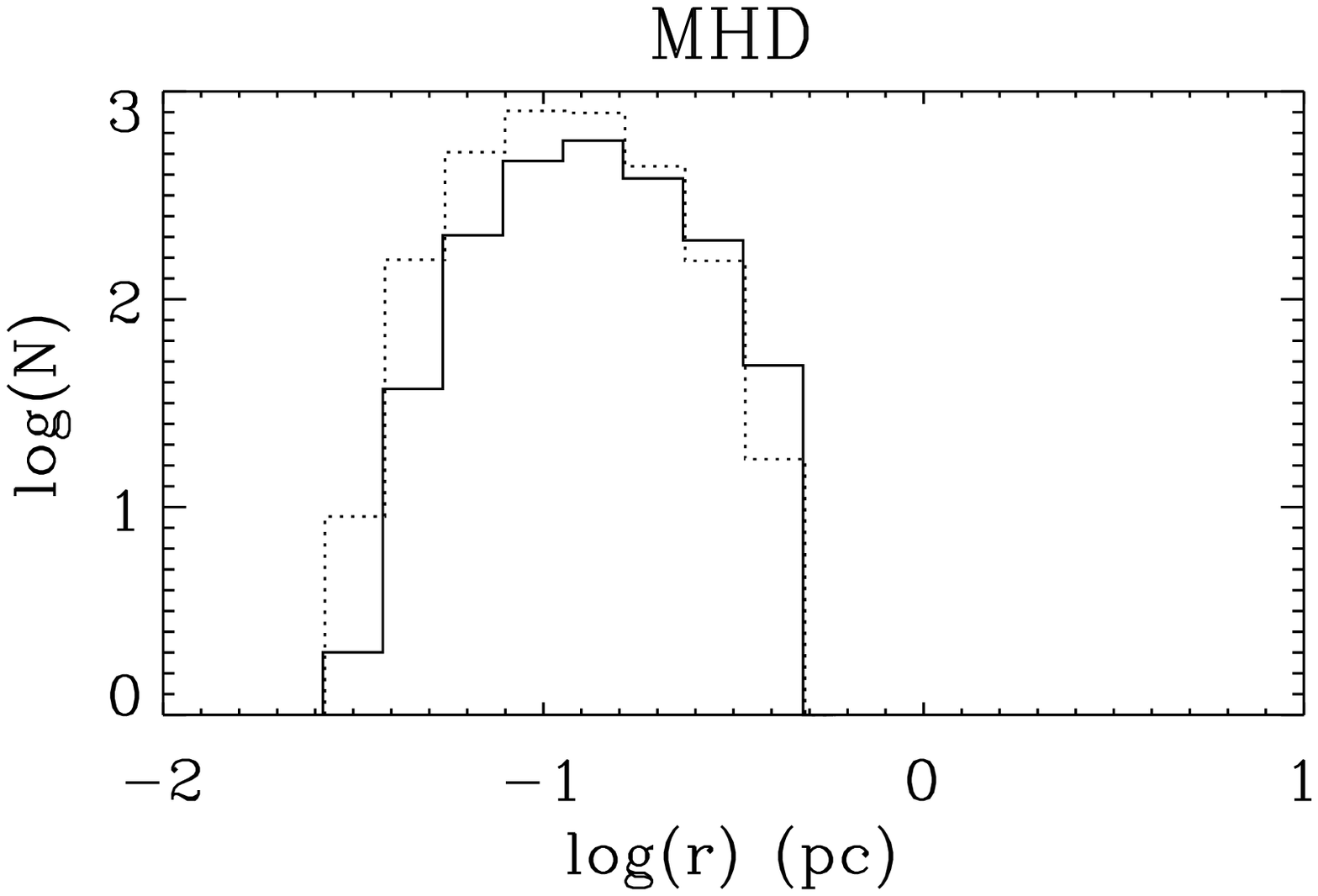}}
\put(0,0){\includegraphics[width=8cm]{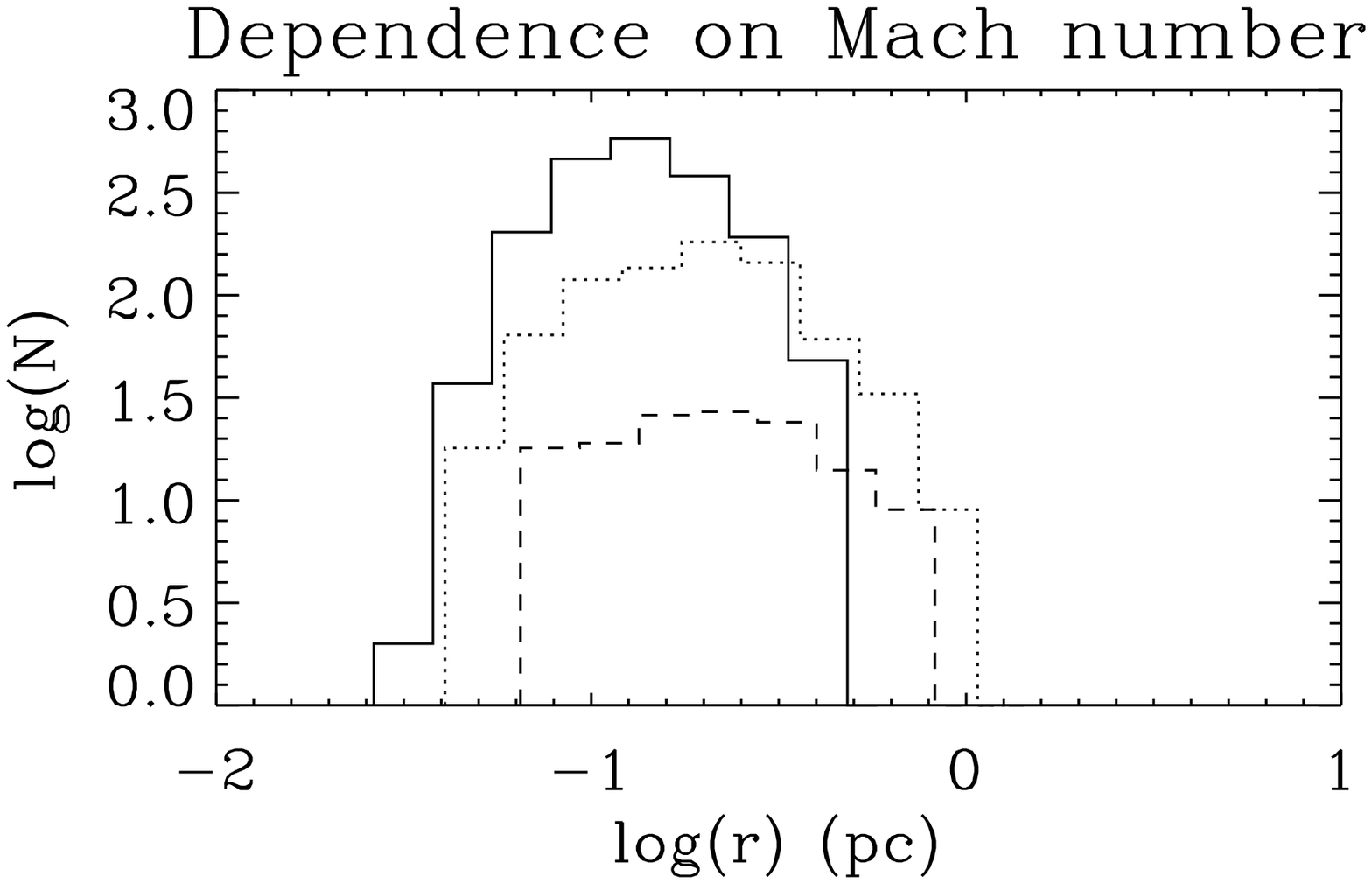}}
\end{picture}
\caption{Same as Fig.~\ref{aspect_ratio_mhd_decay1} for the distribution of the thickness of the clumps.}
\label{rad_mhd}
\end{figure}

Figures~\ref{length_hydro} and~\ref{length_mhd} show the thickness distribution for the Hydrodynamical and MHD 
simulations respectively. The distributions 
of the hydrodynamical and MHD simulations at Mach 10 show very similar behaviours. They peak 
at about $\simeq$0.1 pc for both thresholds (with a small shift toward higher values for the lowest 
threshold 50 cm$^{-3}$). This is  similar to the behaviour displayed by the length distribution
which also present a peak, though shifted toward larger values.
  The high resolution simulation (middle and top panels
of Fig.~\ref{length_mhd}) show again a systematic trend toward smaller values. This is 
again consistent with the peak being due to the numerical diffusion. 

Unlike for the length distribution however, 
there is no powerlaw tail, instead the whole distribution 
is a narrow peak (full width at half maximum of about 0.4). Thus we conclude that the thickness
of the clumps (that also represent the thickness of filaments when only the very elongated ones
are selected) is largely due to the finite resolution of the simulations. This in 
turns means that in order to describe physically the interstellar filaments down to their
thickness, realistic  dissipative processes should be consistently included
(see the discussion section).  It is worth to remind that the numerical algorithm
that is used in this work is very similar to most methods implemented in 
other codes used in the study of the ISM and beyond. This conclusion is therefore not restricted to the 
present work only but seemingly affect the simulations performed with solvers which 
do not explicitly treat the dissipative processes.
 Note that  if instead of the mean radius, the 
distribution of all the local radius is plotted, one finds that it typically extends
to values of about 3-4 times larger though most of the points are still close to a few grid points.

 The thickness of the clumps presents some dependence on the Mach number as
shown in bottom panel of Fig.~\ref{length_mhd}. However, changing the 
Mach number by a factor 10 leads to a shift of the size smaller than a factor 2. 
This contradicts the explanation of the interstellar filaments being entirely determined by  shocks.
A velocity perturbation at scale $L$ has a typical  amplitude $V = V_0 (L/L_0)^{0.5}$ as suggested 
by  Larson relations (Larson 1981, Falgarone et al. 2009, Hennebelle \& Falgarone 2012). 
The Rankine-Hugoniot conditions then 
lead to a density enhancement $\rho_s/\rho \simeq (V / C_s)^2= (V_0/ C_s)^2 \times  (L/L_0)$
and to a thickness $L_s/L = \rho / \rho_s =  {\cal M}^2 $. As can be seen the thickness is 
expected to vary with ${\cal M}^2$ a behaviour that we do not observe in these simulations
instead a more shallow dependence on the Mach number is observed.  It does 
not mean that compression is not playing any role. Indeed since these structures are denser than the  
surrounding medium, compression is occuring. However, the very reason of these structures being 
elongated does not seem to be a mere compression.

\section{Links between geometry, velocity and forces}
Next, we would like to understand why the clumps tend 
to have such low aspect ratio or in other words, why are 
there so many filaments in the simulations ? 
So far we have seen that the MHD simulations tend to be 
more filamentary than the hydrodynamical ones and that 
the aspect ratio does not strongly evolve with the Mach number.
There is only a weak trend for it to decrease when ${\cal M}$
increases (bottom panel of Fig.~\ref{aspect_ratio_mhd_decay1}).
Both facts do not straighforwardly
 agree with the earlier proposition that 
filaments are due to the collision of two shocked sheets (e.g. Padoan et al. 2001) and suggest that the process entails other aspects
than a mere  compression. 
Indeed in this scenario, one would expect high Mach number flows
 to be more filamentary
but also since magnetic field renders the collisions less supersonic, it is unclear 
why magnetized simulations would be more filamentary.
On the other hand, the simple numerical experiment presented in Sect.~\ref{prelimi}
suggests that the filaments could simply be fluid 
particles that have been stretched by the turbulent motions. 
The most important difference with the shock scenario is that 
filaments are not born as elongated objects rather they become elongated 
as time proceeds.   

In this section we investigate whether the mechanism by which filaments form
is indeed the stretch of the fluid particles induced by  turbulence.
It should be kept in mind that shocks or say convergent motions must play 
a role at some stage because the gas within the filaments must accumulate. 
The question is then to try to estimate their respective influence. 

\subsection{Alignment between strain and filament axis}

\setlength{\unitlength}{1cm}
\begin{figure} 
\begin{picture} (0,8.5)
\put(0,0){\includegraphics[width=8cm]{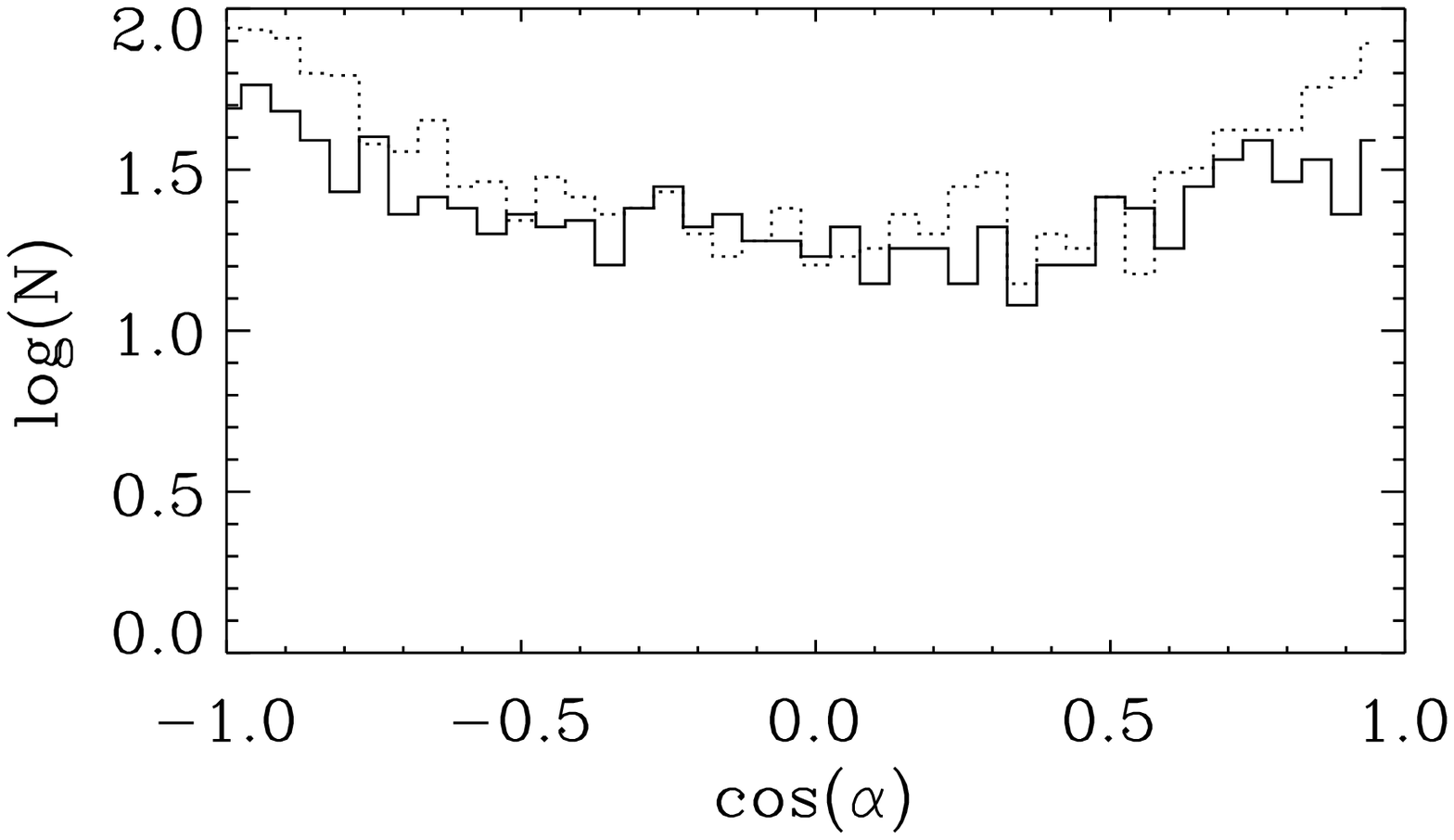}}
\put(0,4){\includegraphics[width=8cm]{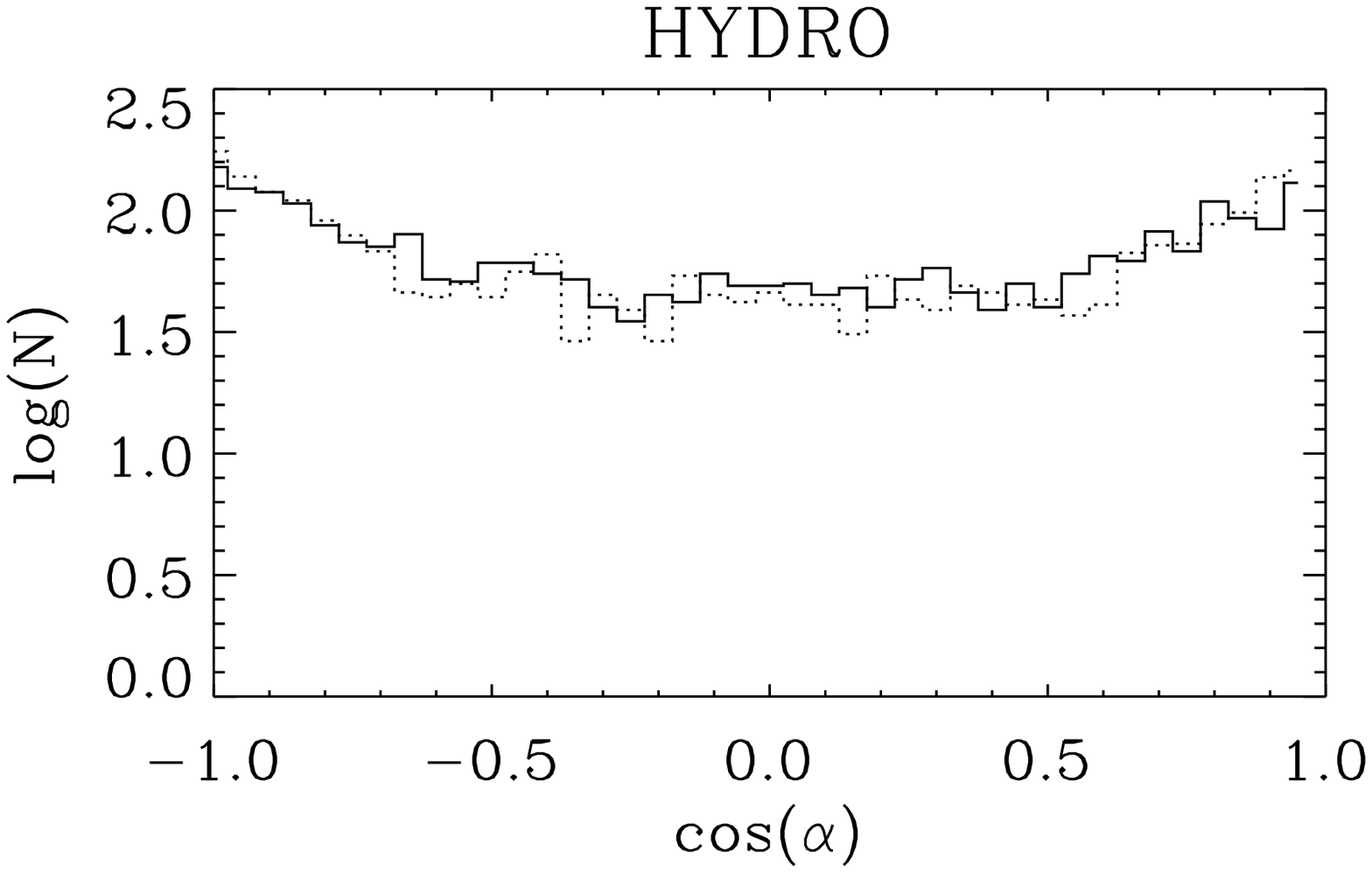}}
\end{picture}
\caption{Same as Fig.~\ref{aspect_ratio_hydro_decay1} for the distribution of 
 $\cos \alpha$ (the angle
between the main axis and the strain) in the clumps.}
\label{cos_as_hydro_decay}
\end{figure}

\setlength{\unitlength}{1cm}
\begin{figure} 
\begin{picture} (0,13.5)
\put(0,5){\includegraphics[width=8cm]{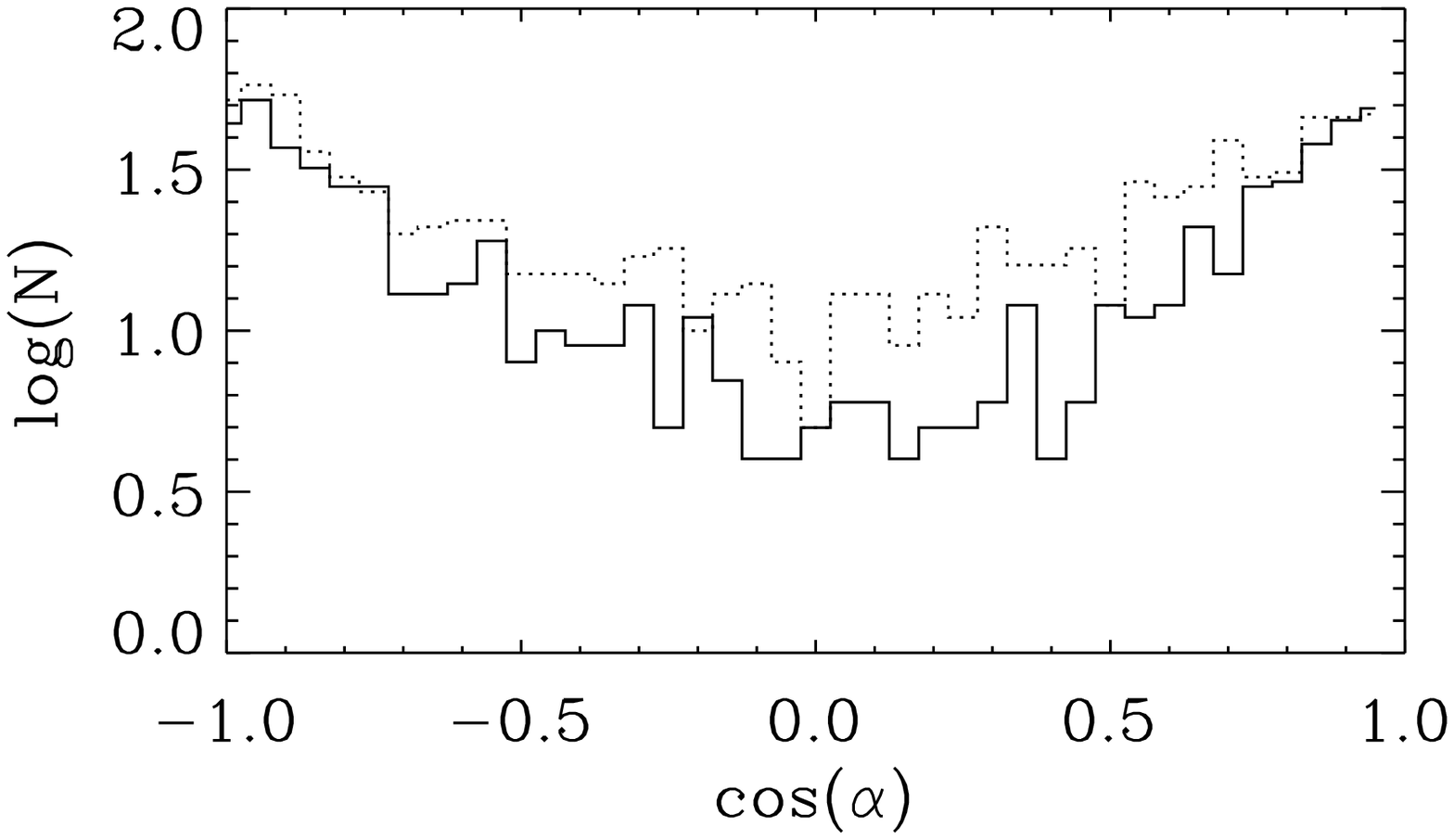}}
\put(0,9){\includegraphics[width=8cm]{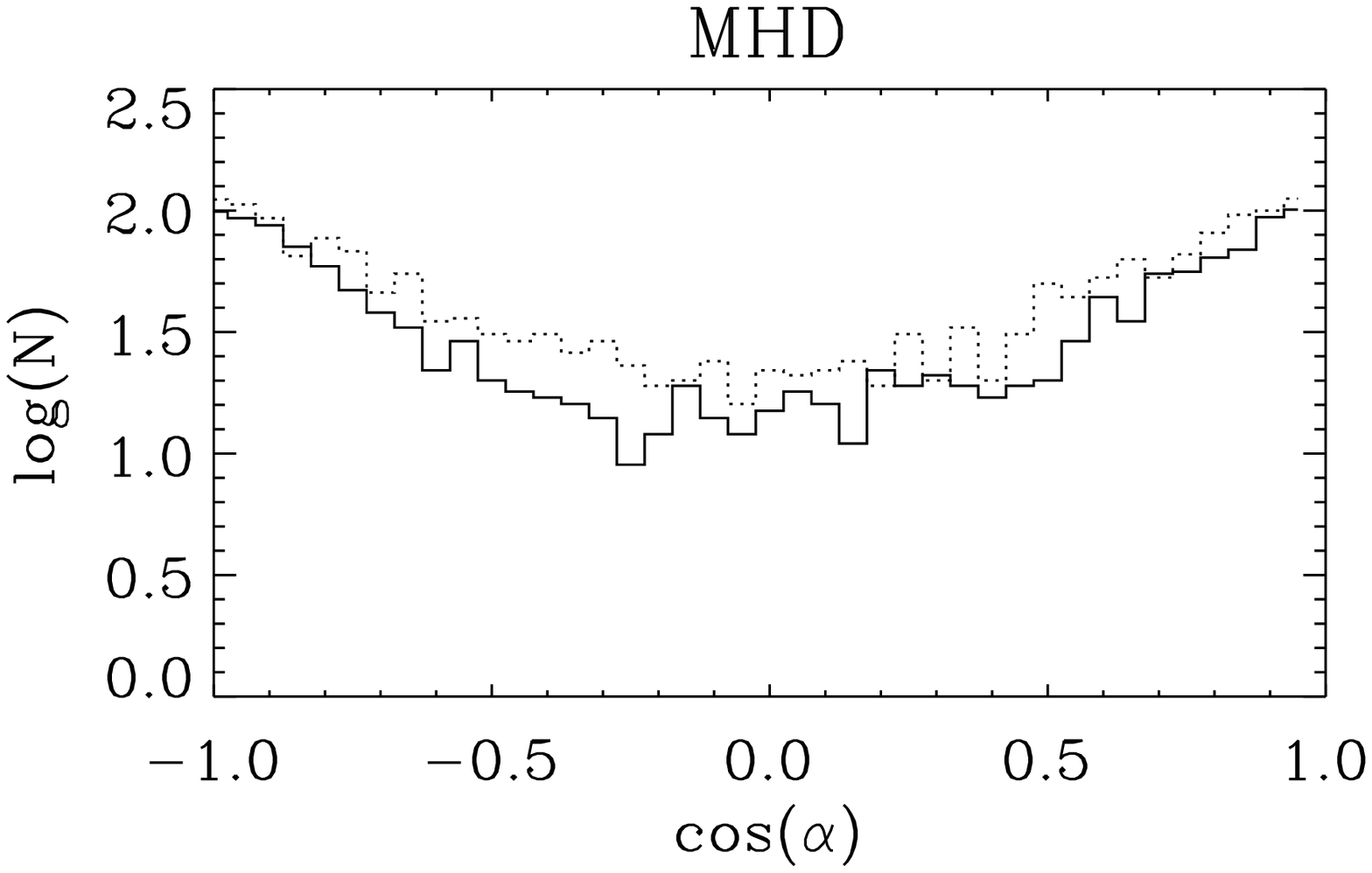}}
\put(0,0){\includegraphics[width=8cm]{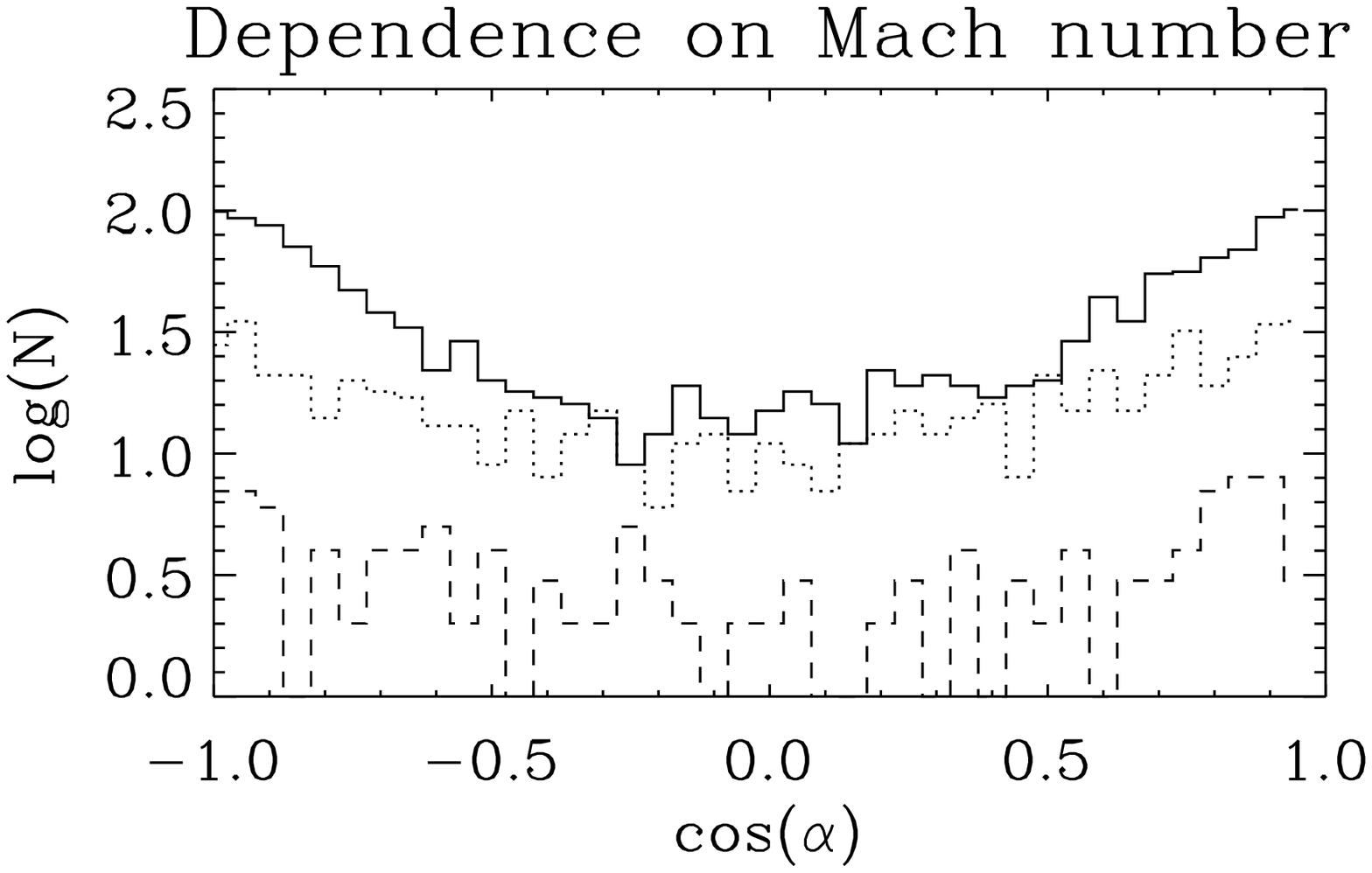}}
\end{picture}
\caption{Same as Fig.~\ref{aspect_ratio_mhd_decay1} for the distribution of 
 $\cos \alpha$ (the angle
between the main axis and the strain) in the clumps.}
\label{cos_as_mhd_decay}
\end{figure}

To investigate whether the filaments are indeed particle fluids which 
have been stretched by the turbulent motions, we 
study the correlation between the eigenvector associated to 
the largest eigenvalue of the inertia matrix that we later 
call the axis of the filament, and the eigenvector of the 
largest eigenvalue of the strain tensor which gives 
the direction along which the clumps are stretched. More 
precisely, we study the distribution of $\cos \alpha$, 
where $\alpha$ is the angle between these two eigenvectors.
Indeed if the two eigenvectors tend to be aligned, this will be 
a clear evidence that filaments are stretched by the 
velocity field.

Figures~\ref{cos_as_hydro_decay} and \ref{cos_as_mhd_decay}
show results for the two density thresholds.
Clearly in all cases there are more clumps for which 
$\cos \alpha$ is close to -1 or 1 than  clumps 
for which $\cos \alpha$ is close to 0. In other words, 
there is a trend for the filament axis and the strain to be 
aligned. This clearly shows that the primary cause of the 
existence of filaments, that is to say the existence 
of elongated clumps,  is the stretching of the 
fluid particles induced by the turbulent motions. This 
does not imply that shocks may not be also forming filaments, 
in particular intersection between two shocked layers
as it has been previously suggested. However, this cannot 
be the dominant mechanism because in such configurations, one 
would expect the filament axis and the strain to be 
randomly distributed.

A comparison between Figs.~\ref{cos_as_hydro_decay} and
 \ref{cos_as_mhd_decay} also reveals that the trend is clearly 
more pronounced for MHD simulations than for hydrodynamical ones. 
For example, there are about 8-10 times more objects having 
$|\cos \alpha|=1$ than objects with $\cos \alpha=0$ in 
the MHD case. In the hydrodynamical case, this ratio is about $3-5$.
Again since the MHD simulation is more filamentary that the hydrodynamical 
one, this is not consistent with the dominant origin of 
filaments being due to shocks since magnetic field tends
to reduce the effective Mach number. Finally we see that 
in lower Mach simulations (dotted and dashed lines of bottom panel in Fig.~\ref{cos_as_mhd_decay}),
this trends is also present but reduced.

\subsection{Comparison between strain and divergence}

\setlength{\unitlength}{1cm}
\begin{figure} 
\begin{picture} (0,8.5)
\put(0,0){\includegraphics[width=8cm]{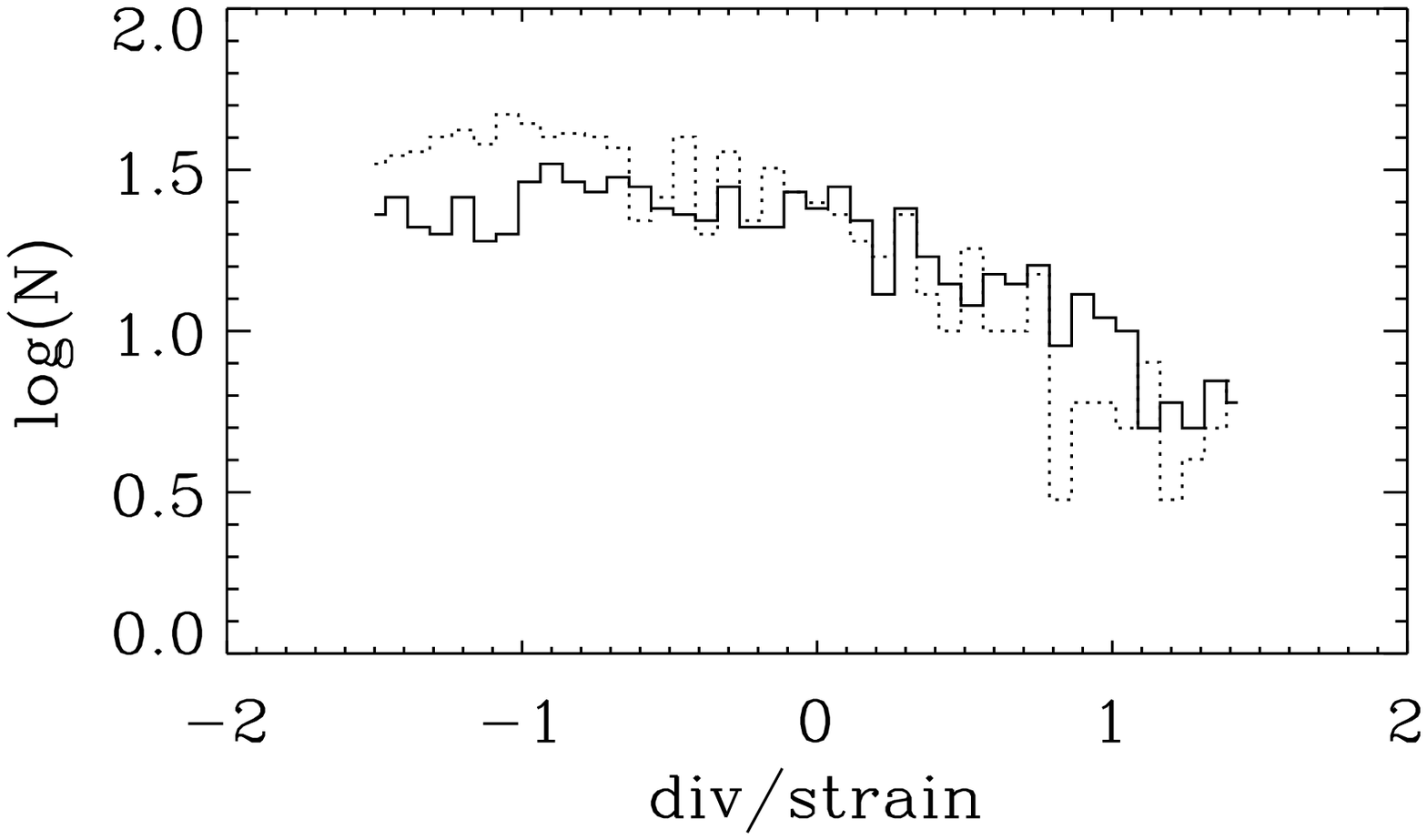}}
\put(0,4){\includegraphics[width=8cm]{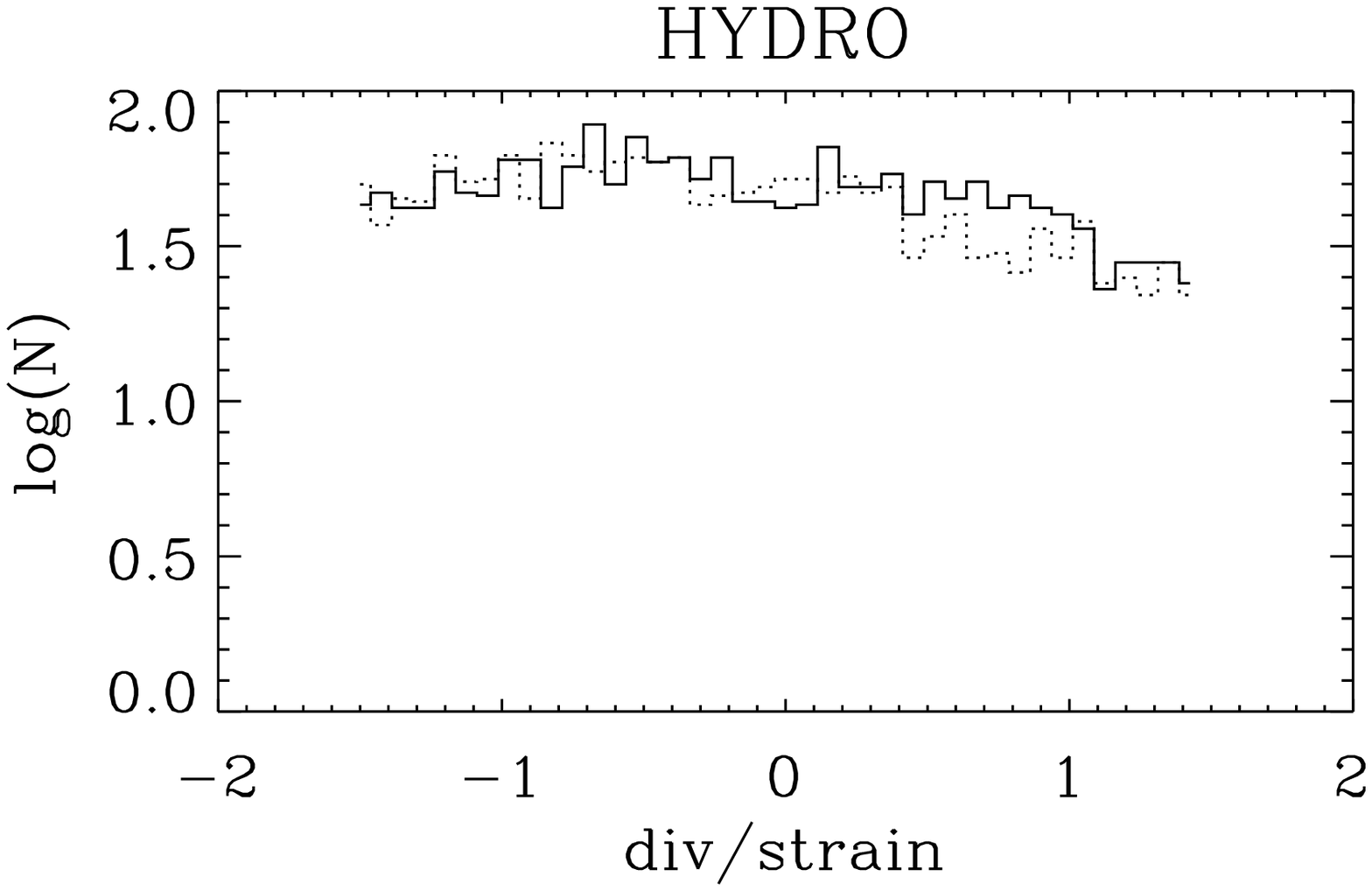}}
\end{picture}
\caption{Same as Fig.~\ref{aspect_ratio_hydro_decay1} for the distribution of 
the ratio of the divergence and strain in the clumps.}
\label{div_strain_hydro_decay}
\end{figure}

\setlength{\unitlength}{1cm}
\begin{figure} 
\begin{picture} (0,13.5)
\put(0,5){\includegraphics[width=8cm]{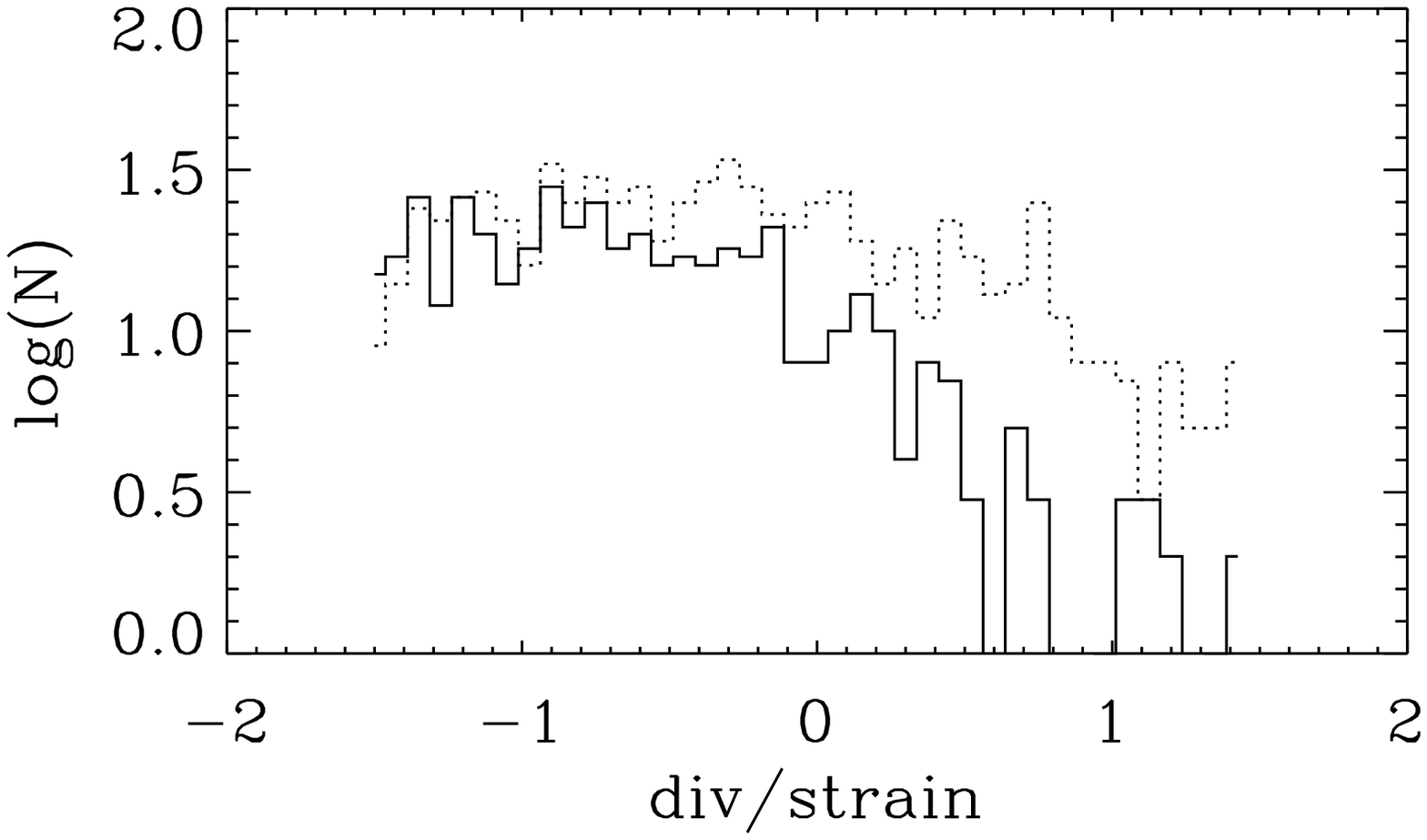}}
\put(0,9){\includegraphics[width=8cm]{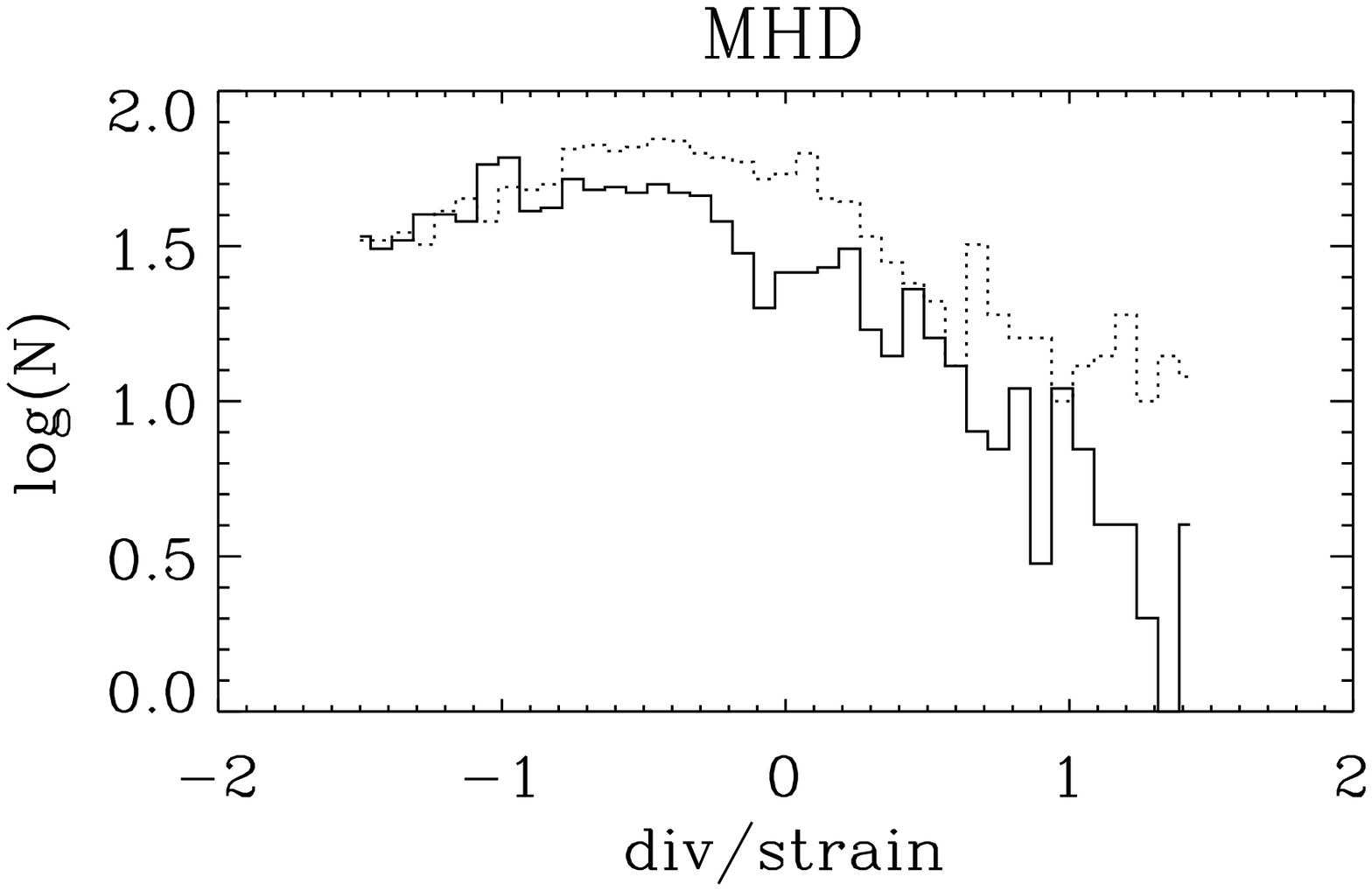}}
\put(0,0){\includegraphics[width=8cm]{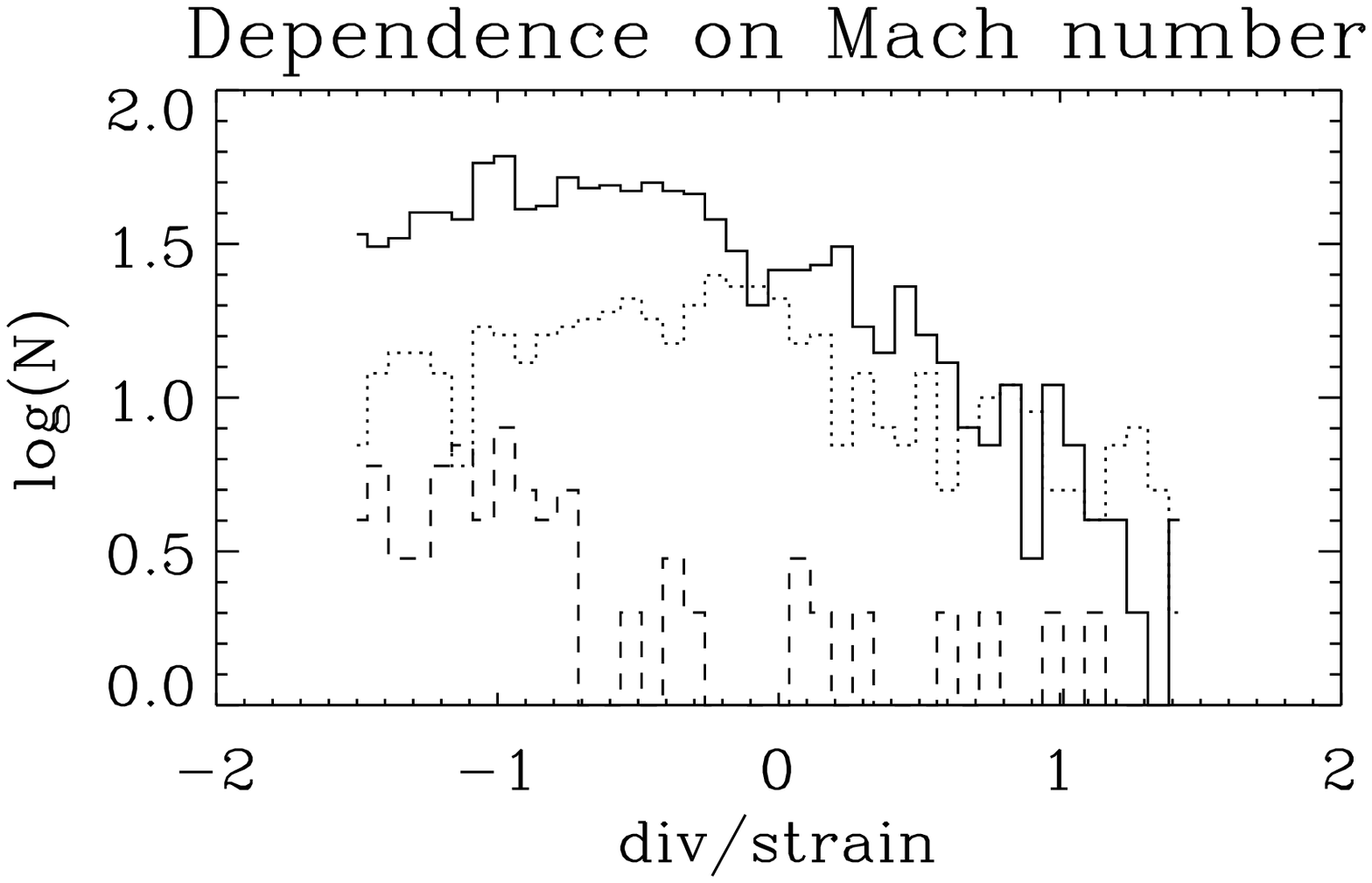}}
\end{picture}
\caption{Same as Fig.~\ref{aspect_ratio_mhd_decay1} for the distribution of 
the ratio of the divergence and strain in the clumps.}
\label{div_strain_mhd_decay}
\end{figure}

The correlation between the strain and the filament axis suggests that 
strain is an important, if not the dominant reason for having filaments 
within the ISM. However, the clumps are  regions of high densities where the gas has been 
accumulated. Therefore it is important to understand the respective role of 
the compressive motions and the straining ones. For this 
purpose, we study the distribution of the ratio of the divergence and the strain, 
$r_{ds}$, 
calculated as described in Sect.~\ref{strain}. This quantity 
allows us to directly estimate whether the structure is globally 
contracting or expanding and whether this global change of volume
dominates or is dominated by the change of shape described by the strain.

Figure~\ref{div_strain_hydro_decay} 
shows results for the hydrodynamical simulation for
the two thresholds (50 cm$^{-3}$ upper panels and 200 cm$^{-3}$
lower panels).  The distribution 
of $r_{ds}$ is pretty flat  and extends
between about $-1.2$ and 1.2. It is roughly symmetrical with respect to 0
particularly for the lowest density threshold.  
This indicates that the contribution of compressive and solenoidal 
motions for the dynamics of the clumps is comparable though most clumps 
have a mean strain larger than their divergence (but few clumps have a small 
$r_{ds}$).  For our highest threshold, the number of clumps that are expanding
is lower than the number of clumps that are contracting.

Figure~\ref{div_strain_mhd_decay} 
shows results for the MHD simulation.
The general trends are qualitatively similar than for the hydrodynamical 
simulations except for the important fact that the distributions are 
much less symmetrical with respect to zero. There are much less expanding clumps
than contracting ones. This suggests that the magnetic field tends 
to confine the clumps and makes their reexpansion more difficult. 

Apart from this, the clear conclusion is that both 
compression/expansion and strain are important with the latter 
being generally dominant over the former.

\subsection{Confinement by Lorentz force}

\setlength{\unitlength}{1cm}
\begin{figure} 
\begin{picture} (0,8.5)
\put(0,0){\includegraphics[width=8cm]{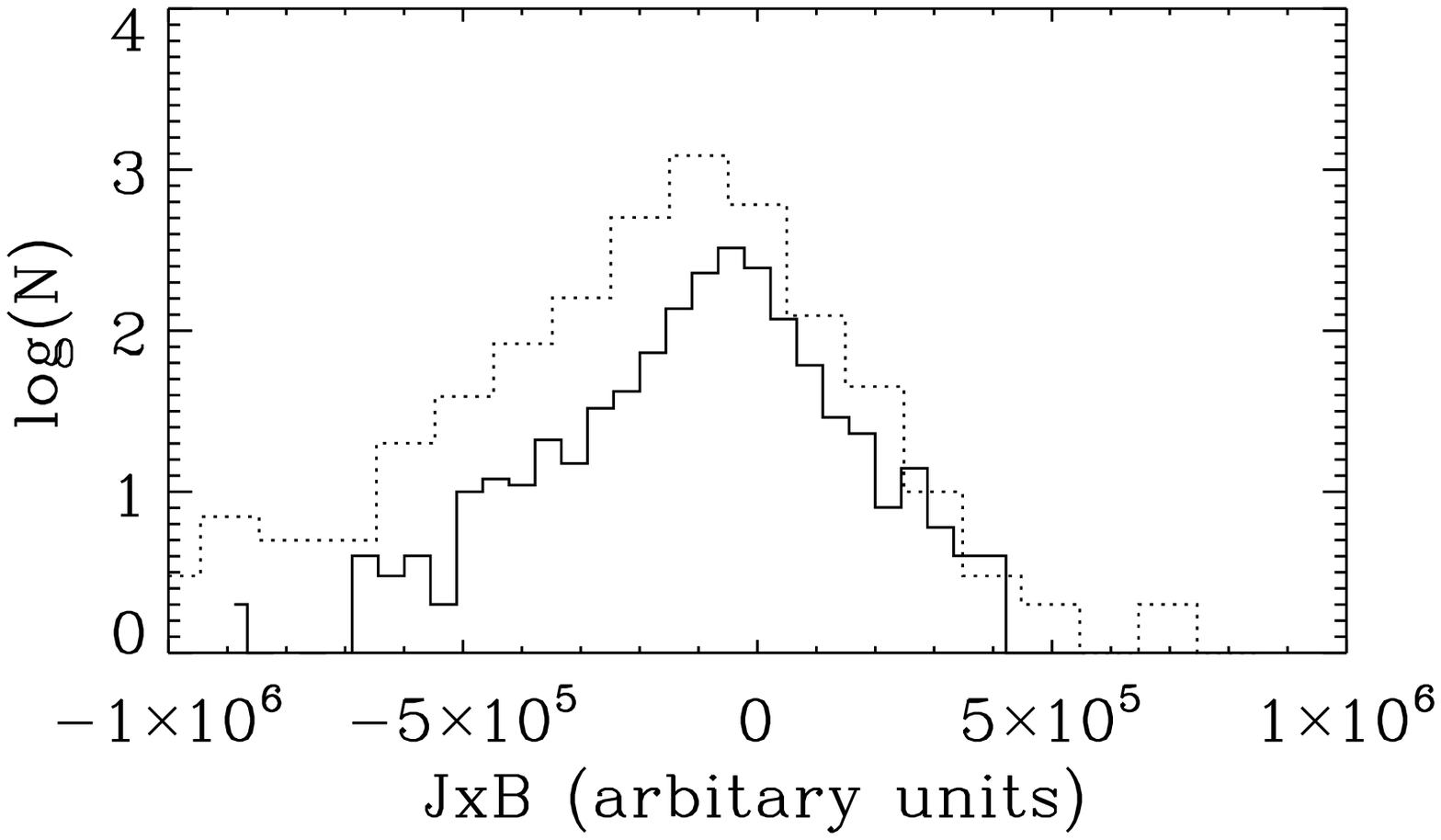}}
\put(0,4){\includegraphics[width=8cm]{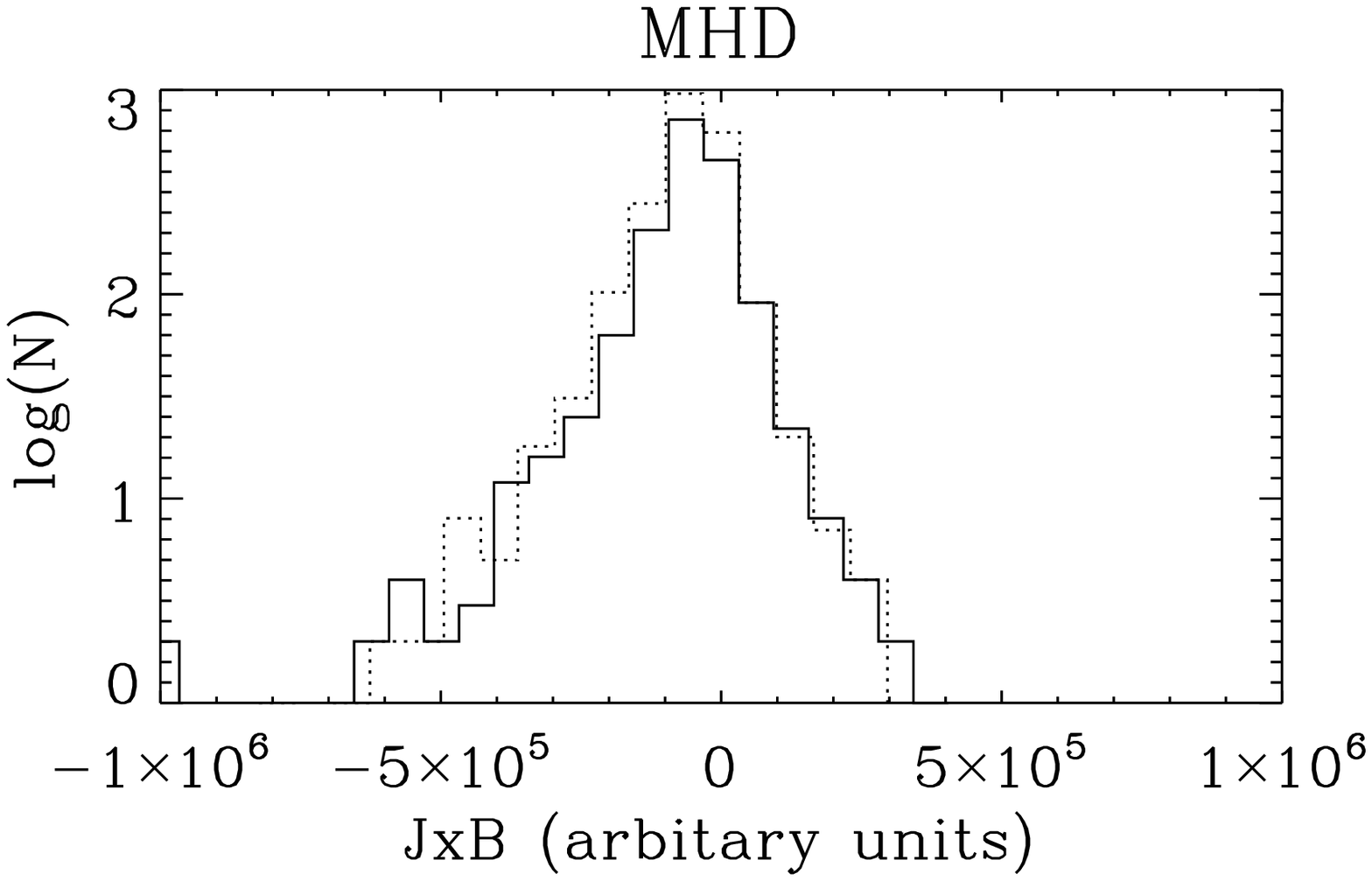}}
\end{picture}
\caption{Mean value of the Lorentz force 
radial component (see text) within the clumps. 
(threshold 50 cm$^{-3}$: upper panel and 200 cm$^{-3}$: lower panel) for the 
high resolution MHD simulation at time $t=2.26$ Myr (dotted line) and the fiducial 
run at time 1.81 Myr (solid line).}
\label{jxb_mhd_decay}
\end{figure}

\setlength{\unitlength}{1cm}
\begin{figure} 
\begin{picture} (0,8.5)
\put(0,0){\includegraphics[width=8cm]{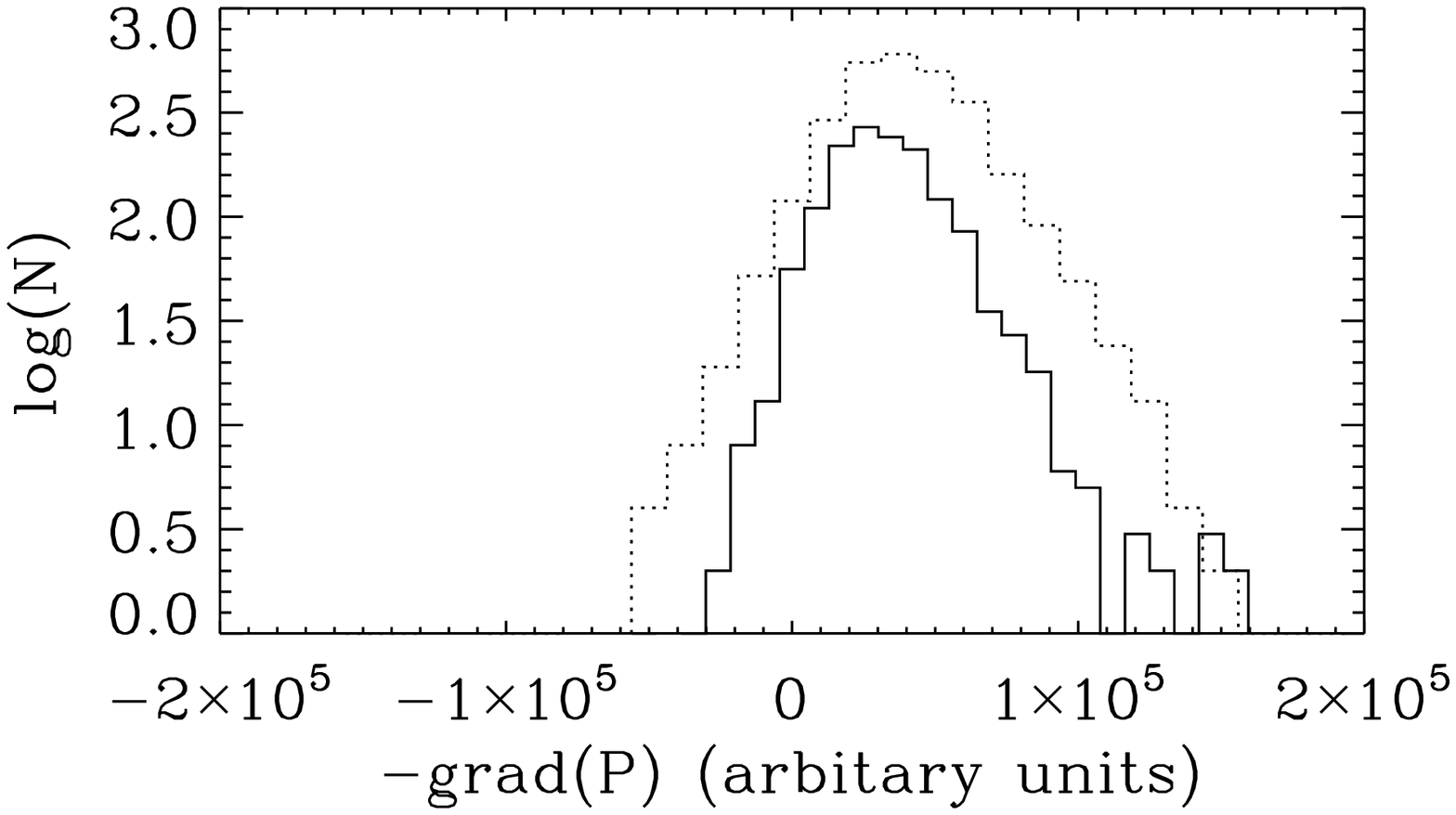}}
\put(0,4){\includegraphics[width=8cm]{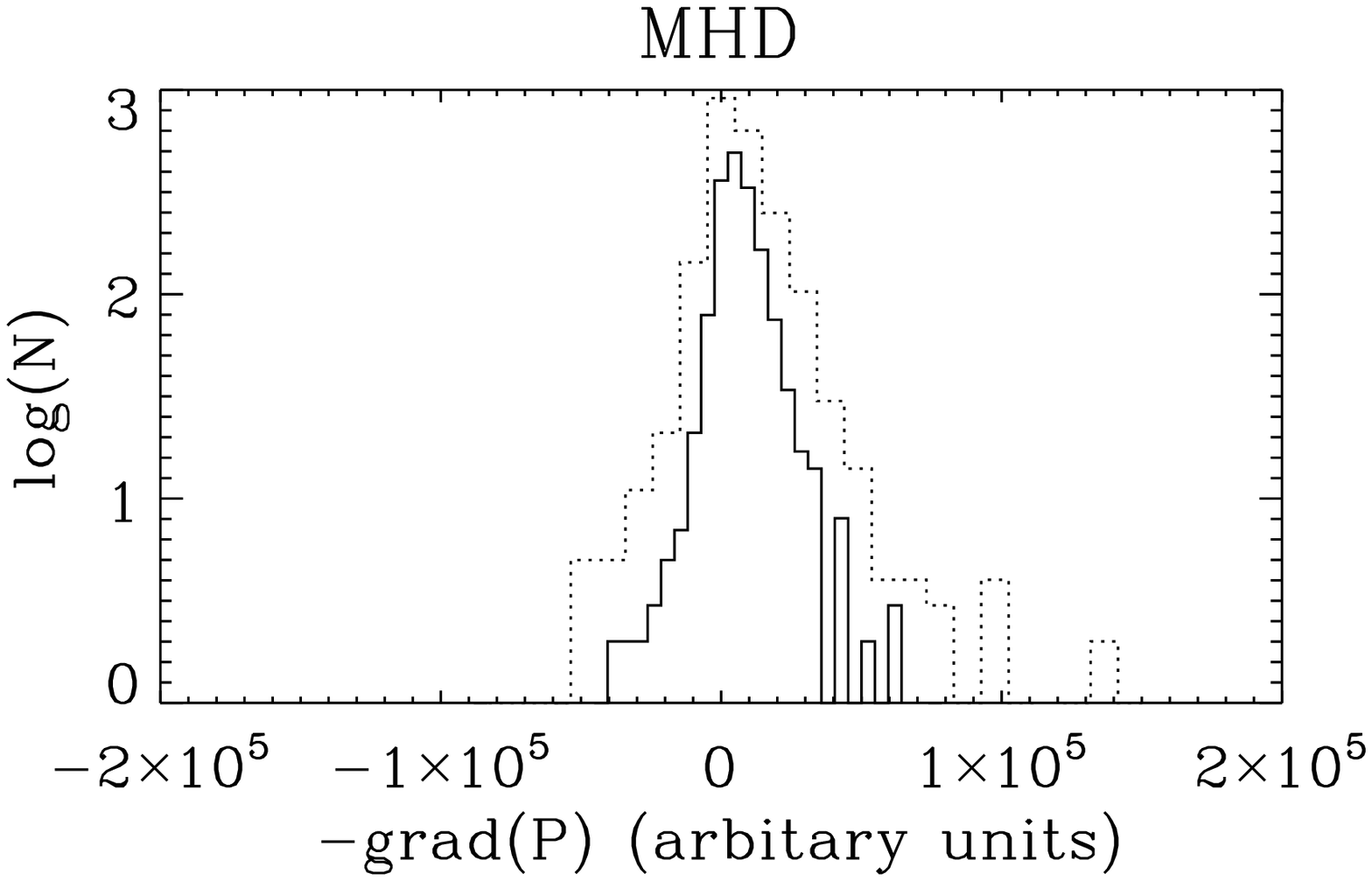}}
\end{picture}
\caption{Same as Fig.~\ref{jxb_mhd_decay} for the mean value of the pressure force 
radial component (see text) within the clumps }
\label{press_mhd_decay}
\end{figure}

To verify that the Lorentz force is indeed reducing the clump expansion, 
we have computed the mean component of the Lorentz force in the direction 
perpendicular to the local direction of the clumps. 
 To accomplish this, given a cell center $M$,
 we first compute the position of $M'$ defined by 
${\bf GM'}= ({\bf u_i^j .GM }) {\bf u_i^j}$, that is to say 
$M$ is projected onto the local axis in $M'$ ($u_i^j$ is defined 
as explicited in Sect.~\ref{skeleton}). 
The component of the  force perpendicular to the  local direction 
of the clump (${\bf u_i^j}$) is thus simply ${\bf F(M).MM'}/|MM'|$.
The total contribution is thus obtained by summing over the whole 
structure
\begin{eqnarray}
{ \int {\bf F(M).MM'}/|MM'|  dm \over \int dm}.
\end{eqnarray}
The result is displayed in Fig.~\ref{jxb_mhd_decay} 
for the two density thresholds. As can be seen there are 
structures that the Lorentz force tends globally to 
confine (integrated component is negative) and structures
that it tends to expand. However, it 
is clearly skewed towards the negative values and in most cases
the Lorentz force tends to confine the structure and maintain 
the coherence of the clumps. More quantitatively, we find that 
depending on the run parameters and thresholds the fraction 
of clumps for which the integrated component of the Lorentz
force is negative, is between 55 and 65$\%$.

For comparison, it is interesting to investigate the effect of the pressure 
force and the corresponding distribution is plotted in 
Fig.~\ref{press_mhd_decay}. As expected it is seen that the 
pressure force tends almost always to expand the structure. We also see
from the amplitude that the Lorentz force largely dominates  
over the pressure force. 

These results are in good agreement with the results inferred from the 
divergence of strain ratio, $r_{ds}$ and also with the simple
numerical experiment presented in section~\ref{prelimi}. The magnetic force tends to 
prevent the clump re-expansion  because the magnetic field lines permeate through 
the clumps and connect the fluid particles together.

\section{Discussion}
The role played by two important processes, namely non-ideal mhd effects and gravity, has not been included or investigated in this work 
and require at least some discussions.

\subsection{The role of dissipative processes}
Although our simulations seem to roughly agree with the Herschel observations about the 
constancy of the filament inferred by Arzoumanian et al. (2011). It is 
important to keep in mind that this thickness is most likely setup in the present 
simulations (and in any similar simulation) by the numerical diffusion which 
operates at the scale of a few computing cells (here about 0.05 pc for the fiducial run). 
Therefore although encouraging this resemblance must be taken with the greatest care
at this stage. Moreover, our filaments are selected using a density threshold which 
is also different from what Arzoumanian et al. (2011) have been doing. 

Nevertheless, the conclusion that the thickness of the filaments, at least  in the way we define
them, is setup by a dissipative process is  intriguing and leads to the 
obvious question: which dissipative mechanism could actually produce a scale 
comparable to a size of 0.1 pc ? The only known dissipative mechanism in the interstellar
medium, which leads to comparable scales is the ion-neutral friction as investigated 
by Kulsrud \& Pearce (1969).

Let us remind that when ion-neutral friction is taken into consideration the 
induction equation can be written as 
\begin{equation}
\partial _t {\bf B}  + \nabla \times  ({\bf B} \times {\bf v}) = 
\nabla \times \left( {1 \over 4 \pi \gamma_{ad} \rho \rho_i}  ( (\nabla \times {\bf B}) \times {\bf B}) \times {\bf B} 
 \right). 
\end{equation}
where $\gamma_{ad}$ is the ion-neutral friction coefficient whose value is about 
$\gamma_{ad} \simeq  3.5 \times 10^{13}$ g$^{-1}$ s$^{-1}$ (e.g. Shu et al. 1987).

Although the right-hand side has not the standard form of a diffusion, it is 
a second order term which  dissipates mechanical energy into heat. 
We can easily compute
a magnetic Reynolds number associated to this equation as 
\begin{equation}
R_{e,m}  = { V(l) l \over \nu },
\end{equation}
where $\nu = B^2/ (4 \pi \gamma_{ad} \rho \rho_i)$.
As it is done in standard approach  of turbulence, we assume that the energy 
flux, $\epsilon = \rho V(l)^3 / l$, is constant through the scales.
Thus we can write
\begin{equation}
R_{e,m}  = { \epsilon^{1/3} \rho^{-1/3} l^{4/3} \over \nu },
\end{equation}
Estimating $\epsilon$ at the integral scale, $L_0$, we get 
\begin{equation}
R_{e,m}  = \left( { \rho _0 \over \rho } \right)^{1/3} {V_0 \over L_0^{1/3}} {  4 \pi \gamma_{ad} \rho \rho_i  \over B^2 } l^{4/3}.
\end{equation}

The smallest scale which can be reached in a turbulent  cascade is typically obtained when the Reynolds number is 
equal to about 1. This leads for $l_{diss}$ the dissipation length the following expression
\begin{equation}
l_{diss}  = \left( {  L_0^{1/3}  \over \rho _0^{1/3} V_0 } \right)^{3/4}  \left( { B^2 \over  4 \pi \gamma_{ad} \rho^{2/3} \rho_i  } 
\right)^{3/4}.
\end{equation}

The impact of non-ideal MHD processes on the density field can be clearly seen in the simulations performed
by Downes \& O'Sullivan (2009, 2011). These authors have run similations of molecular clouds at the scale 
of 0.2 pc. The figures 1 of these two papers display the density field in the ideal MHD case and 
in the non-ideal MHD one. Clearly, many small scale filaments are seen in the ideal MHD  simulation
which evidently are very close to the numerical resolution. This is rather different from what is 
seen in the non-ideal MHD simulations in which only large scale structures (bigger that the cell
size) are produced.  Interestingly, the large scale patern is unchanged but the small scales
are completely differents. 
These simulations demonstrate that indeed 
non-ideal MHD  processes have a very strong 
impact on the clump structure.

To estimate the  dissipation length we  use typical  values for the diffuse interstellar medium. 
The values of $V_0$, $\rho_0=m_p n_0$ and $L_0$ are linked through the Larson relations (Larson 1981).   
We will choose as fiducial values $V_0=2.5$ km s$^{-1}$, $\rho_0=100$ cm$^{-3}$ and $L_0=10$ pc. 
Typical magnetic fields are about 5 $\mu$G in the diffuse gas and 10-20$\mu$G in the 
molecular gas for densities of a few 10$^3$ cm$^{-3}$. The ionization is also important and vary
significantly. 

In the molecular gas the ionization is  of the order of $10^{-6}-10^{-7}$ 
(Le Petit et al. 2006, Bergin \& Tafalla 2007) and the 
ion density $\rho_i$ can be approximated as $C \sqrt{\rho}$ where
 $C=3 \times 10^{-16}$ cm$^{-3/2}$ g$^{1/2}$.
Using this expression, a density of $10^3$ cm$^{-3}$ and a magnetic field
 of $20 \mu$G, we get 
$l_{diss} \simeq 0.2$ pc which is entirely reasonable. Note that assuming 
that the magnetic field 
increases as $\sqrt{\rho}$ (e.g. Crutcher 1999), we find that the density 
dependence is extremely shallow
seemingly suggesting that this scale could indeed be representative of a 
broad range of conditions. These numbers as well as the analysis 
is similar to the results presented in McKee et al. (2010). 

In the diffuse gas, which has a density of only 
a few 100 cm$^{-3}$, the ionization 
is about $10^{-4}-10^{-5}$ (e.g. Wolfire et al. 2003) which leads for a 
density of 200 cm$^{-3}$ and a 
magnetization of 5 $\mu$G to $l_{diss} \simeq 3 \times 10^{-3} - 2  \times 10^{-2} $ pc, that is to 
say much smaller values.  

It is important to stress that these numbers remain indicative 
at this stage and should not be directly interpreted 
as the size of the structures which can certainly be different by a factor 
of a few. Nevertheless a clear consequence of these estimates is that 
the filaments should be much thinner in the weakly shielded 
gas ($A_v<1$), in particular in the HI and at the periphery of 
molecular clouds.

\subsection{Influence of gravity}
Gravity can also play an important role in the formation of massive filaments. 
Indeed gravity is well known to amplify initial anisotropies (Lin et al. 1965)
and has been found in various studies to play an important role in triggering 
the formation of self-gravitating filaments (Hartmann \& Burkert 2007, Peretto et al. 2007). 
Indeed in these studies gravity acts to amplify the initial elongation of a clump which 
could have been induced by turbulence. 

It is therefore likely that gravity can play  a significant role in the formation 
of the most massive filaments and may be even for setting the width of the marginally self-gravitating 
ones as recently advocated by Fischera \& Martin (2012). 

The nature of this elongation 
is  quite different from what has been studied here. It is a selective 
contraction along  two directions rather than a stretching along one direction.
We stress however that  gravity is self-consistently included in the 
collision flow calculations presented in the appendix~\ref{conv} and that as discussed in the appendix 
the results are very similar to what has been found for the more diffuse ISM studied 
in the paper. Thus is seems that except probably for very dense filaments (e.g. integral-shaped Orion  or DR21 filaments),  
gravity may not modify the picture very significantly.

\section{Conclusion}

We have performed a series of numerical simulations to study the 
formation of clumps in the turbulent ISM
paying particular attention to the reason that causes 
the elongation.  We have run both 
hydrodynamical and MHD simulations and have varied the Mach number.
To verify the robustness of our results, we have used two different setup
namely  
decaying turbulence and colliding flows.  
To quantify the structure properties, we first extract the clumps using a 
simple clipping algorithm. We then compute and diagonalize the 
inertia matrix and the strain tensor. We also develop a skeleton-like
approach which allows to infer the mean thickness and to compute whether
forces tend to expand or confine the structures.
We find that in all simulations most of the clumps are significantly 
elongated and the main axis of the structure tends to be aligned 
with the strain particularly in MHD simulations. The proportion of filamentary objects is 
higher in the MHD simulations that in the hydrodynamical ones 
in which  a significant fraction of the clumps are  sheets rather than filaments.
While the pressure force tends as expected to expand the clumps, the Lorentz force
tends on average to confine them allowing the filaments to persist longer.
In all simulations, irrespectively of the magnetic intensity and Mach number, 
we find that the thickness of the clumps, that is to say the mean thickness
of all the clump branches, is always close to a few computing cells
seemingly suggesting that in the ISM dissipative processes are responsible
of setting its value. Performing simple orders of magnitude, we find that the 
ion-neutral friction in regions of sufficient extinction,  
leads to values that are close to what has been recently inferred 
from Herschel observations (Arzoumanian et al. 2011). In unshielded 
regions like HI or in the outskirts of molecular clouds where the ionization 
is larger, this would imply that the filament thickness should be 
at least ten times smaller.  \\ \\

\emph{acknowledgments}
I thank the anonymous referee for insightful comments. 
I thank Pierre Lesaffre, Philippe Andr\'e and Doris Arzoumanian for many 
related and inspiring discussions. 
This work was granted access to HPC resources of CINES under the 
allocation x2009042036 made by GENCI (Grand Equipement National de Calcul Intensif).
PH acknowledge the finantial support of the Agence National pour la Recherche through the 
 COSMIS project.
This research has received funding from the European Research Council under the European
 Community's Seventh Framework Programme (FP7/2007-2013 Grant Agreement no. 306483).

\appendix

\section{Results of decaying turbulence with a 
magnetic field tilted with respect to the mesh}
\label{tilt}
As we found in this paper that in magnetized flows, the 
clumps are more filamentary, it is important to check that no obvious 
numerical artefact is producing this effect. We have thus repeated 
one of the runs (decaying turbulence with 5$\mu$G and one level of AMR) 
with a magnetic field initially tilted to 45$^\circ$ with respect to the mesh. 
Figures~\ref{triaxis_incl} and~\ref{cos_as_mhd_incl}  show respectivelly 
the bidimensional distribution $\mu_1/\mu_2$ vs $\mu_2/\mu_3$ 
and the histogram of $\cos (\alpha)$, the cosine of the angle between 
the filament axis and the strain. As can be seen they are very similar to the 
corresponding results shown for the untilted case.

\setlength{\unitlength}{1cm}
\begin{figure} 
\begin{picture} (0,7)
\put(0,0){\includegraphics[width=8cm]{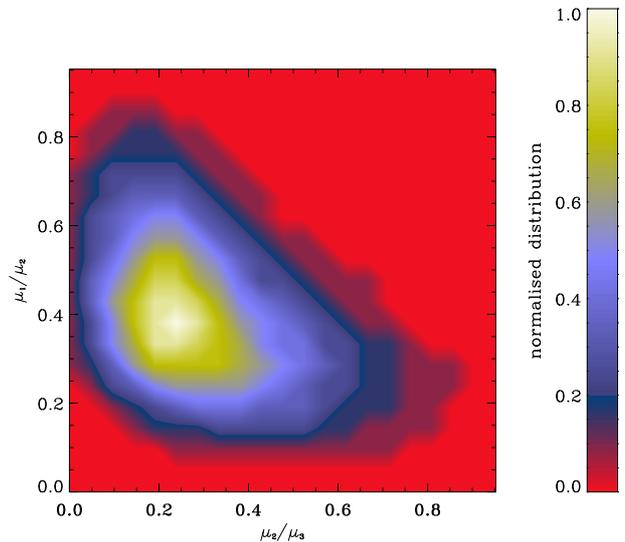}}
\end{picture}
\caption{Normalised bidimensional histogram displaying
$\mu_1/\mu_2$ as a function of $\mu_2/\mu_3$
for the  decaying 
turbulence simulation with inclined magnetic field at time 1.32 Myr.
 }
\label{triaxis_incl}
\end{figure}

\setlength{\unitlength}{1cm}
\begin{figure} 
\begin{picture} (0,4)
\put(0,0){\includegraphics[width=8cm]{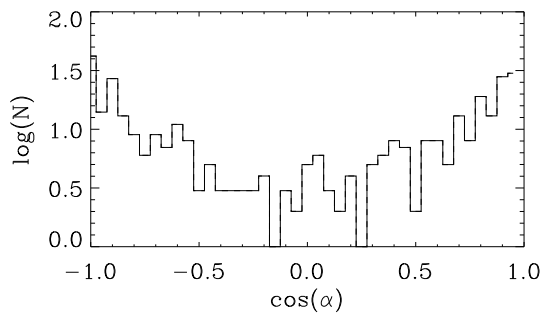}}
\end{picture}
\caption{ Distribution of 
 $\cos \alpha$ (the angle
between the main axis and the strain) in the clumps for the  decaying 
turbulence simulation with inclined magnetic field  at time 1.32 Myr.}
\label{cos_as_mhd_incl}
\end{figure}

\section{Results of colliding flow simulations}
\label{conv}
Here we present the results obtained for the colliding flow 
simulations. These simulations are described in Sect.~\ref{initial}. 
The purpose of this appendix is to demonstrate 
that the results obtained in the paper are not a consequence 
of a particular choice of initial and boundary conditions.
We use timesteps which represent a comparable  evolution. 
For the four runs, significant
masses of gas (typically 10$^4$ M$_\odot$ of gas denser than 100 cm$^{-3}$)
have been accumulated and at several places, collapse has proceed or 
is still proceeding.  
We restrict our attention to the most important quantities 
that have been studied namely the  clump aspect ratio 
computed using the inertia matrix and the skeleton-like approach, 
the length and thickness of the clumps, the cosine of 
the angle between the clump axis and the strain and the divergence
over strain ratio. In the six plots the solid line represents
the high resolution intermediate magnetization run, the 
dotted line is the standard resolution intermediate magnetization one,
the dashed line represents the highly magnetised run while 
the dot-dashed displays the hydrodynamical simulation.

\setlength{\unitlength}{1cm}
\begin{figure} 
\begin{picture} (0,4.5)
\put(0,0){\includegraphics[width=8cm]{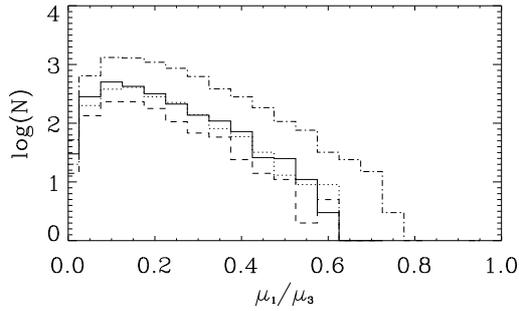}}
\end{picture}
\caption{Distribution of aspect ratio, $\mu_3/\mu_1$ of the clumps 
(threshold 500 cm$^{-3}$).
Solid line: high resolution intermediate magnetization run at time 12.94  Myr.
Dotted line: standard resolution intermediate magnetization run 
at time 16.82 Myr. Dashed line: highly magnetised run at time 18 Myr.
Dot-dashed line: hydrodynamical run at time 15.1 Myr. }
\label{aspect_ratio_colli1}
\end{figure}

\setlength{\unitlength}{1cm}
\begin{figure} 
\begin{picture} (0,4.5)
\put(0,0){\includegraphics[width=8cm]{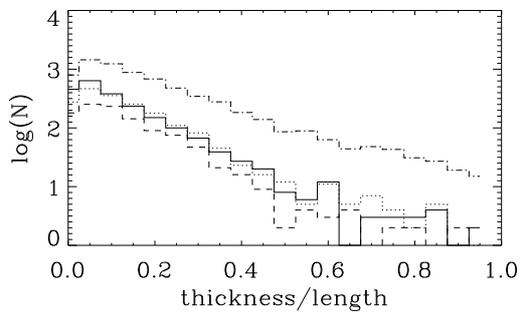}}
\end{picture}
\caption{Same as Fig.~\ref{aspect_ratio_colli1} for the
distribution of aspect ratio, $R/L$ of the clumps 
(threshold 500 cm$^{-3}$).}
\label{aspect_ratio_colli2}
\end{figure}

Figures~\ref{aspect_ratio_colli1} and~\ref{aspect_ratio_colli2}  
show very similar trends with Figs.~\ref{aspect_ratio_hydro_decay1},
~\ref{aspect_ratio_mhd_decay1},~\ref{aspect_ratio_hydro_decay2} and
~\ref{aspect_ratio_mhd_decay2}. In particular, in the hydrodynamical 
simulation the clumps have larger aspect ratios than in the 
MHD simulations.

\setlength{\unitlength}{1cm}
\begin{figure} 
\begin{picture} (0,4.5)
\put(0,0){\includegraphics[width=8cm]{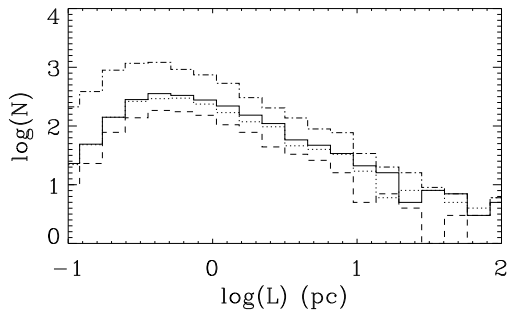}}
\end{picture}
\caption{Same as Fig.~\ref{aspect_ratio_colli1} for the distribution of the length of the clumps.}
\label{length_colli}
\end{figure}

\setlength{\unitlength}{1cm}
\begin{figure} 
\begin{picture} (0,4.5)
\put(0,0){\includegraphics[width=8cm]{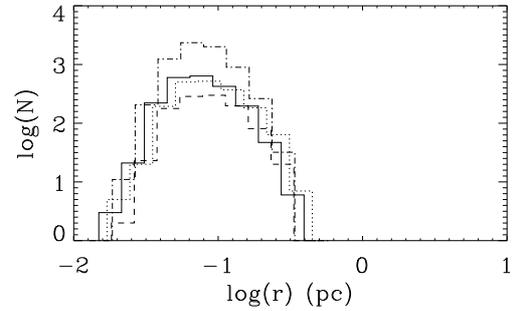}}
\end{picture}
\caption{Same as Fig.~\ref{aspect_ratio_colli1} for the distribution of the thickness of the clumps.}
\label{thickness_colli}
\end{figure}

The length and the thickness of the clumps are very 
similar to what has been inferred in the decaying simulations. 
The peaks are also located at the same position of about 
0.5 pc for the length and 0.1 pc for the thickness.

\setlength{\unitlength}{1cm}
\begin{figure} 
\begin{picture} (0,4.5)
\put(0,0){\includegraphics[width=8cm]{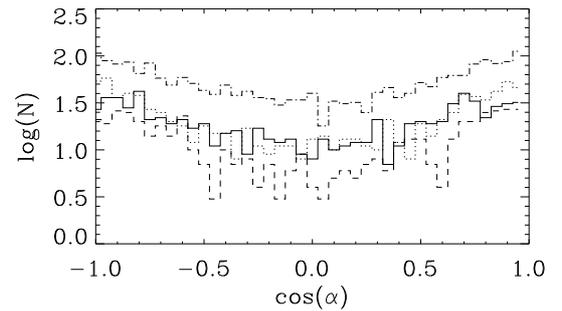}}
\end{picture}
\caption{Same as Fig.~\ref{aspect_ratio_colli1} for the distribution of 
 $\cos \alpha$ (the angle
between the main axis and the strain) in the clumps.}
\label{cos_as_colli_decay}
\end{figure}

In Fig.~\ref{cos_as_colli_decay} the trends of
the filament axis and the strain to be preferentially aligned
is also clear. As for the decaying turbulence simulations,  
it is more pronounced for the magnetised runs than 
for the hydrodynamical runs. 

\setlength{\unitlength}{1cm}
\begin{figure} 
\begin{picture} (0,4.5)
\put(0,0){\includegraphics[width=8cm]{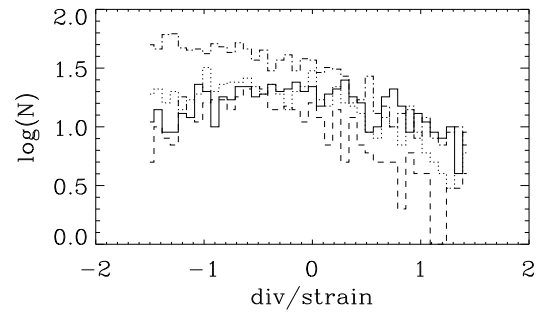}}
\end{picture}
\caption{Same as Fig.~\ref{aspect_ratio_colli1} for the distribution of 
the ratio of the divergence and strain in the clumps.}
\label{div_strain_colli_decay}
\end{figure}

Figure~\ref{div_strain_colli_decay} is also very similar to the trends
inferred in Figs.~\ref{div_strain_hydro_decay} and~\ref{div_strain_mhd_decay}
and discussed previously.

\end{document}